%% file: 3470.tex
\documentclass[twocolumn,longauth]{aa} 	

\usepackage{txfonts}
\usepackage{graphicx}
\usepackage{epsfig}
\usepackage{ulem}
\usepackage{longtable}

\begin{document}

   \title{The pulsation modes of the pre-white dwarf  PG~1159-035}

   \author{J.~E.~S.~Costa\inst{1}
           \and S.~O.~Kepler\inst{1}
           \and D.~E.~Winget\inst{2}
	   \and M.~S.~O'Brien\inst{47}
	   \and S.~D.~Kawaler\inst{4}
	   \and A.~F.~M.~Costa\inst{1}
	   \and O.~ Giovannini\inst{1,5}
	   \and A.~ Kanaan\inst{6}
	   \and A.~S.~Mukadam\inst{42}
	   \and F.~ Mullally\inst{2}
	   \and A.~ Nitta\inst{3}
	   \and J.~L.~Proven\c{c}al\inst{8}
	   \and H.~Shipman\inst{8}
           \and M.~A.~Wood\inst{9}
	   \and T.~J.~Ahrens\inst{9}
	   \and A.~Grauer\inst{10}
	   \and M.~Kilic\inst{41} 
           \and P.~A.~Bradley\inst{11}
	   \and K.~Sekiguchi\inst{12}
	   \and R.~Crowe\inst{13}
	   \and X.~J.~Jiang\inst{14}
	   \and D.~Sullivan\inst{15}
	   \and T.~Sullivan\inst{15}
	   \and R.~Rosen\inst{15}
	   \and J.~C.~Clemens\inst{16}
	   \and R.~Janulis\inst{17}
	   \and D.~O'Donoghue\inst{18}
	   \and W.~Ogloza\inst{19}
	   \and A.~Baran\inst{19}
	   \and R.~Silvotti\inst{20}
	   \and S.~Marinoni,\inst{21}
           \and G.~Vauclair\inst{22}
	   \and N.~Dolez\inst{22}
	   \and M.~Chevreton\inst{23}
	   \and S.~Dreizler\inst{24,25}
	   \and S.~Schuh\inst{24,25}
	   \and J.~Deetjen\inst{24}
	   \and T.~Nagel\inst{24}
	   \and J.-E.~Solheim\inst{26,27}
	   \and J.~M.~Gonzalez Perez\inst{26,28}
	   \and A.~Ulla\inst{29}
	   \and Martin Barstow\inst{30}
	   \and M.~Burleigh\inst{30}
	   \and S.~Good\inst{30}
	   \and T.S.~Metcalfe\inst{31}
	   \and S.-L.~Kim\inst{32}
	   \and H.~Lee\inst{32}
	   \and A.~Sergeev\inst{33}
	   \and M.C.~Akan\inst{34}
           \and \"{O}.~\c{C}ak{\i}rl{\i}\inst{34}
	   \and M.~Paparo\inst{35}
	   \and G.~Viraghalmy\inst{35}
	   \and B.~N.~Ashoka\inst{36}
	   \and G.~Handler\inst{37}
	   \and \"Ozlem  H\"urkal\inst{38}
	   \and F.~Johannessen\inst{26}  
           \and S.~J.~Kleinman\inst{3}
	   \and R.~Kalytis\inst{17}
	   \and J.~Krzesinski\inst{19}
	   \and E.~Klumpe\inst{39}
	   \and J.~Larrison\inst{39}
	   \and T.~Lawrence\inst{4}
	   \and E.~Mei\v{s}tas\inst{17}
	   \and P.~Martinez\inst{18}
           \and R.~E.~Nather\inst{2}
	   \and J.-N.~Fu\inst{48}
	   \and E.~Pak\v{s}tien\.e\inst{17}  
	   \and R.~Rosen\inst{26}
	   \and E.~Romero-Colmenero\inst{18}
	   \and R.~Riddle\inst{44}
           \and S.~Seetha\inst{37}
	   \and N.~M.~Silvestri\inst{42}
	   \and M.~Vu\v{c}kovi\'c\inst{4,43}
	   \and B.~Warner\inst{18}
	   \and S.~Zola\inst{40}
           \and L.~G.~Althaus\inst{45,46}
           \and A.~H.~C\'orsico\inst{45,46}
           \and M.~H.~Montgomery\inst{2}
          }

   \offprints{costajes@gmail.com}

   \institute{Instituto de F\'{\i}sica, Universidade Federal do Rio Grande do Sul, 
	91501-970 Porto Alegre, RS, Brazi --- \email{costajes@gmail.com} 
   \and Department of Astronomy \& McDonald Observatory, University of Texas, Austin, TX 78712, USA 
   \and Gemini Observatory, Hilo, Hawaii, 96720, USA 
   \and Department of Physics and Astronomy, Iowa State University, Ames, IA 50011, USA 
   \and Universidade de Caxias, Caxias do Sul, RS, Brazil 
   \and Departamento de F\'{\i}ica, Universidade Federal de Santa Catarina, CP 476, 
	CEP 88040-900, Florian\'opolis, SC, Brazil --- \email{kanaan@fsc.ufsc.br}
   \and Sloan Digital Sky Survey, Apache Pt. Observatory, PO Box 59, Sunspot, NM 88349, USA 
   \and Department of Physics and Astronomy, University of Delaware,  Newark, DE 19716, USA 
   \and Dept. of Physics and Space Sciences \& The SARA Observatory, 
	Florida Institute of Technology, Melbourne, FL 32901, USA 
   \and Department of Physics and Astronomy, University of Arkansas at Little Rock, USA
   \and Los Alamos National Laboratory, X-2, MS T-085 Los Alamos, NM 87545, USA 
   \and Subaru National Astronomical Observatory of Japan, Mitaka, Tokyo 181, Japan --- \email{kaz@saburu.naoj.org} 
   \and University of Hawaii, Hilo, Hawaii, USA 
   \and Beijing Astronomical Observatory, Academy of Sciences, Beijing 100080, 
	P.R. China --- \email{jiang@astro.as.utexas.edu}
   \and University of Victoria, Wellington, New Zealand 
   \and University of North Carolina, Chapell Hill, NC 27599, USA 
   \and VU Institute of Theoretical Physics and Astronomy, Go\v{s}tauto 12, 01108 Vilnius, Lithuania  
   \and South African Astronomical Observatory 
   \and Mt. Suhora Observatory, Cracow Pedagogical University, UI. Podchorazych 2, 
	30-084 Cracow, Poland 
   \and INAF - Osservatorio Astronomico di Capodimonte, Napoli, Italy
   \and INAF - Osservatorio Astronomico di Bologna, Bologna, Italy 
   \and Universit\'e Paul Sabatier, Observatoire Midi-Pyr\'en\'ees, CNRS/UMR5572, 
	14 av. E. Belin, 31400 Toulouse, France 
   \and Observatoire de Paris-Meudon, DAEC, 92195 Meudon, France ---
	\email{chevreton@obspm.fr} 
   \and Institut f\"ur Astrophysik,Friedrich-Hund-Platz 1, D-37077 G\"ottingen, Germany 
   \and Institut f\"ur Astronomie und Astrophysik T\"ubingen, Universitat T\"ubingen,
	Sand 1, D--72076 T\"ubingen, Germany 
   \and Institutt for Fysikk, Universitetet i Troms\o, N-9037 Troms\o, Norway --- 
          \email{j.e.solheim@astro.uio.no} 
   \and Institut for Teoretisk Astrofysikk, Universitetet i Oslo, 
	pb 1029-Blindern, N-0315 Oslo, Norway
   \and Instituto de Astrofisica, C/ Via Lactea s/n, E-38200 La Laguna, Tenerife, Spain
   \and Universidade de Vigo, Depto. de F\'{\i}sica Aplicada, Facultade de Ciencias do Mar, 
	Campus Lagoas-Marcosende s/n, 36200 Vigo, Spain
   \and Department of Physics and Astronomy, University of Leicester, UK 
   \and High Altitude Observatory, National Center for Atmospheric Research, 
        3080 Center Green Dr (CG1/3164), USA --- \email{travis@hao.ucar.edu}
   \and Korea Astronomy and Space Science Institute, Daejeon, 305-348, Korea 
   \and Terskol Observatory, Ukraine 
   \and Ege University Observatory, Bornova 35100, Izmir, Turkey
   \and Konkoly Observatory, P.O. Box 67, H-1525 Budapest, Hungary
   \and Indian Space Research Organization, India 
   \and Institut f\"ur Astronomie, Universit\"at Wien, Turkenschanzstrasse 17, A-1180 
	Wien, Austria 
   \and Ege  University  Faculty  of  Science Astronomy and Space Sciences Department, 
	Izmir, Turkey 
   \and Middle Tennessee State University, Department of Physics and Astronomy Murfreesboro, 
	TN 37132, USA 
   \and Astronomical Observatory, Jagiellonian University, ul. Orla 171, 30-244 Krakow, Poland --- \email{szola@oa.uj.edu.pl} 
   \and Department of Astronomy, The Ohio State University, 140 W. 18th Avenue, Columbus, OH 43210, USA ---
        \email{kilic@astronomy.ohio-state.edu}
   \and Department of Astronomy, University of Washington, Box 351580, Seattle, WA 98195, USA ---
        \email{nms@astro.washington.edu} 
   \and Instituut voor Sterrenkunde, Celestijnenlaan 200B, 3001 Leuven, Belgium ---
        \email{Maja.Vuckovic@ster.kuleuven.be}
   \and Thirty Meter Telescope Project, 2632 E. Washington Blvd, Pasadena, CA 91107, USA ---
         \email{email: riddle@tmt.org}  
   \and Facultad de Ciencias Astron\'omicas y Geof\'{\i}sicas, 
        Universidad Nacional de La Plata, Paseo del Bosque S/N, (1900) La Plata, Argentina
        --- \email{althaus@fcaglp.unlp.edu.ar}  --- \email{acorsico@fcaglp.unlp.edu.ar} 
   \and Instituto de Astrof\'{\i}sica La Plata, IALP, CONICET-UNLP, Argentina  
    \and Department of Astronomy, Yale University, New Haven, Connecticut,  CT 06851, USA ---
               \email{obrien@astro.yale.edu} 
    \and Department of Astronomy, Beijing Normal University, Beijing, P.R.China  
   }

   \date{2007 Aug 28 Received ---; accepted 2007 Oct 31}

   \abstract{}{}{}{}{} 

   \abstract
     { \object{PG~1159-035}, a pre-white dwarf with 
       $T_{\mathrm{eff}}\simeq 140\,000$~K,
       is the prototype of both two classes: the PG~1159 spectroscopic class and the DOV
       pulsating class.
       Previous studies of \object{PG~1159-035} photometric data obtained
       with the Whole Earth Telescope (WET) showed a rich frequency spectrum
       allowing the identification of 122 pulsation modes. 
       Analyzing the periods of pulsation,
       it is possible to measure the stellar mass, the rotational
       period and the inclination of the rotation axis, to estimate an upper 
       limit for the magnetic field, and even to obtain information about 
       the inner stratification of the star.
     }
     { 
	We have three principal aims: to increase the number of detected and identified pulsation modes
	in \object{PG~1159-035}, study trapping of the star's pulsation modes, and to improve or constrain
	the determination of stellar parameters.
     }
     {
       We used all available WET photometric data from
       1983, 1985, 1989, 1993 and 2002 
       to identify the pulsation periods.
     }
     { 
       We identified 76 additional pulsation modes, 
       increasing to 198 the number of known pulsation modes 
       in  \object{PG~1159-035}, the largest number of modes detected in any star besides the Sun.
       From the period spacing we estimated a mass $M/M_\odot = 0.59\pm 0.02$ for \object{PG~1159-035},
	with the uncertainty dominated by the models, not  the observation.
       Deviations in the regular period spacing suggest that 
       some of the pulsation modes are trapped,
       even though the star is a pre-white dwarf and the gravitational settling is ongoing.
       The position of the transition zone that causes the mode trapping
       was calculated at $r_c/R_\star = 0.83\pm 0.05$.
       From the multiplet splitting, we calculated the rotational period 
       $P_{\rm rot} = 1.3920\pm 0.0008$ days and an upper limit for the magnetic field, $B< 2000$ G.
       The total power of the pulsation modes at the stellar surface changed less than 30\%
       for $\ell=1$ modes and less than 50\% for $\ell=2$ modes.
       We find no evidence of linear combinations between the 198 pulsation mode frequencies.
       \object{PG~1159-035} models have not  significative convection zones,
       supporting  the hypothesis that nonlinearity arises in the
       convection zones in cooler pulsating white dwarf stars.
     }
     {}


   \authorrunning{Costa et al.}  

   \titlerunning{Pulsation modes of the star PG~1159-035} 

   \maketitle

\section{Introduction}   

The star \object{PG~1159-035} was identified by R. F. Green in 1977 in a survey for
objects with ultraviolet excess, known as the Palomar-Green Survey (Green et al.  1986).
The presence of lines of He II in the \object{PG~1159-035} spectrum suggested a high
superficial temperate (McGraw et al.  1979). The analysis of the far
ultraviolet flux distribution --- from $\sim 1200$~\AA\ to the Lyman limit
at $912$~\AA\ --- obtained with the {\it Voyager 2} ultraviolet spectrophotometer
indicated an effective temperature above $100\,000$ K 
(Wegner et al.  1982). Later analysis with the {\it IUE} and {\it EXOSAT} 
show that \object{PG~1159-035}  is one of the hottest stars known 
(Sion et al.  1985, Barstow et al.  1986); the current estimated temperature for \object{PG~1159-035} is
$140\,000\pm 5\,000$ K (Werner et al.  1991, Dreizler et al.  1998 and Jahn et al.  2007)
and $\log g(cgs) = 7.0\pm 0.5$ (Werner et al. 1991), 
placing it in the class of the pre-white dwarf stars.

McGraw et al.  (1979) 
discovered that \object{PG~1159-035} is a variable star and identified at least two
pulsation periods. The Fourier transform of more extensive light curves
obtained in the following years, between 1979-1985, allowed the detection
of eight pulsation modes (Winget et al. 1985);
the highest amplitude mode has a period of $516\,s$. 

A long light curve is necessary
to resolve two nearby frequencies in the  Fourier transform (FT) of a multiperiodic
pulsating star.  The Fourier transform resolution is
roughly proportional to the inverse of the light curve length.
For instance, if the difference between two frequencies is equal to 100 $\mu$Hz, 
just three hours of photometric data are needed to resolve them,
but, more than 10 days are needed  if the difference between them is of 1 $\mu$Hz.
On other hand, the presence of gaps in the light curve  introduces
in the FT an intricate structure of side-lobes, which may hinder
the detection and identification of real pulsation frequencies.

With the establishment of the WET ({\it Whole Earth Telescope}) 
in 1988 (Nather et al.  1990), 
\object{PG~1159-035} was observed for about 12 days with an effective 
coverage around $60\%$, resulting in a quasi-continuous 228 hours
of photometric data. The high-resolution Fourier transform of the light curve
allowed the detection and identification of 122 peaks (Winget et al.  1991).

In white dwarf and pre-white dwarf stars, gravity plays the role of the restoring
force in the oscillations. Any radial displacement of mass suffers 
the action of the gravitational
force causing the displaced portion of mass to be scattered inwards and sideways. 
This type of nonradial pulsation modes are called {\it g-modes}. 

General nonradial pulsations are characterized by three
integer numbers: $k$, $\ell$, $m$. The number $k$ is called
{\it radial index} 
and is related with the number of {\it nodes} in the
radial direction of the star. The number $\ell$ is called {\it index of the
spherical harmonic} (or {\it degree of the pulsation mode}). 
For nonradial modes, $\ell > 0$, while a radial pulsation has $\ell = 0$. 
In white dwarf stars the pulsations are dominated by temperature
variations (Robinson, Kepler, and Nather 1982). The index $\ell$ is related
with the total number of hotter and colder zones relative to the mean effective
temperature on the stellar surface.
Finally, the number $m$ is a number between $-\ell$ and $+\ell$
and is called the {\it azimuthal index}. 
The degeneracy of modes with different $m$ is broken when the spherical symmetry
is broken, for example, by rotation of the star, or the presence of magnetic fields.
The absolute value of the 
azimuthal index, $|m|$, is related with the way  the cold zones
and hot zones are arranged on the stellar surface. The sign
of the index $m$ indicates the direction of the temporal pulsation
propagation. We adopted the convention used by                   
Winget et al.  (1991): $m$ is positive if the pulsation and the
rotation have the same direction and negative if they have 
opposite directions. 

The \object{PG~1159-035} Fourier transform published in 1991 also revealed the presence of triplets
and multiplets, caused by {\it rotational splitting}, allowing the
determination of the rotational period of the star ($P_{rot}=1.38$ days).
Stellar rotation causes the g-modes with $m\ne 0$ to appears in the
FTs as frequencies slightly higher ($m>0$) 
or lower ($m<0$) than the frequency
of the $m=0$ mode, depending on whether the pulsation is travelling in
the same direction of the rotation of the star (higher frequency) or
in opposite direction (lower frequency). 
Slow rotation splits a mode in a multiplet of $2(\ell+1)$ peaks.
For  $\ell=1$ modes, the multiplets have three peaks 
(triplets) and for 
$\ell=2$ modes they have five peaks (quintuplets). 
But not necessarily all components are
seen in the FTs, because some of them might be excited with
amplitudes bellow the detection limit.
Besides stellar rotation, a weak magnetic field 
can also break the degeneracy and cause an observable splitting of the pulsation modes
into $(\ell+1)$ components in first order. However, no notable 
magnetic splitting has been observed in \object{PG~1159-035} (Winget et al. 1991).

The immediate goal of this work was to detect and identify a larger
number of pulsating modes in \object{PG~1159-035} from the analysis and
comparison of the FTs of photometric data obtained in different 
years. A consequence is the improvement in the determination
of the spacing between the periods of the pulsation modes used in
the calculation of the stellar mass and in the determination of the
inner stratification of the star. The analysis of the splitting in frequency
in the multiplets of the combined data allows the calculation of the
rotation period with higher accuracy and a better estimate of a upper
limit for the strength of the star's magnetic field. We are also
interested in the search for possible linear combination of frequencies,
as an indication of nonlinear behavior.

This paper is organized as follow:
in next Section we present some basic background in pulsation theory.
In \S3 and \S4 we discuss the observational data and the data reduction
process used in this work. In \S5 we discuss the detection of pulsation modes
from the \object{PG~1159-035} FTs. Then, in \S6, we present the calculation
of the period spacing for the detected pulsation modes. The mode identification,
i.e., the determination of the numbers $k$, $\ell$ and $m$ of the detected
pulsation modes is discussed in \S7 and  in \S8 \S9
we calculate the rotational and the magnetic splittings 
and the rotation period of the star. An 
estimate of the inclination angle of the rotational axis of the star
is done in \S10 and in \S11 we use the  splitting to obtain an
upper limit for the \object{PG~1159-035} magnetic field.
In \S12 we present  the estimate of the mass of \object{PG~1159-035}
from the period spacing in comparison with the masses calculated from
spectroscopic models. The analysis of a possible trapping of pulsation
modes in \object{PG~1159-035} is presented in \S13 and in \S14 we use the results
to calculate the position of a possible trapping zone inside the star.
In \S15 we comment on the absence of linear combination
of frequencies in \object{PG~1159-035} and in \S16 on the energy conservation
of the pulsation modes in the star surface. Finally, in \S17 we summarize our main
results.

\section{Some Background}   

The periods of g-modes for a given $\ell$ must increase monotonically with
the number of radial nodes, $k$. This occurs because the restoring
force is proportional to the displaced mass, which is smaller when the number
of radial nodes, $k$, is larger. 
For white dwarfs and pre-white dwarfs stars, 
a weaker restoring force implies in a longer period. 
One of the known methods to calculate the oscillation periods inside a
resonant cavity is the WKB (Wentzel-Kramers-Brillouin)  approximation, 
well know in Quantum Mechanics [see, for instance, Sakurai (1994)].
In the case of pulsating stars, this approximation is based on the hypothesis that
the wavelength of the radial wave is much smaller than the length scales
in which the relevant physical variables (density, for example) are
changing inside the star. This is approximately true for g-modes with
large values of $k$ ($k \gg \ell$). In this asymptotic limit, 
e.g. Kawaler et al.  (1985) 
the WKB
result approaches
a simple expression:
\begin{equation}
  P_{\ell, k} \simeq \frac{P_o}{\sqrt{\ell (\ell + 1)}} \, k + \epsilon
  \label{eq:A} 
\end{equation}
where, $P_{\ell k}$ is the period with index $\ell$ and $k$ and $P_o$ and
$\epsilon$ are constants (in seconds). The mean spacing between two
consecutive periods ($P_{\ell,{k+1}}-P_{\ell,k}$) of same $\ell$ is:
\begin{equation}
  \Delta P_\ell \simeq \frac{P_o}{\sqrt{\ell(\ell+1)}} 
  \label{eq:C} 
\end{equation}
The constant $P_o$ in Eq.\ref{eq:A} strongly depends on the stellar mass (Kawaler and Bradley 1994)
and, therefore, the determination of $\Delta {P_\ell}$ 
allows us to measure the mass of the star. 
On the other hand, the internal stratification of the star causes
the differences $P_{\ell,{k+1}}-P_{\ell,k}$ to have 
small deviations relative to the mean spacing, $\Delta {P_\ell}$. 
The analysis of these deviations can give us relevant information about 
the internal structure of the star.

\section{The Observational Data}  

\begin{table*}
	\caption{Observational campaigns of \object{PG~1159-035} between 1979-2002.}
	\label{table:1} 
	\centering
	\begin{tabular} {cccccccc} 
		\hline\hline
		Year &  Number    &    Length     & Hours of   & Effective & Overlapping & Spectral \\
     		&    of      &    (days)     & photometry & coverage  & rate        & resolution  \\ 
     		&  datum    &               &   (h)      &           &             &  ($\mu$Hz) \\ 
		\hline 
		1979 &  $523$     &     0.1       &   2.9      & 100.0\%   & ---         &   95.0      \\ 
		1980 &   $1\,294$ &     5.1       &   7.2      &   5.9\%   & ---         &    2.3      \\
		1983 &  $11\,758$ &    96.0       &  64.5      &   2.8\%   & ---         &    0.2      \\
		1984 &   $2\,658$ &     1.3       &  14.8      &  47.4\%   & ---         &    5.0      \\
		1985 &   $4\,380$ &    64.6       &  48.1      &   3.0\%   &     0.1\%   &    0.2      \\
		1989 &  $82\,471$ &    12.1       & 228.8      &  65.4\%   &    13.4\%   &    1.0      \\
		1990 &  $11\,690$ &     7.7       &  16.2      &   8.8\%   & ---         &    1.5      \\
		1993 & $248\,162$ &    16.9       & 345.2      &  64.3\%   &    20.8\%   &    0.7      \\
		2000 &  $14\,794$ &    10.3       &  24.5      &   9.2\%   &     0.7\%   &    1.1      \\
		2002 &  $33\,770$ &    14.8       & 116.5      &  27.7\%   &     5.1\%   &    0.8      \\ 
		\hline
	\end{tabular}
\end{table*}

\object{PG~1159-035} has been observed with time series photometry
at McDonald Observatory since 1979, soon after being identified
as a pulsating star by McGraw et al.  (1979). In 1983 the star was observed several times during
three months, revealing the presence of at least eight pulsation frequencies.
New observations were obtained in 1984 and 1985 (Winget et al. 1985), confirming the
persistence of the previously detected pulsation modes. 
Campaigns of quasi-continuous
observations were carried out with WET in 1989, 1990, 1993, 2000 and 2002;
however, in  1990, 2000 and 2002 \object{PG~1159-035} was observed as a 
secondary target.

Details about the observational campaigns are given in Table~\ref{table:1}. 
The {\it overlapping rate}, in column six, is the fraction  of time 
in which two telescopes carry out simultaneous observations of the star
causing an overlap of photometric measurements in the
total light curve. 
The {\it spectral resolution}, in the last column, 
is the approximate mean width of the 
frequency peaks (in $\mu$Hz) in the FT of the total light 
curve of each yearly data set. 
Logs and additional information about the observational campaigns 
are presented in Winget et al.  (1985), Winget et al.  (1991),
Bruvold (1993) and Costa et al.  (2003).

\section{Data Reduction}    

The reduction of the photometric data was based on the process described by
Nather et al.  (1990) and Kepler et al.  (1995), 
but with some additional care in the atmospheric correction.

Most of the observations were obtained with three channel photometers.
While one of the channels is used to observe the target star, another
channel observes a non-variable star used as comparison star, and
measurement of the adjacent sky are taken with the third channel. 
After discarding bad points in the light curve of the three channels, 
the measurements are calibrated and the
sky level is subtracted from the light curves of the two stars (target and comparison).
To correct by atmospheric extinction to first order, the light curve of the target star is
divided, point-by-point, by the light curve of the comparison star. 

\begin{figure}[ht]
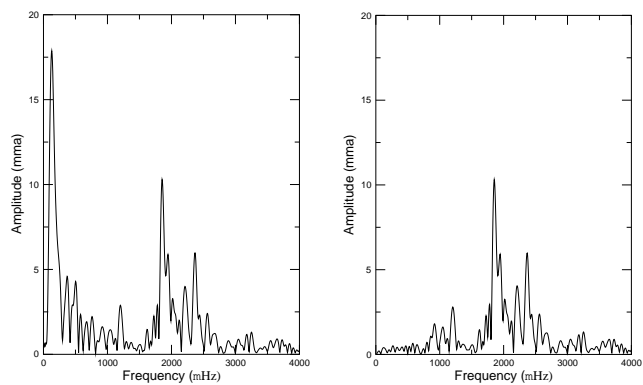
 
	\vspace{0.4cm}
	\centering
	\begin{tabular}{cc}
		\psfig{figure={3470fg1a.eps},width=4.0cm} &
		\psfig{figure={3470fg1b.eps},width=4.0cm} 
	\end{tabular} 
	\vspace{0.4cm}
	\caption{Left: Fourier transform of the light curve of an individual night
         	with peaks of low frequency and high amplitude. 
         	Right: Fourier transform of the same light curve after the use 
         	of a high-pass filter.} 
	\label{fig:1} 
\end{figure}

The most critical step in the data reduction is the atmospheric correction.  
During the night, the sky transparency changes on different timescales,
affecting the light curves of the two stars.
The division of the light curve of the target star by the light curve
of the comparison star does not completely eliminate the effect of atmospheric
extinction in the resulting light curve, because the atmospheric extinction effect
is dependent of the star color and in most of the cases the two stars
do not have the same color (\object{PG~1159-035} is blue). 
This implies that some residual signal due to the
atmospheric extinction remains in the resulting light curve, appearing
in the FTs of the individual nights as one or more peaks of low frequency
 ($f<300\,\mu$Hz) and relative high amplitude,  as shown in the left graph
in Fig.~\ref{fig:1} (see also Breger and Handler 1993). 

 We performed numerical simulations to study the effect of signals
with low frequency and high amplitude (LFHA) on the determination of the 
parameters of the pulsation modes (frequency, amplitude and phase). 
Our results show that the LFHA can introduce {\it significant}  
errors in the determinations  of the frequencies, amplitudes and phases
of the pulsation modes. For pulsating stars, as \object{PG~1159-035}, with many 
pulsation modes with low amplitudes ($A < 1\,mma$), this interference
can represent a serious problem.

To minimize this effect, we fitted a polynomial of $4^{\rm th}$ order 
to the light curve of each individual night, but even so,
 residual frequencies with considerable amplitude persisted in 
the residual light curve. To eliminate them, we used a {\it high-pass filter},
an algorithm that detects and eliminates signals with high amplitudes and 
frequencies lower than $300\,\mu$Hz, as illustrated in  Fig.~\ref{fig:1}.
Note that the limit of $300\,\mu$Hz is far less  than our
frequency range of interest, $1000-3000\,\mu$Hz, where we see the pulsation
modes. We note that {\it all} signals with frequencies lower than $300\,\mu$Hz, even if
they are present in the star, are eliminated.

\section{Detection of Pulsating Periods}  

\begin{figure}
  \centering
  \begin{tabular}{cc}
  	\epsfig{figure={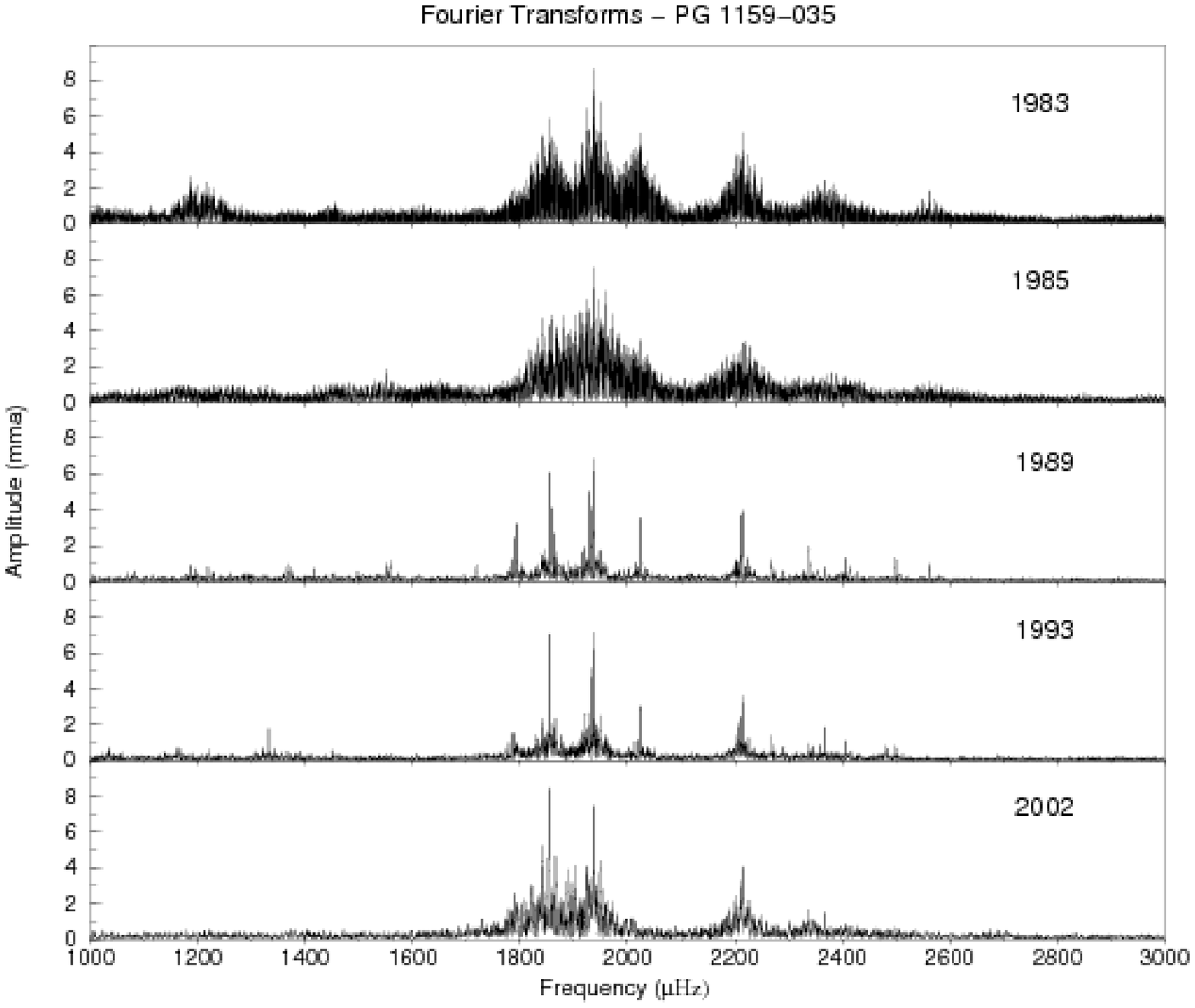},height=10cm,width=6cm} &
  	\epsfig{figure={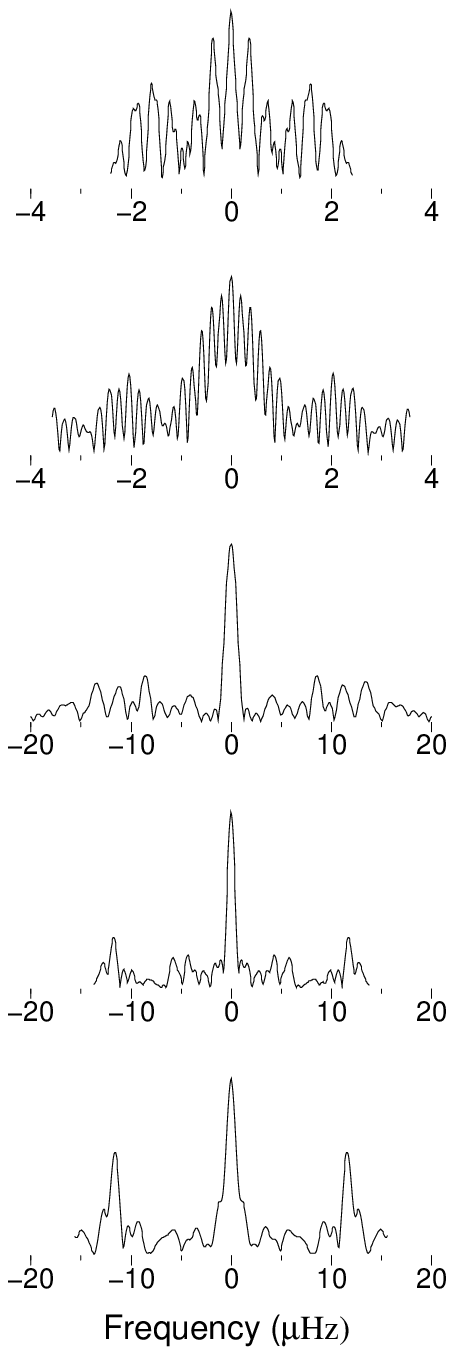},height=9cm,width=1.5cm} 
  \end{tabular} 
  \caption{Periodograms of \object{PG~1159-035} of the years of 1983, 1985, 1989, 1993
           and 2002. The respective spectral windows are shown on the right.}
  \label{fig:2} 
\end{figure}

Figure \ref{fig:2} shows the FTs for each one of the 
annual light curves of \object{PG~1159-035} for the frequency range of
interest ($1000-3000\,\mu$Hz). 
Frequency is in $\mu$Hz and amplitude is in units of 
$mma$ (milli-modulation amplitude). 
The respective spectral  windows are  on the right side, with the
same scale in amplitude, but different scale in frequency.

To find pulsation frequencies, we used an  iterative approach: 
(0) starting with an empty list of candidate frequencies
and with the FT of the original light curve; 
(1) identify, inside the
range of interest in the FT, the peaks with amplitudes
above the detection limit (taking care to discard aliases).
If there is no peak above the detection limit, the algorithm stops.
(2) Put the detected frequencies in the list of candidate frequencies, and
(3) using a nonlinear method, fit sinusoidal curves using {\it all}
frequencies from the list based on the original light curve.
The fitting refines the values of the initial frequencies and
calculates their amplitudes and phases. 
(4) The fitted sinusoidals are subtracted
from the original light curve and the FT of the
residual light curve is calculated. Then, the algorithm
returns to the step (1) to search for other possible pulsation 
frequencies.

Usually, the detection limit is based on the
local average amplitude of the peaks in the  FT, $\bar{A}$.
Kepler (1993) and Schwarzenberg-Czerny (1991, 1999),  following Scargle (1982), 
demonstrated that non-equally spaced data sets of multiperiodic 
light curves do not follow a normal noise distribution, 
because the residuals are correlated. 
They conclude that the probability of a peak
in the FT above $4\,\bar{A} $ 
has a $1/1000$ chance of being due
to noise (therefore, not a real signal) for a large frequency range of
interest (see also Breger and Handler 1993 and Kuschnig et al. 1997 
for similar estimates).

The comparison of the FTs of the light curves of
the different years shows that a mode can appear with an amplitude
above the limit of $4\,\bar{A}$ in one FT and
have a amplitude below this limit in the
FT of another year. To detect a larger number of 
pulsation modes we used a lower detection limit.
The presence of a same peak in different FTs reinforces
the probability of it being a real pulsation mode.

A lower detection limit was empirically estimated from the
following Monte Carlo simulation: 
(1) the light curve is randomized and (2)
its FT is calculated for the frequency range of
interest. (3) The highest peak in the FT, $A_{\rm max}$ 
is found and computed. The sequence above is repeated 1000 times
and (4) the average amplitude for the higher peak, $\langle {A_{\rm max}} \rangle$
and its standard deviation, $\sigma$ are calculated.
Then, (5) the detection limit is defined as 
$A_{\rm detect} = \langle {A_{\rm max}}\rangle + 3.5 \sigma $.
In all our cases, the factor $3.5 \sigma$ is
$\sim 20\%$ of $\langle {A_{\rm max}}\rangle$, therefore, 
$A_{\rm detect} = 1.2 \, \langle {A_{\rm max}} \rangle$.

This way to define the detection limit doesn't take into account
that the real noise is not  white, {\it but} the calculation
uses the same temporal sampling of the original light curve and the
same frequency range used in the frequency analysis.

\begin{figure} 
\centerline{\epsfig{figure={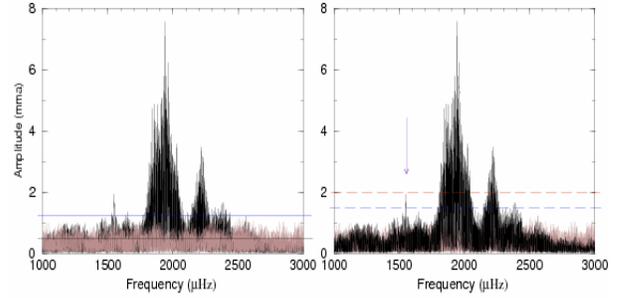}, width=8cm, height=4cm}}
\caption{Periodogram of the 1985 data set of \object{PG~1159-035}
(dark) and the FT of the same data set randomized (light).
Note that the randomization destroys all signals. The horizontal solid lines in
the {\it left} graph show the levels: 
$\bar{A}$ (upper) and ${\bar{A}}_m$ (lower);
while the graph {\it right} show the detection limits of 
$A_1 = 4\bar{A}$ (upper) and $A_0$ (lower).
The arrow shows a peak that is not detected when the limit of 
$A_1$ is used.}
\label{fig:3}
\end{figure}

\begin{figure}  
	\centering
	\epsfig{figure={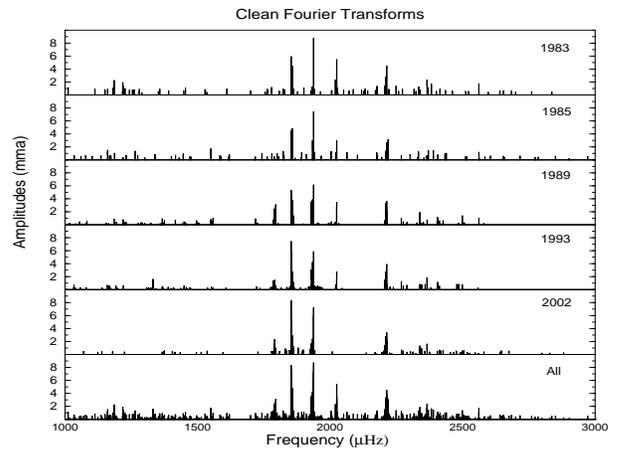},width=8cm,height=6cm} 
	\caption{The clean Fourier transform  for each annual light curve of \object{PG~1159-035}.
         	The bottom graph is a merge of the clean Fourier transforms of all years.}
	\label{fig:4} 
\end{figure}

We classified the peaks of each FT into the six probability levels
listed in Table~\ref{table:2}. Initially, we selected all peaks with probability
levels 1-4. Of course, with the inclusion of peaks with lower probability levels
the chance of including false pulsation frequencies increases, but we hope to be
able to discard the major part of them analyzing their places
into multiplets, as discussed in the next sections.
Figure~\ref{fig:4} shows the ``clean Fourier transforms'' for each year, 
with only the selected peaks. The bottom FT is a merger of all of them. 
The detected pulsation periods are listed in 
Table~13, Table~14, Table~15, Table~16 and Table~17.               
The {\it time of maximum}
($T_{\rm max}$) in the tables' last column is an instant when the pulsation reaches a maximum
in amplitude. The times of maximum are computed in seconds from the BCT (Barycentric Coordinate Time)
date $T_o$, given in the tables' caption.

The comparison of the clean FTs shows that most of the peaks with
high amplitudes are persistent, appearing in all five FTs, but in
all cases their amplitudes change, even taking into account their uncertainties (see Fig.~\ref{fig:9}).  
This shows that the amplitude of the pulsations
modes are changing with time and sometimes their amplitude decrease below the 
detection limits. For this reason,
to identify a large number of pulsation modes,
it is necessary compare the clean FTs of several years.

\begin{table}
	\caption{Confidence levels of the found peaks.} 
	\label{table:2}
	\centering
	\scriptsize
	\begin{tabular}{lp{5cm}} 
		\hline \hline
		Level   & Description \\
		\hline
		1 & 	Peak with amplitude $A\ge 4\bar{A}$ and appearing
	          	in one or more of the FTs \\
		2 & 	Peak with amplitude $4\bar{A} >A  \ge A_{\rm detect}$ 
	          	and appearing in two or more FTs \\
		3 & 	Peak with amplitude $4\bar{A} > A \ge A_{\rm detect}$,
		  	but appearing only in one of the FTs \\
		4 & 	Peak with amplitude $ A < A_{\rm detect}$, but appearing
		  	in two or more FTs with an amplitude greater
		  	than the nearest peaks. \\
		5 & 	Peak with amplitude $ A < A_{\rm detect}$, appearing
		  	in only one of the FTs with an amplitude greater
		  	than the nearest peaks. \\
		6 & 	Peak with amplitude $ A < A_{\rm detect}$ in all FTs,
		  	with amplitudes not higher than the amplitude of
		  	the nearest peaks. \\ 
		\hline
	\end{tabular} 
\end{table}

\section{The Period Spacing }  
\label{sect:6}

The way  the spacing in period is calculated and 
the identification of the pulsating mode is done follows a classical circular argument:
first, we assume an initial period spacing for the ($\ell=1$) modes, then we look for periods in the
overlapped clean FT that fit it, then the period spacing is again calculated
refining its initial value. 
The found periods are assumed be ($\ell=1$, $m=0$) modes and then we look for
the lateral components ($\ell=1$, $m=\pm 1$)  of the triplets, consistent with the
expected spacing caused by the rotational splitting. The peaks corresponding to identified
modes are removed from the clean FTs and then we apply the same process to the remaining
peaks to looking for ($\ell=2$) pulsation modes. The remaining peaks that are not identified 
either as ($\ell=1$) or as ($\ell=2$) pulsation modes are discarded (in all cases, these peaks
had low probability levels). Then, we look for peaks with lower probability
levels (5 or 6) in the original FTs that fit the absent expected frequencies.

An initial value for the period spacings or, $\Delta P_\ell$, can be calculated from
the  Kolmogorov-Smirnov (K-S) test.  Kawaler (1988) used the K-S test
to study the spacing in the first eight periods
detected by Winget et al.  (1985) in the \object{PG~1159-035} data. 
Later, Winget et al.  (1991) also used the K-S test to estimate the mean spacing
between the 122 detected periods in the 1989 WET data and found
$\Delta P_1 = 21.50\pm 0.03\,s$ (for $\ell=1$ modes), 
and $\Delta P_2 = 12.67\pm 0.03\,s$ (for $\ell=2$ modes). 

\begin{figure}
	\centerline{ 
	\epsfig{figure={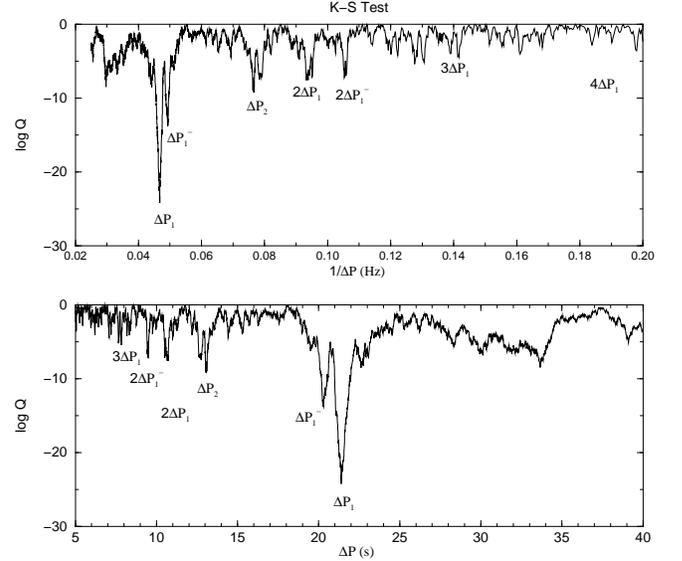},width=8.6cm} 
	} 
	\caption{The K-S applied to list of candidate  pulsation periods.}
	\label{fig:5}
\end{figure}

We applied the K-S test to our list of candidate pulsation periods. 
The result is shown in Fig.~\ref{fig:5}.
The upper graph shows the confidence level (log Q) versus $1/\Delta P$,
making the identification of harmonics in period spacings easier. 
The lower graph shows (log Q) versus $\Delta P$ (in seconds). 
The spacing $\Delta P_1$ is $21.39\,s$, 
while $\Delta P_2$ is $13.06\,s$ (see Table~\ref{table:3}).
The ratio between the two values is $\Delta P_1 / \Delta P_2 \simeq 1.638$, 
close to the expected $\sqrt{3}$. 
The difference around $5\%$ is due, mainly, 
to the overlapping of the two sequences ($\ell=1$ and $\ell=2$).

\begin{table} 
	\caption{Spacing found by the K-S Test.} 
	\label{table:3} 
	\centering
	\begin{tabular} {lccr} 
		\hline \hline
		Spacing            &  (s)    &  (Hz)   & $\log Q$ \\ 
		\hline
		$\Delta P_1$       &  21.39  &  0.047  & $-21.4$  \\
		$\Delta P_1^-$     &  20.29  &  0.050  & $-20.3$  \\
		$\Delta P_1^{--}$  &  19.59  &  0.051  & $- 3.2$  \\ 
		$\Delta P_1^{+}$   &  22.65  &  0.044  & $- 8.1$  \\ 
		$\Delta P_1^{++}$  &  23.06  &  0.043  & $- 6.8$  \\ 
		\hline
		$\Delta P_2$       &  13.06  &  0.077  & $-9.1$   \\
		$\Delta P_2^-\;$ ? &  12.80  &  0.078  & $-7.4$   \\ 
		\hline
	\end{tabular}
\end{table}

\begin{figure} [ht]
	\centerline{ \epsfig{figure={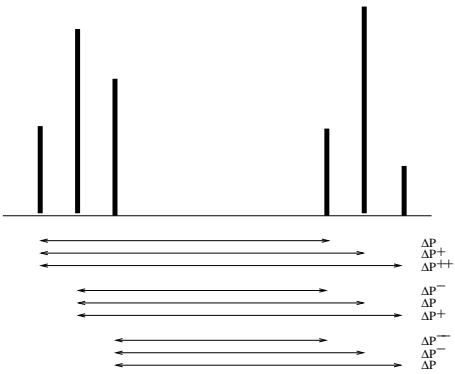},width=6cm} } 
	\caption{Possible spacings between the components of two triplets.}
	\label{fig:6}
\end{figure}

An explanation for the structures of valleys (minima) revealed in the
K-S test of Fig.~\ref{fig:5} is illustrated in Fig.~\ref{fig:6}, 
where we can see all possible spacings
between the peaks of two consecutive triplets (mode $\ell=1$). 
The spacing between peaks of same $m$ is $\Delta P$, but there are greater
spacings ($\Delta P^+$ and $\Delta P^{++}$) and shorter spacings ($\Delta P^-$
and $\Delta P^{--}$). We must take into account that not all the the triplets
frequencies are excited to detectable amplitudes.
This can explain the lower and asymmetrical valleys around the valleys of
$\Delta P_1$ and  $\Delta P_2$ in Fig.~\ref{fig:5} .

If the spacing between periods with same $\ell$ were exactly constant, the correct
values for $\Delta P_1$ and $\Delta P_2$ would appear as sharp valleys in the
K-S Test. The non-negligible width of the valleys indicate that
the spacing is not exactly constant, having a certain deviation around 
 $\Delta P_\ell$, as is theoretically expected and discussed in Sect.~\ref{sect:13}.

\section{Mode Identification}   

\begin{table*} 
	\caption{Identified $\ell=1$ pulsation modes.} 
	\label{table:4} 
	\centering
	{\scriptsize
	\input{3470tb4.tex} }
\end{table*}

In the FTs, the $\ell=1$ and $\ell=2$ sequences 
overlap. The identification of the periods with $\ell=1$ is easier and 
more secure, because the spacing between them is large and they appear as 
triplets and not as higher multiplets. 
We used  Eq.\ref{eq:A} and the rotation period of the star,
$P_{rot}=1.38$ days, found by Winget et al.  (1991),
to calculate the {\it approximate} position  of the peaks of the triplets.
The identification of the  $\ell=1$ pulsation modes is
done by comparing the peaks in the FTs with the predicted positions.
All peaks identified as $\ell=1$ modes are listed in Table~\ref{table:4}. 
For peaks present in more than one FT, the periods, frequencies and
amplitudes  in Table~\ref{table:4} are the average values.

\begin{figure}
	\centerline{\epsfig{figure={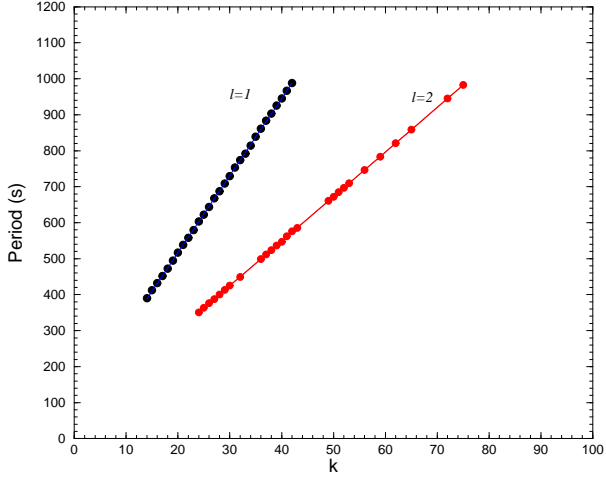},width=8.0cm}}
	\caption{Observed periods sequences for the modes $\ell=1$ and $\ell=2$.} 
	\label{fig:7} 
\end{figure}

We set the index $k$ of each triplet assuming $k=20$ for the triplet of 517 s,
as calculated by Winget et al.  (1991). Comparing
the observed period spacing for $\ell=1$ modes in the 1989 data set with models for 
pulsating PG1159 stars calculated by Kawaler and Bradley (1994) (hereinafter KB94),
Winget et al.  (1991) calculated that the triplet of 517 has index $k=20\pm 2$.
The plot of period versus $k$ is shown in Fig.~\ref{fig:7}. 
Fitting a straight line to the points, we can refine $\Delta P_1$ 
and calculate $\epsilon$ [Eq.~1]:
\begin{equation}
  \Delta P_1  =  21.43 \pm 0.03\,s
  \label{eq_kl2} 
\end{equation}

\begin{equation}
  \epsilon = 88.05 \pm 21.43\,s
  \label{eq_kl1} 
\end{equation}
where, $\epsilon$ is the period for $k=0$ (radial mode). Our result for
$\Delta P_1$ differs in $\sim 2\sigma$ from the value calculated by 
Winget et al. (1991).

\begin{figure}
	\centerline{ \epsfig{figure={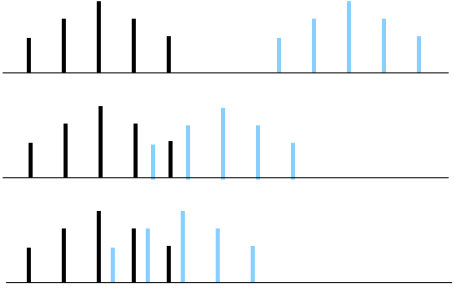},width=6cm} }
	\caption{Overlapping of two multiplets  ($\ell=2$). For periods less than 600 s,
 	 the overlappings do not occur. Between $600-750$ s, there is the overlapping of
 	 only a peak and up to 750 s, the overlapping of two peaks.} 
	\label{fig:8} 
\end{figure}

\begin{figure}[ht]
        \vspace{0.8cm}
	\epsfig{figure={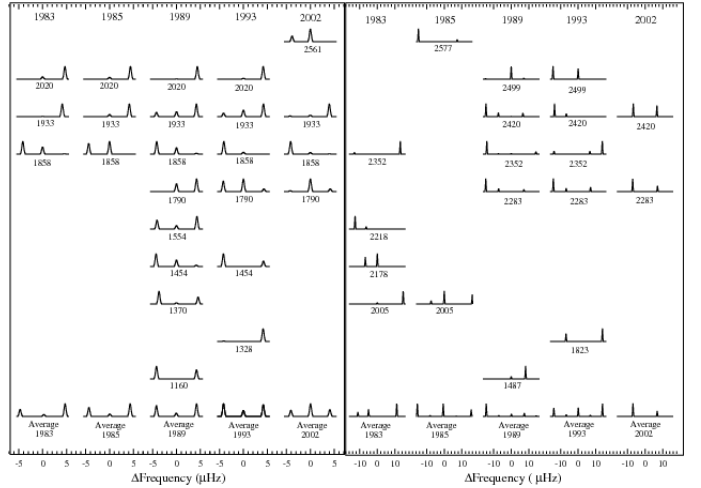}, width=9.0cm}    
	\caption{Representation of $\ell=1$ (left panel) and $\ell=2$ (right panel) 
  		detected multiplets for the indicated years in the top of each panel. 
  		All the multiplets have at least two components. The number bellow
  		each multiplet is its frequency in $\mu$Hz. At the bottom of each panel are the average
  		multiplets for each year. The heights of  each multiplet peaks are normalized by the
  		power of its highest peak.} 
	\label{fig:9} 
\end{figure}

Using the value for $\epsilon$ above
and the initial estimate for $\Delta P_2$ we started
the investigation of the $\ell=2$ sequence of pulsation modes. The indexes $k$
for each $\ell=2$ mode are calculated from Eq.~\ref{eq:A} with an
uncertainty of $\pm 2$.
The identified $\ell=2$ modes are in Table 12 and the sequence of $P$ as a function   
of $k$ is shown in Fig.~\ref{fig:7}. The new computed value for $\Delta P_2$ is:
\begin{equation}
  \Delta P_2 = 12.38\pm 0.01\,s
\end{equation}
differing in $\sim 9 \sigma$ from the value found by Winget et al.  (1991),
but the uncertainty in $\Delta P_2$ can be underestimated, as explained later in this section.
It is important
to note that if the true index for the 517 s triplet is $k\ne 20$, 
$\epsilon$ and the indexes $k$ for the $\ell=2$ sequence will need to be recalculated, 
but not $\Delta P_1$ and  $\Delta P_2$.
The ratio between the two period spacing is now closer to $\sqrt{3}$,
 the expected theoretical value,
\begin{equation}
  \frac{\Delta P_1}{\Delta P_2} = \frac{21.43}{12.38} \simeq 1.731 
\end{equation}
differing by less than $0.06\%$. 
After the mode identification from the combined data, we used the modes with $m=0$
to calculate $\Delta P_1$ and $\Delta P_2$ from each annual dataset. The results
are shown in Table 12. (Note that some lines with no data were omitted       
only to short the table.)

The amplitudes of most $\ell=2$ modes are very low and the absence of
the multiplet components hinders the identification of the azimuthal index,
$m$, of the other components.
An additional complication is the overlapping of multiplets (see Fig.~\ref{fig:8})
which is more serious when $k$ (and the period) increases. The multiplets with
periods less than $\sim 600\,s$ appear isolated in the FTs of \object{PG~1159-035}. 
But, between $\sim 600$ and $\sim 750\,s$,
the overlap of the more external components ($m=\pm 2$) occurs, and a component
of one multiplet ``invades'' the space of the neighbor multiplet, and vice-versa. 
From periods of $\sim 750\, s$, there are overlaps of two components 
($m=\pm 2$ and $m=\pm 1$). 
In the overlap regions, the mode identification is specially difficult and sometimes impossible,
which explains the lack of continuity of period sequence in the curve for the $\ell=2$ 
shown in Fig.~\ref{fig:7}. 
The overlap can also lead to the misidentification of the pulsation modes. For example,
an ``invader'' $m=-1$ or $m=-2$ peak can have a period near to the expected period for the
local $m=0$ mode, being identified as the $m=0$ mode of the local multiplet.
This can explain why we obtained a better fitting for the $\ell=2$ sequence than for the
$\ell=1$ sequence. In this case, the uncertainty in $\Delta P_2$ is underestimated.

The columns ``W91'' in Table~\ref{table:4} and Table 12                    
show the mode identification, $\ell,\,m$, as reported in Winget et al. (1991).
Colons (:) after $m$ indicate that other identification are possible. Modes in parenthesis,
indicate a possible alternative identification. The symbol \verb+?+ indicates when the
index $m$ is unknown. 
Most of the identifications obtained by us are the same as those by Winget et al.  (1991).

\section{The Splitting in Frequency}  

The observed splittings in frequency are caused by a combination of 
effects of the stellar rotation and the star's magnetic field. 
The magnetic splitting depends on the strength $B$ and the geometry of the magnetic field of the star
(Jones et al.  1989). For a symmetric magnetic field aligned to the pulsation symmetry axis, in first order,

\begin{equation}
  \delta\nu_{\rm mag}  \propto m^2 \gamma_k \, \, B^2
\end{equation}

\noindent
where $\gamma_k$ is a proportionality constant which depends on the internal structure
of the star,  on the index $k$ (and so, on the period), and on the shape of the magnetic field.
If the rotation is slow  ($P_{\rm rot} \gg P_{\rm puls}$) and if the rotation axis and the
pulsation symmetry axis are approximately aligned, the rotational splitting is given by (Hansen et al.  1977):

\begin{equation}
  \delta\nu_{\rm rot}  = m\,(1-C-C_1) \,\Omega_{\rm rot} \, + \, {\cal O}(\Omega_{\rm rot}^2)  + ...
	\label{eq:B}
\end{equation}

\noindent
where $C=C(k,\ell)$ is the uniform rotation coefficient while $C_1=C_1(k,\,\ell,\,|m|)$ 
contains the nonuniform rotation effects and depends on the adiabatic pulsation properties,
the equilibrium structure of the star, and the rotation law. In the asymptotic limit of high radial
overtones, i. e., large values of $k$ (Brickhill 1975),  $C \approx 1/ \ell\,(\ell+1)$; and,
if we assume uniform rotation as a first approximation, $C_1 = 0$. If the second order terms 
in Eq.\ref{eq:B} are neglected related to $\delta\nu_{\rm mag}$, then 
$\delta\nu_{\rm rot} \approx m\, \Omega_{\rm rot}$. 
While the rotation splits a g-mode in ($2\ell+1$) components, an aligned magnetic field 
splits it only in ($\ell+1$) components.

Figure~\ref{fig:9} shows  the triplets (left panel) and multiplets
(right panel) with at least two detected components 
found in the FT of each annual data set. 
Assuming that the above mentioned conditions are true
for \object{PG~1159-035}, the presence of multiplets with ($2\ell+1$)
peaks in its FT indicate that the rotational splitting is the dominant. 
To estimate $\delta \nu_{\rm rot}$ and $\delta\nu_{\rm mag}$ we calculated
the $\delta\nu$ spacing in frequency between consecutive peaks of the 
Fig.~\ref{fig:9} multiplets, and fitted  to

\begin{equation}
  \delta\nu \approx m\, \Omega_{\rm rot} + m^2 \bar{\gamma} \, B^2
\end{equation}
where $ \bar{\gamma}$ is the average of $\gamma_k$.
The spacings in frequency for the combined data $\ell=1$ modes  are
$\delta\nu_{\rm rot,1} =4.134\pm 0.002$ $\mu$Hz 
and $\delta\nu_{\rm mag,1} =0.007\pm 0.002$ $\mu$Hz. 
The contribution of the magnetic splitting is less than 1\%. 
For the $\ell=2$ modes we found $\delta\nu_{\rm rot,2} =6.90\pm 0.01$ $\mu$Hz. 
Unfortunately, the absence of peaks in the multiplets does not made it possible 
to estimate $\delta\nu_{\rm mag,2}$. 
Winget et al. (1991) analyzing only the \object{PG~1159-035} 1989 data set, found: 
$<\delta\nu_{\rm rot,1}>=4.22\pm 0.04$ $\mu$Hz 
and $<\delta\nu_{\rm rot,2}>=6.92\pm 0.07$ $\mu$Hz.
Table~\ref{table:5} shows the rotational spacing in frequency for each data set.

\section{The Rotational Period}  

For  uniform rotation and  asymptotic overtone limit in $k$, the rotation period
in the region of period formation, $P_{\rm{rot}}$, can be calculated from the frequency 
spacing (Kawaler et al. 1999) as
\begin{equation}
  P_{\rm{rot}} = \frac{1- [\ell\,(\ell + 1)\,]}{\delta\nu_{\rm rot,\ell}}
\end{equation} 

Calculating the $P_{\rm{rot}}$ mean value for $\ell=1$ and $\ell=2$,
Winget et al. (1991) obtained $P_{\rm{rot}} = 1.388\pm 0.013$ days. 
From our combined data, we obtained $P_{\rm{rot}} = 1.3930\pm 0.0008$ days
for $\ell=1$ and $P_{\rm{rot}} = 1.3973\pm 0.0022$ days. 
The two periods' average  is
$P_{\rm{rot}} = 1.3935\pm 0.0008$ days, consistent with the previous value, but with a 
significative larger accuracy. The rotational periods calculated for each data set are
shown in Table~\ref{table:5}.

\begin{table*}[ht] 
	\caption{Results for each annual data set.} 
	\label{table:5} 
	\centering
	{\tiny
	\input{3470tb5.tex} }
\end{table*}

\section{Inclination of the Rotational Axis}   

Theoretically, if the pulsational symmetry axis and the rotational axis are approximately
aligned and 
if the amplitudes of all the pulsation modes of a multiplet are the same,
the multiplets appear in the FT with a symmetrical design and the relative amplitudes
of the components depend on the inclination angle, $i$, of the rotational axis (Pesnell 1985).

\begin{figure}[ht] 
	\centering
	\vspace{0.5cm}
	\epsfig{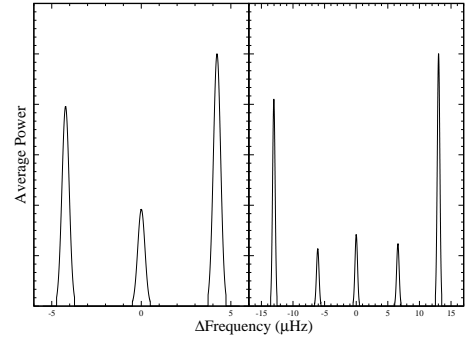}
	\vspace{0.5cm}
	\caption{Average multiplet for $\ell=1$ (left panel) and for $\ell=2$ (right panel) pulsation modes
  		of PG~1159-035.} 
	\label{fig:10} 
\end{figure}

As noted by Winget et al. (1991), the \object{PG~1159-035} multiplets do not have a symmetrical design and,
as shown in Fig.~\ref{fig:9}, the relative powers (and amplitude) of the multiplets components 
change in time, but the average multiplets, shown in the bottom of the
panels in Fig.~\ref{fig:9}, are approximately symmetrical relative to the central
peak. From the average multiplets for the 1989 data set, Winget et al. (1991) estimated
an inclination angle, $i\simeq 60^o$. In Fig.~\ref{fig:10} the mean multiplets
calculated from all the multiplets shown at Fig.~\ref{fig:9} are shown. The relative powers
of the peaks of the $\ell=1$ {\it and} $\ell=2$ average multiplets 
suggest a bit larger inclination angle, $i\simeq 70^o\pm 6^o$, but consistent with the
previous result.

\section{The Magnetic Field}   

If we assume any asymmetries in the splittings are due to magnetic filed effects,
we can estimate an upper limit to the magnetic field.
From the asymmetric  in frequency splitting within a multiplet,  $\delta\nu_{\rm mag}$, we are able
to estimate an upper limit to the strength of the magnetic field, $B$, 
since $\delta\nu_{\rm mag} \approx  m^2\, \gamma_k\, B^2$.
We calculated the proportionality constant by scaling the results of Jones et al. (1989)
for $\ell=1$ modes (see their Fig.1), and obtained an upper limit $B < 2000$G, with a
average value $B\sim 1200$G. Our upper limit for $B$ is three times less than the one
found by Winget et. al. (1989), of $6000$ G. 
Vauclair et al. (2002) found the limit $B<500$ for another PG~1159 star,
the hot \object{RXJ~2117+3212}. 
As observed by Vauclair too, the estimates of the upper limit for $B$
are taken from the calculations for a pure carbon white dwarf star by Jones et al. (1989),
and it can only be an approximated value when scaled to \object{PG~1159-035}.

\section{Mass Determination}  

The mass is the stellar parameter with largest impact on the internal structure
and evolution of the stars. However, with exception of a small fraction of stars belonging
to binary systems, the mass cannot be obtained by direct observation. 
The mass of (pre-)white dwarf stars can be spectroscopically estimated, from the comparison
of the observed spectrum with theoretical spectra predicted by stellar atmospheric models
(spectroscopic mass). For the pulsating stars, the mass can also be asteroseismologically
derived by way of the comparison of the observed spacing in the star's pulsation periods
and the ones predicted by pulsation models (seismic mass).

\subsection{KB94 Parameterization}

From $\Delta P_1$ and  $\Delta P_2$ derived in \ref{sect:6}, the proportionality constant 
$P_o$ in Eq.\ref{eq:C} may be calculated:
\begin{equation}
  P_o = \sqrt{\ell (\ell+1)}\, \Delta P_\ell 
\end{equation} 
For $\ell=1$, $P_o=30.31\pm 0.04$~s and for $\ell=2$, $P_o = 30.32\pm 0.03$~s.
The weighted average is:
\begin{equation}
  P_o = 30.32\pm 0.03\quad (s)
\end{equation}
The previous result (KB94) is $P_o = 30.5\pm 3.0$.
The constant $P_o$ depends on the internal structure of the star (see, e.g., Shibahashi 1988):
\begin{equation}
  P_o = 2\pi^2 \left( \int \frac{N(r)}{r} dr \right)^{-1} 
\end{equation}
where, $N(r)$ is the {\it Brunt-V\"ais\"al\"a frequency} and the integration is done 
over all the region of propagation of the g-modes inside the star.
From the parameterization 
of a grid of models,
KB94 found an expression for $P_o$ 
as a function of three stellar parameters,
the stellar mass (in $M_\odot$), $M$; 
the luminosity (in $L_\odot$), $L$; 
and the  fractional mass of the helium superficial layer, $q_{\scriptscriptstyle \mathrm Y}$:

\begin{equation}
  P_o = z\, \left(\frac{M}{M_\odot}\right)^a\, \left(\frac{L}{L_\odot}\right)^b\, q_{\scriptscriptstyle \mathrm Y}^c 
\label{eq:D}
\end{equation}
where, $z$, $a$, $b$ and $c$ are constants.
Knowing $P_o$, $L$, $q_{\scriptscriptstyle \mathrm Y}$ and the four constants above, the
stellar mass can be determined:
\begin{equation}
  \frac{M}{M_\odot} = \left(\frac{P_o}{z}\right)^{1/a} \,
				\left(\frac{L}{L_\odot}\right)^{-b/a}\,
				q_{\scriptscriptstyle \mathrm Y}^{-c/a}
\label{eq_mass}
\end{equation}
The general equation to estimate the uncertainty, $\sigma_{\scriptscriptstyle \mathrm  M}$, in the mass
determination is

\begin{eqnarray}
  \sigma_{\scriptscriptstyle \mathrm M}
	  & = & {M\over |a|} \,
                 \left[ 
                 \left( {\sigma_{\scriptscriptstyle \mathrm P_o} \over P_o} \right)^2 +
                 \left( b {\sigma_{\scriptscriptstyle \mathrm L} \over L} \right)^2   +
                 \left( c {\sigma_{\scriptscriptstyle \mathrm {q_{\scriptscriptstyle \mathrm Y}}} \over q_{\scriptscriptstyle \mathrm Y}} \right)^2 +
                 \left( {\sigma_{\scriptscriptstyle \mathrm z} \over z} \right)^2 +
                 \right.                                    \label{eq_sigmaMassa} \\
           & + & \left.
                 \left( \ln \left(M/M_\odot\right) \, \sigma_{\scriptscriptstyle \mathrm a}\right)^2 +
                 \left( \ln \left(L/L_\odot\right) \, \sigma_{\scriptscriptstyle \mathrm b}\right)^2 +
                 \left( \ln \left(q_{\scriptscriptstyle \mathrm Y}\right) \, \sigma_{\scriptscriptstyle \mathrm c}\right)^2
                 \right]^{1/2}                            \nonumber     
\end{eqnarray}
The equation above take into account the contribution of all 
parameters of Eq.\ref{eq:D}, but the last term is the dominant one
and all other terms can be neglected. Then,

\begin{equation}
  \sigma_{\scriptscriptstyle \mathrm M}  \simeq M \, \left| \frac{ln\,(M/M_\odot)\,}{a} \right|\, \sigma_{\scriptscriptstyle \mathrm a}
   \label{eq_sM} 
\end{equation}

For PG1159 stars, KB94 calculated $z=18.196$ sec, 
$a=-1.3$, $b=-0.035$ and $c=-0.00012$ with $L=(195\pm 3)\, L_\odot$, 
$q_{\scriptscriptstyle \mathrm Y}=0.039$
and obtained $M/M_\odot = 0.59\pm 0.01$ for \object{PG~1159-035}.
The uncertainty for $a$ was not published, but if we assume
that the $\sigma_{\scriptscriptstyle \mathrm a}$ is of the same order
of the last significant digit of $a$, $\sigma_{\scriptscriptstyle \mathrm a}\simeq 0.1$, and use our
result for $P_o$, we obtain $M/M_\odot = 0.59\pm 0.02$, while
Winget et al. (1991) found $M/M_\odot = 0.586\pm 0.003$. The difference
in the uncertainties for $M/M_\odot$ is probably because Winget assumed a
smaller value for $\sigma_{\scriptscriptstyle \mathrm a}$, in spite of our higher accuracy in 
the measured $\sigma_{\scriptscriptstyle \mathrm P_o}$.
The dominant uncertainty in the mass determination is the theoretical
models, not the observations.

\subsection{New Asteroseismological Models} 
\label{sect:12_2} 

C\'orsico  et al. (2006) performed an extensive
g-mode stability analysis on PG1159 evolutionary models, considering
the complete evolution of their progenitors, obtaining 
$M/M_\odot \simeq 0.536$ for \object{PG~1159-035}. They point out that for this
mass and at the effective temperature of \object{PG~1159-035}, their analysis predicts
that the model is pulsationally unstable, but with a period spacing of
$\Delta P_1 \simeq 23$s, which is in conflict with the observed $\Delta P_1 = 21.43$s.
To have a $\Delta P_1$ compatible with the observed one, the 
mass of \object{PG~1159-035} should be $M/M_\odot \simeq 0.558$, $1.6\sigma$ less than our 
result. They suggest that improvements
in the evolutionary codes for the thermally pulsing AGB phase and/or for the
helium burning stage and early AGB could help to alleviate the discrepancies
between the spectroscopic mass and the mass calculated from the period spacing.

Preliminary results of a detailed asteroseismological study on \object{PG~1159-035}
on the basis of an enlarged set of full PG1159 evolutionary models 
(C\'orsico  et al.   2007 in preparation) 
indicate that the PG~1159-035  stellar mass is either $\approx 0.585 M_{\odot}$  
(if the  star is  on  the rapid  contraction phase  before
reaching  its   maximum  effective  temperature)    or  $\approx  0.577 M_{\odot}$   
(if  the  star   has  just   settled  onto   its  cooling
track).  These inferences are  derived from  a comparison  between the
observed period spacing and  the asymptotic period spacing. This range
in mass is in agreement with the value of $M_\star \approx 0.59 M_{\odot}$
derived by Winget  al. (1991) and KB94 --- and  also in agreement with
the value derived in the  present paper from the KB94 parameterization
--- on the basis of an asymptotic analysis.

We must  emphasize, however, that  the derivation of the  stellar mass
using the  asymptotic period spacing  is not entirely reliable  in the
case of PG1159 stars.  This  is because the asymptotic predictions are
strictly valid for chemically homogeneous stellar models, while PG1159
stars  are  expected to  be  chemically  stratified, characterized  by
pronounced  chemical gradients  built  up during  the progenitor  star
life. A  more realistic approach to  infer the stellar  mass of PG1159
stars  is to  compare  the  average of  the  computed period  spacings
($\overline{\Delta  P_{\ell}}$) with the  observed period  spacing. To
this  end, we  computed  adiabatic nonradial  $g$-modes and  evaluated
$\overline{\Delta P_{\ell}}$ by  averaging the computed forward period
spacings ($\Delta P_{k}= P_{k+1}-  P_{k}$, $k$ being the radial order)
in the appropriate range of the observed periods in PG 1159$-$035.  At
the observed  effective temperature we  find two solutions for  the \object{PG~1159-035} 
stellar mass:  $\approx 0.586-0.587 M_{\odot}$ and $\approx
0.56-0.57   M_{\odot}$,  depending   on   its  location   on  the   HR
diagram\footnote{Note  that these values  are somewhat  different from
the mass derived in C\'orsico et  al. (2006) because in that paper the
authors used a different range of periods to compute $\overline{\Delta
P_{\ell}}$,   and  older values for the period  spacing.}. 
We can safely discard the solution before the knee $M_*
\approx 0.586-0.587 M_{\odot}$ because the predicted surface gravity
is much lower ($\log g \approx 6$) than the spectroscopically inferred
value $\log g=  7.0 \pm 0.5$. Thus, our best estimate  for the mass of
\object{PG~1159-035} is $\approx 0.56-0.57 M_{\odot}$. We postpone to a later
publication  (C\'orsico et  al. 2007)  a  detailed asteroseismological
study  based on a  fitting to  the individual  observed periods  in \object{PG~1159-035}.

\subsection{Spectroscopic Mass} 

The stellar mass can also be spectroscopically estimated from the comparison
of optical and/or UV spectra of the star with results of atmospheres models.
Using line blanketed NLTE model atmospheres, Dreizler and Heber (1998) found
$M/M_\odot = 0.54$ for \object{PG~1159-035}. The same result was obtained by
Miller Bertolami and Althaus (2006) from full evolutionary models for
post-AGB PG1159 stars. 

\begin{figure}[h]
	\label{fig:11}
	\centerline{\epsfig{figure={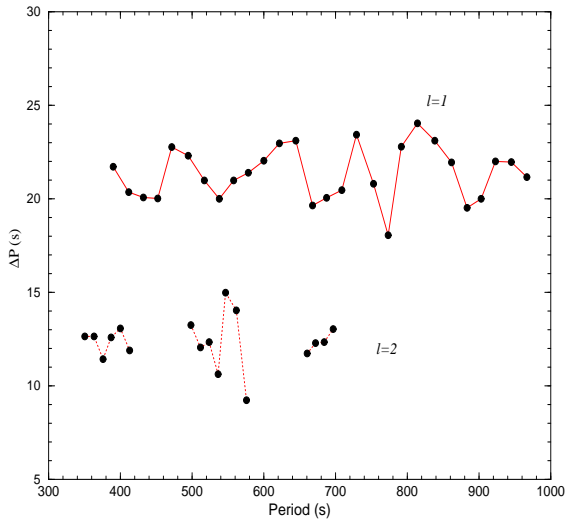},width=7.5cm,height=7cm}} 
	\caption{ $\Delta P$ diagram for \object{PG~1159-035}.}
\end{figure}

\section{Trapped Modes}  
\label{sect:13}

The asymptotic approximation of the Eq.\ref{eq:A} 
calculates with excellent precision the
pulsation periods for completely homogeneous models for PG1159 stars. 
However, the inner composition of pre-white dwarfs changes
between the core of C/O and the superficial layer, rich in helium. 
The regions where the gradient of the inner composition drastically changes 
are called {\it transition zones}. 
These zones can work as a reflection wall for certain pulsation modes, 
isolating them into specific regions of the star, as in a resonance box. 
This way, the modes can be ``trapped'' inside the core of C/O, as well as, 
between the stellar surface and the transition zone.

\begin{table} 
	\caption{Points in the $\Delta P$ diagram.}
	\label{table:6} 
	\centering
	\begin{minipage}{\textwidth} 
		{\scriptsize
		\begin{tabular}{lll|llr} 
			\hline \hline 
			\multicolumn{3}{c|}{$\ell=1$} & \multicolumn{3}{c}{$\ell=2$} \\ 
			\hline
			$k\pm 2$  &  $P$ (s) & $\Delta P$ (s)& $k\pm 2$  &  $P$ (s) & $\Delta P$ (s) \\ 
			\hline
			14 & 390.30   &  21.71 & 24 & 350.75   & 12.64 \\
			15 & 412.01 *\footnote{The asterisk * indicates unsure mode identification.} &  20.36 & 25 & 363.39   & 12.64 \\
			16 & 432.37 *  &  20.07 & 26 & 376.03   & 11.44 \\
			17 & 452.44   &  20.02 & 27 & 387.47   & 12.59 \\
			18 & 472.08 *  &  22.77 & 28 & 400.06   & 13.08 \\
			19 & 494.85   &  22.31 & 29 & 413.14 * & 11.90 \\
			20 & 517.16   &  20.98 & 36 & 498.73   & 13.25 \\
			21 & 538.14   &  20.00 & 37 & 511.98   & 12.05 \\
			22 & 558.14   &  20.98 & 38 & 524.03   & 12.34 \\
			23 & 578.61   &  21.39 & 39 & 536.37   & 10.63 \\
			24 & 600.00 *  &  22.03 & 40 & 547.00   & 14.99 \\
			25 & 622.03 * &  22.96 & 41 & 561.99   & 14.03 \\
			26 & 644.99   &  23.10 & 42 & 576.02   &  9.24 \\
			27 & 668.09   &  19.65 & 49 & 660.46   & 11.74 \\
			28 & 687.74   &  20.06 & 50 & 672.20   & 12.28 \\
			29 & 709.05 * &  20.46 & 51 & 684.48   & 12.35 \\
			30 & 729.51   &  23.43 & 52 & 696.83   & 13.04 \\ 
			31 & 752.94   &  20.80 &    &          &       \\
			32 & 773.74   &  18.06 &    &          &       \\
			33 & 791.80   &  22.78 &    &          &       \\
			34 & 814.58   &  24.04 &    &          &       \\
			35 & 838.62   &  23.10 &    &          &       \\
			36 & 861.72 * &  21.95 &    &          &       \\
			37 & 883.67   &  19.52 &    &          &       \\
			38 & 903.19 * &  20.00 &    &          &       \\
			39 & 923.19 * &  22.00 &    &          &       \\
			40 & 945.01   &  21.97 &    &          &       \\
			41 & 966.98   &  21.15 &    &          &       \\  
			\hline
		\end{tabular} 
		}
	\end{minipage}
\end{table}

\begin{table} 
	\caption{Trapped modes ($\ell=1$).} 
	\label{table:7} 
	\centering
	\begin{tabular}{cc} 
		\hline \hline
    		$i$ &  Period (s)              \\
     		\hline
     		1  &  $452.431\pm 0.002$   \\
     		2  &  $538.154\pm 0.003$   \\
     		3  &  $668.071\pm 0.024$   \\
    		 4  &  $773.744\pm 0.064$   \\
     		5  &  $883.637\pm 0.122$   \\ 
		\hline
  	\end{tabular} 
\end{table} 

The trapping of pulsation modes changes the spacing between consecutive
periods relative to the uniform spacing, $\Delta P_\ell$. 
The analysis of the deviations from the uniform period spacing
allows us to study the trapping of pulsation modes 
inside the star and obtain some relevant information about
its internal structure. The main tool in this analysis is the
$\Delta P$ diagram, a graph of the local spacing, 
$\Delta P = P_{k+1} - P_k$, versus the period $P_k$ of the $m=0$ modes.
Figure~11 shows the $\Delta P$ diagram for \object{PG~1159-035}
using the $m=0$ modes listed in Table~\ref{table:6}.  The $\ell=1$ sequence between
$k=14$ and $k=41$ is complete, while the $\ell=2$ sequence is incomplete.
Fortunately, the most significant information can be obtained from the 
$\ell=1$ sequence alone. 
The ($\ell=1$) trapped modes appears as points of minimum in the $\Delta P$ diagram 
(KB94) and are listed in Table~\ref{table:7}. We note that, with exception of the mode of
452s which has amplitude $\simeq 3$ mma, all other trapped modes have relatively low amplitudes
($\lesssim 0.6$ mma).

\section{Transition Zone}  

As the trapping of the pulsation modes depends on the resonance between the
eigenfunctions for the pulsations and the depth of the transition zone, the
periods for which the modes are trapped are sensitive to the geometric
depth, $r_c$, of the transition zone. 
The periods of the trapped modes can be calculated by the 
analytical approximation, originally proposed by Carl J. Hansen:
\begin{equation}
  P_i^2 \simeq 4 \pi^2 \lambda_i^2 \left[ \left( 1 - {r_c \over R_\star} \right) \,
    \,\ell(\ell+1)
     \, {G M \over {R_\star}^3} \right]^{-1} \,\,\, ,
\label{eq:E} 
\end{equation}
where, $\lambda_i$ are constants related with the zeros of the
Bessel functions; $r_c$ is the position of the transition zone; and $P_i$ is
the period of the trapped mode which has $i$ nodes between the stellar 
surface ($r=R_\star$) and the transition zone ($r=r_c$).

The constants $\lambda_i$ are called {\it trapping coefficients} and are empirically
calculated from models for white dwarfs and pre-white dwarfs by several groups.
Table~\ref{table:8} shows the coefficients $\lambda_i$ for $i=0-9$ for the models 
of PG~1159 calculated by Bradley and Winget (1991).

\begin{table} 
	\caption{Trapping coefficients for PG1159 models. } 
	\label{table:8}
	\centering
	\begin{minipage}{\textwidth}
		\begin{tabular}{cr} 
			\hline \hline 
			$i$ &   $\lambda_i\footnote{\footnotesize Source: Kawaler and Bradley (1990).}\quad \quad$      \\ 
			\hline   
			0   &   $ 3.33\pm 0.10$  \\ 
			1   &   $ 4.92\pm 0.16$  \\ 
			2   &   $ 6.15\pm 0.18$  \\ 
			3   &   $ 7.45\pm 0.16$  \\ 
			4   &   $ 8.76\pm 0.06$  \\ 
			5   &   $10.17\pm 0.18$  \\ 
			6   &   $11.46\pm 0.18$  \\ 
			7   &   $12.77\pm 0.19$  \\ 
			8   &   $13.82\pm 0.35$  \\ 
			9   &   $15.26\pm 0.26$  \\ \hline
		\end{tabular}
	\end{minipage}
\end{table}

From the Eq.\ref{eq:E}, the ratio between two periods, $P_i > P_j$, is given by:
\begin{equation}
  {P_i \over P_j} = {\lambda_i \over \lambda_j} \quad \mbox{with $i > j$ .}
\end{equation}
The first step is to identify the indexes $i$ and $j$. Table~\ref{table:9} shows all ratios for
$P_k/P_{k'}$, with $k > k'$, for the trapped modes of \object{PG~1159-035}, while Table~\ref{table:11}  
shows the ratios between $\lambda_i/\lambda_j$ with $i > j$. From the comparison of
the two tables, we found that the sequence of indexes that best fit is $i=$ 1, 2, 3, 4, 5.
We point out that the identification of the trapped indexes, $i$, 
(fortunately) does not  require exact identifications of the
radial indexes, $k$.

\begin{table}
	\caption{Ratios between the observed trapped periods.}
	\label{table:9}
	\centering
	\begin{tabular} {c|cccc} 
		\hline \hline 
        	&  538.14s &  668.09s  &  773.74s  &  883.67s  \\ 
		\hline
		452.44s  &  1.189   &   1.477  &   1.710  &   1.953  \\
		538.14s  &   ---    &   1.241  &   1.438  &   1.643  \\
		668.09s  &   ---    &   ---    &   1.158  &   1.323  \\
		773.74s  &   ---    &   ---    &   ---    &   1.142  \\ 
		\hline
	\end{tabular}
\end{table}

\begin{table}
	\caption{Values for $r_c/R_\star$ for each trapped mode of
	PG 1159-035 ($R/R_\odot = 0.025\pm 0.005$).} 
	\label{table:10}
	\centering
	\begin{tabular}{cccc} 
		\hline \hline
		$i$ &   $P_i$ (s)   &  $\lambda_i$  &  $r_c/R_\star$   \\ 
		\hline
 		1  &  452.431002   &     4.92      &  $0.84\pm 0.05$  \\
 		2  &  538.153994   &     6.15      &  $0.83\pm 0.06$  \\
 		3  &  668.070986   &     7.45      &  $0.83\pm 0.05$  \\
 		4  &  773.744021   &     8.76      &  $0.83\pm 0.02$  \\
 		5  &  883.635988   &    10.17      &  $0.82\pm 0.06$  \\  
		\hline
	\end{tabular}
\end{table}

\begin{table*}
\caption{Ratios $\lambda_i/\lambda_j$, with $i > j$.}    
\label{table:11}
\centering
\begin{minipage}{\textwidth}
\begin{tabular}{c|ccccccccc} 
            & $\lambda_1$ &$\lambda_2$ & $\lambda_3$ & $\lambda_4$ & $\lambda_5$ & $\lambda_6$ & $\lambda_7$ &  $\lambda_8$ & $\lambda_9$ \\ \hline
$\lambda_0$ &   1.477     &  1.847     &   2.237     &    2.630    &    3.054    &    3.441    &    3.835    &     4.150    &   4.583     \\ 
$\lambda_1$ &     ---     &\fbox{1.331}\footnote{The rations in the boxes are those ones than best fit to 
rations between the observed trapped periods, shown in  Table~\ref{table:9}.}
						  &\fbox{1.514} &\fbox{1.780} &\fbox{2.067} &\fbox{2.329} &    2.596    &     2.809    &   3.102     \\
$\lambda_2$ &     ---     &   ---      &\fbox{1.211} &\fbox{1.424} &\fbox{1.654} &\fbox{1.863} &    2.076    &     2.247    &   2.481     \\
$\lambda_3$ &     ---     &   ---      &   ---       &\fbox{1.176} &\fbox{1.365} &\fbox{1.538} &    1.714    &     1.855    &   2.048     \\
$\lambda_4$ &     ---     &   ---      &   ---       &    ---      &\fbox{1.161} &\fbox{1.308} &    1.458    &     1.578    &   1.742     \\ 
$\lambda_5$ &     ---     &   ---      &   ---       &    ---      &    ---      &\fbox{1.127} &    1.256    &     1.358    &   1.500     \\ 
$\lambda_6$ &     ---     &   ---      &   ---       &    ---      &    ---      &    ---      &    1.114    &     1.206    &   1.332     \\
$\lambda_7$ &     ---     &   ---      &   ---       &    ---      &    ---      &    ---      &    ---      &     1.802    &   1.195     \\
$\lambda_8$ &     ---     &   ---      &   ---       &    ---      &    ---      &    ---      &    ---      &     ---      &   1.104     \\ \hline 
\end{tabular}
\end{minipage} 
\end{table*}

For $\ell=1$, the expression in Eq.\ref{eq:E} becomes:
\begin{equation}
  P_i^2 = 2 \pi^2 \lambda_i^2 \left[ \left( 1 - {r_c \over R_\star} \right) \,
    {G M \over {R_\star}^3} \right]^{-1} \,\,\, ,
\label{eq_pi2b} 
\end{equation}
and we obtain:
\begin{equation}
  r_c/R_\star = \left( 1 - { {2\pi^2 \lambda_i^2}\over {G M P_i^2}}
        {R_\star}^3 \right) 
\label{eq_rc} 
\end{equation}
For positive values for $r_c/R_\star$, the radius of the star, $R_\star$, 
must be $R_\star < 0.045\, R_\odot$. Using $M/M_\odot=0.59$ and
assuming $R/R_\odot = 0.025\pm 0.005$
(KB94), we can calculate $r_c$ for each trapped mode
of \object{PG~1159-035} , as shown in Table~\ref{table:10}. All results
are concentrated around $0.83\, R_\star$, with small dispersion:
\begin{equation}
  r_c/R_\star = 0.83\pm  0.05 \;\;\; .
\end{equation}

Theoretical models calculated by Paul Bradley
for PG~1159 stars  derived from standard post-asymptotic giant branch
(AGB) stellar models  (see KB94 for a detailed description), and fitted to
\object(PG~1159-035) data with radio $R_\star = 0.026\,R_\odot$, 
calculate that the position of 
the transition zone between the core of C/O and the He layer 
is between 0.60 and 0.65~$R_\star$, 
which differ from our value by a factor of $\sim 4\sigma$.
This discrepancy suggest that the assumed
parameters in the models are not the best ones for 
\object{PG~1159-035}, if the trapping occurs at the He to C/O transition. 


\section{Linear Combinations of Frequencies}  

There is no evidence of linear frequency combination involving the almost
200 pulsation frequencies in \object{PG~1159-035}. 
The presence of peaks in the FT resulting of linear
combinations of frequencies indicates a nonlinear behavior. 
Linear combinations of frequencies
have been observed in DAVs and in DBVs. 
It is the case of the DAV star
\object{GD~154} (Pfeiffer et al. 1996) and the DBVs stars
\object{GD~358} (Vuille et al.  2000 and Kepler et al.  2003).

The physical causes of the nonlinear behavior are not  well understood. 
But, perhaps, PG~1159-059 is giving us a hint:
a difference between  PG1159 stars, and all other pulsating white dwarf 
stars is that DAVs and DBVs have convective layers, while PG1159 stars
do not have significant ones.
This fact is an indirect support to the hypothesis that the origin of the nonlinear behaviors
be the convection, as was proposed by Brickhill (1990, 1992), 
Goldreich and Wu (1999a, 1999b), Weidner and Koester (2003), and
Montgomery (2005).

\section{Power Conservation}  

The kinetic energy of oscillation, $E_{\rm kin}$, is defined by
\begin{eqnarray}
E_{\rm kin} 
    & = & \frac{\sigma^2}{2} \int_0^{M_\star} |\delta \vec{r}|^2 dM_r \\
    & = & \frac{\sigma^2}{2} I_{k \ell}
\end{eqnarray}
where $\sigma=2\pi \nu$ is the angular eigenfrequency, $\delta\vec{r}$
is the Lagrangian displacement vector, $I_{k \ell} \equiv
\int_0^{M_\star} |\delta \vec{r}|^2 dM_r$, and the relative radial
displacement is normalized to $\delta r/r=1$ at the stellar surface.
Since modes with the same surface amplitude can have very different 
$I_{k \ell}$ values, we can only calculate total kinetic energies if
we have a numerical model of the star with which to calculate these
quantities. 

If we only want to compare the \emph{surface} kinetic energy densities
of different modes, however, we can do a bit better than this.
Ignoring geometric factors throughout this derivation, we have 
$E_{\rm  kin} \propto \sigma^2 \xi_h^2$, where $\xi_h$ is the horizontal
displacement at the surface. From Robinson, Kepler, \& Nather (1982),
we find that
\begin{eqnarray}
\frac{\delta F}{F} \propto \frac{1}{\sigma^2} \frac{\delta r}{r}
\end{eqnarray}
Assuming the Cowling approximation and that the oscillations are
adiabatic at the surface, we have (e.g., Unno et al. 1989)
\begin{eqnarray}
\frac{\xi_h}{\delta r} \propto \frac{1}{\sigma^2},
\end{eqnarray}
which, combined with the previous result yields
\begin{eqnarray}
\frac{\delta F}{F} \propto \frac{\xi_h}{r}.
\end{eqnarray}
Thus, 
\begin{eqnarray}
E_{\rm kin} \propto \sigma^2 \xi_h^2 \propto \sigma^2 
\left(\frac{\delta F}{F}\right)^2
\end{eqnarray}
\begin{eqnarray}
\propto \nu^2 A^2.
\end{eqnarray}

The total surface kinetic energy of the oscillations is the sum of the
individual surface kinetic energies of all the pulsation modes:
\begin{equation}
E_{\rm kin} \propto \sum_{i} {\nu_i}^2 A_i^2
\end{equation}
where $\nu_i$ are the frequencies (in units of $\times 10^{-8} \mu$Hz)
and $A_i$ are the observed photometric mode amplitudes. The total
kinetic oscillation energy for the observed $\ell=1$ and $\ell=2$
pulsation modes in PG~1159-035 is shown in Table~\ref{table:5}, in
terms of $\times 10^{-8} \mu\rm{Hz}^2$. 
The arithmetic means of the kinetic oscillation energy for the 
$\ell=1$ and $\ell=2$ pulsation modes are $E_1 = 2.4\pm 0.4$ and 
$E_2=1.1\pm 0.4$. and $E_{\rm  tot}=E_1 + E_2 = 3.5\pm 0.6$.  
The deviations relative to the mean
value year to year are less than $30\%$ for $E_1$, less than $50\%$
for $E _2$ and less than $35\%$ for $E_{\rm tot}$. 
With exception of 1989 and 1993 (our best datasets) the deviations 
are larger than the uncertainties in $E_1$ and $E_2$, suggesting that
the total surface kinetic energies are not  conserved. 
For all data sets,
$E_1$ corresponds to $65 - 75\%$ of the detected modes total energy,
$E_{\rm tot}$, not correcting for geometrical effects.

\section{Conclusions and Comments}  

Winget et al. (1991), analyzing the WET 1989 data set of \object{PG~1159-035}, found 122 pulsation
modes, with frequencies between 1000 and 3000 $\mu$Hz in the star's FT,
with a spacing in period of $\Delta P_1=21.50 \pm 0.03\,$s and $\Delta P_2 = 12.67 \pm 0.03\,$s.
The seismological mass calculated from the spacing in period and using a theoretical
model for a PG~1159 star was $M/M_\odot = 0.586\pm 0.003$. The analysis of the fine structure
of the multiplets shown a spacing in frequency $\delta\nu_{\rm rot,1}=4.22\pm 0.04\mu$Hz
and   $\delta\nu_{\rm rot,2}=6.92\pm 0.07\mu$Hz for $\ell=1$ and $\ell=2$ modes, respectively,
which allowed the calculation of the star's rotation period, 
$P_{\rm rot}=1.38\pm 0.01$ days;
of the rotation axis inclination, $i\sim 60^o$; and to estimate an upper limit
for the magnetic strength, $B<6000\,$G. 

In this work, we followed the same steps of Winget, but using a larger number of data sets
and improved  data reduction and  data analysis techniques.
The combination of the Fourier transforms of the data sets from different years (1983, 1985, 1989,
1993 and 2002) allows us to refine the determination or put new constraints over several
stellar parameters of \object{PG~1159-035}. Our main results are:

\begin{enumerate}
  \item We identified 76 new pulsation modes, increasing to 198 the total number of known
        pulsation modes for \object{PG~1159-035}. Only 14 of them (all with $\ell=1$) are 
        present in the FTs of all the years, but with different amplitudes.
        The comparison of the annual FTs shows that the amplitudes of the pulsation
        modes are changing in time, and can reach amplitudes below our detection limits.
        No evidence of $\ell>2$ modes was found in the combined FTs of all the years.

  \item The period spacings are $\Delta P_1=21.43\pm 0.03\,s$ for $\ell=1$ modes and
        $\Delta P_2 = 12.38\pm 0.01\,s$ for $\ell=2$ modes. The period constant,
        $P_o=\sqrt{\ell (\ell+1)}\, \Delta P_\ell$, is $P_o=30.32\pm 0.03$ sec.

  \item We found a mass $M/M_\odot = 0.59\pm 0.02$ from the KB94 parameterization. 
        The apparent lower accuracy in the mass
        determination relative to the Winget determination,
        although we have substantially decreased the uncertainty in $P_o$,
        is because we took into account the dominant uncertainty 
        in the theoretical models. As pointed out by Winget, the stated error in their
        determination of the \object{PG~1159-035} mass reflects only the uncertainty in the measured period
        spacing, and not the systematic errors associated to the models. In this sense, 
        even though our result is formally less accurate, it is more realistic.

  \item Analyzing the spacing in frequency inside the multiplets we found that the
        splitting due to the stellar rotation is $\delta\nu_{\rm rot,1}=4.154\pm 0.002\mu$H\ for
        $\ell=1$ modes and $\delta\nu_{\rm rot,2}=6.90\pm 0.01\mu$Hz for $\ell=2$ modes.

  \item We also estimated that the splitting in frequency due to the magnetic field for
        $\ell=1$ modes is $\delta\nu_{\rm mag,1}=0.007\pm 0.002\mu$Hz, contributing with less
        than 1\% of the total splitting. Unfortunately, it was not possible to calculate 
        the magnetic spacing in frequency for the $\ell=2$ modes, due to the absence of several
        components in the multiplets.

  \item From the rotational  frequency splitting we calculated the rotational period 
         $P_{\rm rot} = 1.3920\pm 0.0008$ days. 

  \item The magnetic splitting in frequency suggests a upper limit for the magnetic strength
        lower than the previous estimates: $B\lesssim 2000\,$ G.

  \item The analysis of the fine structure of the combined data multiplets 
	($\ell=1$ {\it and} $\ell=2$) suggests
        that the inclination angle of the rotational axis is $i\sim 70^o\pm 6^o$.

  \item The $\Delta P$ diagram of \object{PG~1159-035} for $\ell=1$ modes suggests that
        \object{PG~1159-035} is already a stratified star. The $\Delta P$ diagram presents five minima 
        that can be interpreted as  the indication of trapped modes with periods of
        452.43s, 538.12s, 668.1s, 773.7s and 883.6s.

  \item For this sequence of trapped modes, we calculated the position of
        the transition zone that causes the trapping mode at 
        $r_c/R_\star = 0.83\pm 0.05$ for a star with $M/M_\odot = 0.59$ and 
        $R_\star/R_\odot = 0.025$. 

  \item There is no evidence of linear combinations of frequencies in \object{PG~1159-035}. 
        As models of PG 1159-035 do not have any significant convective layer, this provides indirect 
        support to the hypothesis that nonlinearity arises in the convection zone.

  \item Comparing the total power of the pulsation modes of the different years between
        1983 and 2002, we observe that the differences relative to the mean value are less 
        than $30\%$ for the $\ell=1$ modes and less than $50\%$ for the $\ell=2$ modes,
	indicating that the total surface kinetic energies are not conserved.  

\end{enumerate}

In continuation to the present work, we measured the temporal changing of the pulsation period 
($\dot P$) of several pulsation modes in \object{PG~1159-035}. As PG~1159-035 is a very hot
star, it is quickly evolving and its pulsation periods are changing in time. The period changes
are large enough to be directly measured and for some we can  derive the
second order temporal variation ($\ddot P$). The results are presented in a separated paper 
(Costa et al.  2007). 

We are currently performing a detailed asteroseismological study on \object{PG~1159-035} 
based on an enlarged set of full PG1159 evolutionary models. Preliminary results, obtained
from the comparison of the average period spacings of the models and of the observed periods,
suggest that the mass of PG~1159-035 is $\approx 0.56 - 0.57\,M_\odot$ (see Sect.\ref{sect:12_2}). 
The next step is to fit the models to the individual observed periods in PG~1159-035. 
This study will be published in a separated paper (C\'orsico et al.  2007 in preparation).

\begin{acknowledgements}
This work was partially supported by CNPq-Brazil, NSF-USA and MCyT-Spain.
In particular, M.~H. Montgomery was supported by NSF grant AST-0507639.
\end{acknowledgements}



\Online
\begin{appendix} 

\longtab{12}{\label{table:12} \input{3470tb12.tex} }

\longtab{13}{\label{table:13} \input{3470tb13.tex}  }

\longtab{14}{\label{table:14} \input{3470tb14.tex}   }

\longtab{15}{\label{table:15} \input{3470tb15.tex}  }

\longtab{16}{\label{table:16} \input{3470tb16.tex}   }

\longtab{17}{\label{table:17} \input{3470tb17.tex}  }

\end{appendix}

\end{document}

%% file: 3470tb4.tex

 
\begin{tabular}{crlllcc||crlllcc} 
\hline \hline 
$k\pm 2$ & $m$ & Period &    Freq.    &  Ampl. & Confid. & W91 &
$k\pm 2$ & $m$ & Period &    Freq.    &  Ampl. & Confid. & W91 \\
          &     &  (s)   & ($\mu$Hz) & (mma)& Level   & ($\ell,\,m$) &
          &     &  (s)   & ($\mu$Hz) & (mma)& Level   & ($\ell,\,m$) \\ 
\hline 
   & +1 & 389.72 & 2565.94 & 0.2 & 5 & 29   &        & +1 & 705.32 & 1417.80 & 0.8 & 1 & 1,+1   \\
14 & 0  & 390.30 & 2562.13 & 1.0 & 1 & 2,-2 &   29   & 0  & 709.05 & 1410.34 & 0.3 & 5 & 1,0 \\
   & +1 & 390.84 & 2558.59 & 0.2 & 5 & 29   &        & -1 & 711.58 & 1405.32 & 0.4 & 3 &    \\
\hline
  & +1 &        &         &     &   &    &     & +1 & 727.09 & 1375.36 & 0.7 & 1 & 1,+1 \\
15 & 0 & 412.01 & 2427.13 & 0.6 & 1 &    &  30 &  0 & 729.51 & 1370.78 & 0.3 & 2 & 1,0: \\
  & -1 & 413.14 & 2420.49 & 0.2 & 3 &    &     & -1 & 731.45 & 1367.15 & 1.0 & 1 & 1,-1: \\
\hline
  & +1 & 430.38 & 2323.53 & 0.3 & 5 &    &     & +1  & 750.56 & 1332.34 & 1.6 & 1    &   \\
16 & 0 & 432.37 & 2312.83 & 0.5 & 3 &    &  31 &   0 & 752.94 & 1328.13 & --- & 6    & 1,-1    \\
  & -1 & 434.15 & 2303.35 & 0.5 & 3 &    &     &  -1 & 755.31 & 1323.96 & 0.3 & 2    &   \\
\hline
  & +1 & 450.83 & 2218.13 & 3.5 & 1 & 1,0:  &     & +1  &        &         &     &   &       \\
17 & 0 & 452.06 & 2212.10 & 3.0 & 1 &       & 32  &   0 & 773.74 & 1292.42 & 0.3 & 3 & 1,0   \\
  & -1 & 453.24 & 2206.34 & 1.0 & 1 & (1),? &     & -1  & 776.67 & 1287.55 & 0.4 & 3 & 1,-1 \\
\hline
  & +1 &        &         &     &   &    &     & +1  & 790.26 & 1265.41 & 1.4 &   &         \\
18 & 0 & 472.08 & 2118.29 & 0.4 & 3 &    &  33 &   0 & 791.80 & 1262.95 & --- & 6 &        \\
  & -1 & 475,45 & 2103.27 & 0.3 & 3 &    &     &     & 793.34 & 1260.49 & 0.8 & 1 & 1,-1   \\
\hline
  & +1 & 493.79 & 2025.15 & 1.5 & 1 & 1,+1 &     & +1 & 812.57 & 1230.66 & 0.4 & 2 & 2,? \\
19 & 0 & 494.85 & 2020.81 & 0.7 & 1 & 1,0 &   34 & 0  & 814.58 & 1227.61 & 0.4 & 3 & 1,+1 \\
  & -1 & 496.00 & 2016.13 & 0.2 & 3 & 1,-1 &     & -1 & 817.40 & 1223.39 & 0.2 & 3 & 1,0 \\
\hline
  & +1 & 516.04 & 1937.83 & 7.2 & 1 & 1,+1 &     & +1 & 835.34 & 1197.12 & 0.3 & 3 &  \\
20 & 0 & 517.16 & 1933.64 & 4.2 & 1 & 1,0 &   35 & 0  & 838.62 & 1192.44 & 0.6 & 1 & 1,0 \\
  & -1 & 518.29 & 1929.42 & 3.2 & 1 & 1,-1 &     & -1 & 842.88 & 1186.41 & 1.0 & 1 & 1,-1 \\
\hline
  & +1 & 536.92 & 1862.47 & 0.5 & 1 & 1,+1 &     & +1 & 857.37 & 1166.36 & 0.4 & 3 &  \\
21 & 0 & 538.14 & 1858.25 & 0.6 & 1 & 1,0 &   36 &  0 & 861.72 & 1160.47 & 0.5 & 3 &  \\
  & -1 & 539.34 & 1854.12 & 1.0 & 1 & 1,-1 &     & -1 & 865.08 & 1155.96 & 0.7 & 1 &  \\
\hline
  & +1 & 557.13 & 1794.91 & 2.0 & 1 & 1,+1 &     & +1 & 877.67 & 1139.38 & 0.4 & 5 &  \\
22 & 0 & 558.14 & 1791.67 & 2.4 & 1 & 1,0 &   37 & 0  & 883.67 & 1131.65 & --- & 6 &  \\
  & -1 & 559.71 & 1786.64 & 1.0 & 1 & 1,-1 &     & -1 & 889.66 & 1124.02 & 0.3 & 1 &  \\
\hline
  & +1 & 576.03 & 1736.02 & 0.1 & 5 &       &     &     & 898.82 & 1112.57 & 0.9 & 1 &    \\
23 & 0 & 579.12 & 1726.76 & 0.1 & 5 & 2,-1: &  38 & 0   & 903.19 & 1107.19 & 0.7 & 1 &     \\
  & -1 & 581.67 & 1718.18 & 0.1 & 5 &       &     &     &        &         &     &   &    \\
\hline
  & +1 & 601.44 & 1662.66 & 0.3 & 5 & 1,+1 &     & +1 & 923.19 & 1083.20 & 0.5 & 1 & 2(1),? \\
24 & 0 & 603.04 & 1658.25 & 0.2 & 5 & 1,0 &   39 &  0 & 925.31 & 1080.72 & 0.3 & 2 &  \\
  & -1 & 604.72 & 1653.66 & 0.2 & 5 & 1,-1 &     & -1 & 927.58 & 1078.07 & 0.5 & 3 &  \\
\hline
  & +1 & 621.45 & 1609.07 & 0.2 & 5 & 1,+1 &     & +1 & 943.01 & 1060.43 & 0.5 & 3 &  \\
25 & 0 & 622.00 & 1607.72 & 0.3 & 3 & 1,0 &   40 &  0 & 945.01 & 1058.19 & 0.3 & 3 &  \\
  & -1 & 624/36 & 1601.64 & 0.3 & 5 & 1,-1 &     & -1 & 947.41 & 1055.51 & 0.5 & 1 &  \\
\hline
  & +1 & 641.54 & 1558.75 & 1.0 & 1 & 1,+1 &     & +1 & 962.07 & 1039.43 & 0.3 & 3 &  \\
26 & 0 & 643.31 & 1554.46 & 0.5 & 1 & 1,0 &   41 &  0 & 966.98 & 1034.15 & 0.9 & 1 & 2(1),? \\
  & -1 & 644.99 & 1550.41 & 0.8 & 1 & 1,-1 &     & -1 &        &         &     &   &  \\
\hline
  & +1 & 664.43 & 1505.34 & 0.3 & 3 & 1,+1 &     & +1     &         &         &     &   &            \\
27 & 0 & 668.09 & 1496.80 & 0.3 & 3 & 1,-1 &  42 &   0    & 988.13  & 1012.01 & 0.2 & 3 & 2(1),-1:   \\
  & -1 & 672.21 & 1487.63 & 0.3 & 3 &      &     & -1     & 994.12  & 1005.91 & 0.1 & 5 & 2(1),-2:  \\
\hline
  & +1 & 685.79 & 1458.17 & 0.3 & 2 & 1,+1 &     &  &  &  &  &  &  \\
28 & 0 & 687.74 & 1454.04 & 0.4 & 1 & 1,0  &     &  &  &  &  &  &  \\
  & -1 & 689.75 & 1449.80 & 0.5 & 1 & 1,-1 &     &  &  &  &  &  &  \\
\hline
\end{tabular}
 

%% file: 3470tb5.tex
 
\tiny
\begin{tabular}{lccccccc} 
\hline \hline
Data set                     &      1983        &        1985      &        1989      &      1993        &      2002        & Combined          & W91 \\ 
\hline
$\Delta P_1$ (s)            & $21.31\pm 0.06$  & $21.44\pm 0.04$  & $21.36\pm 0.04$  & $21.49\pm 0.03$  & $21.24\pm 0.06$  & $21.43\pm 0.03$   & $21.50\pm 0.03$ \\
$\Delta P_2$ (s)            & $12.41\pm 0.05$  & $12.36\pm 0.04$  & $12.38\pm 0.02$  & $12.38\pm 0.03$  & $12.38\pm 0.04$  & $12.38\pm 0.01$   & $12.67\pm 0.03$ \\ \hline
$\delta \nu_1$ ($\mu$Hz)    & $4.133\pm 0.003$ & $4.133\pm 0.005$ & $4.167\pm 0.006$ & $4.177\pm 0.007$ & $4.152\pm 0.014$ & $4.153\pm 0.002$  & $4.22\pm 0.04$\\
$\delta \nu_2$ ($\mu$Hz)    & $6.80\pm 0.02$   &      ---         & $7.01\pm 0.03$   & $6.96\pm 0.02$   & $6.81\pm 0.08$   & $6.903\pm 0.011$  & $6.92\pm 0.07$\\ \hline
$P_{{\rm rot}{,\,\ell=1}}$  (d)& $1.400\pm 0.001$ & $1.400\pm 0.002$ & $1.389\pm 0.002$ & $1.386\pm 0.002$ & $1.394\pm 0.005$ & $1.3934\pm 0.0008$& $1.371\pm 0.13$\\
$P_{{\rm rot}{,\,\ell=2}}$  (d)& $1.418\pm 0.004$ &     ---          & $1.376\pm 0.005$ & $1.385\pm 0.003$ & $1.417\pm 0.016$ & $1.3973\pm 0.0022$& $1.388\pm 0.013$\\
$P_{{\rm rot}}  $         (d)& $1.401\pm 0.001$ & $1.300\pm 0.002$ & $1.387\pm 0.002$ & $1.386\pm 0.002$ & $1.396\pm 0.005$ & $1.3939\pm 0.0008$ & $1.38\pm 0.01$\\ \hline
$E_{\ell=1}\,(\times 10^{-8}\,\mu\rm{Hz}^2$)  & $3.10\pm 0.07$   & $2.06\pm 0.08$   & $2.37\pm 0.02$   & $2.38\pm 0.03$   & $2.17\pm 0.03$   & ---                & --- \\
$E_{\ell=2}\,(\times 10^{-8}\,\mu\rm{Hz}^2$)  & $1.630\pm 0.002$ & $1.015\pm 0.002$ & $0.867\pm 0.007$ & $0.873\pm 0.006$ & $1.011\pm 0.010$ & ---                & --- \\
$E_{\rm tot} \,(\times 10^{-8}\,\mu\rm{Hz}^2$)& $4.730$          & $3.077$          & $3.240$          & $3.257$          & $3.176$          & ---                & --- \\ 
$E_{\ell=1}/E_{\rm tot}$                      & 0.66             & 0.67             & 0.73            & 0.73              & 0.68            & ---          & --- \\ \hline
\end{tabular}

%% file: 3470tb12.tex
 
\scriptsize
\begin{longtable}{crlllcc |crlllcc} 

\caption{\label{tab_L2} Identified $\ell = 2$ pulsation modes.} \\

\hline \hline 
$k\pm 2$ & $m$ & Period &    Freq.    &  Ampl. & Confid. & W91 &
$k\pm 2$ & $m$ & Period &    Freq.    &  Ampl. & Confid. & W91 \\
          &     &  (s)   & ($\mu$Hz) & (mma)& Level   & ($\ell,\,m$) &
          &     &  (s)   & ($\mu$Hz) & (mma)& Level   & ($\ell,\,m$) \\ 
\hline 
\endfirsthead
\caption{continued.}\\
\hline \hline
$k\pm 2$ & $m$ & Period &    Freq.    &  Ampl. & Confid. & W91 &
$k\pm 2$ & $m$ & Period &    Freq.    &  Ampl. & Confid. & W91 \\
          &     &  (s)   & ($\mu$Hz) & (mma)& Level   & ($\ell,\,m$) &
          &     &  (s)   & ($\mu$Hz) & (mma)& Level   & ($\ell,\,m$) \\ 
\hline 
\endhead
\hline
\endfoot

   & +1 &        &         &     &   &     &      & +1 & 668.52 & 1495.84 & 0.3 & 5 & --- \\
24 &  0 & 350.75 & 2851.03 & 0.6 & 3 & --- &   50 &  0 & 672.20 & 1487.65 & 0.1 & 5 & --- \\
   & -1 & 352.48 & 2837.04 & 0.5 & 5 & --- &      & -1 &        &         &     &   &     \\
   & -2 & 353.39 & 2829.73 & 0.2 & 5 & --- &      & -2 & 680.33 & 1469.88 & 0.3 & 3 & --- \\
\hline
   & +1 & 362.20 & 2760.91 & 0.6 & 3 & --- &      & +1 &        &         &     &   &     \\
25 &  0 & 363.39 & 2751.86 & 0.5 & 3 & 2,-2&   51 &  0 & 684.48 & 1460.96 & 0.1 & 5 & --- \\
   & -2 &        &         &     &   &     &      & -2 & 693.29 & 1442.40 & 0.2 &   & --- \\
\hline
   & +2 &        &         &     &   &     &      & +2 & 689.77 & 1449.76 & 0.5 & 2 & 1,-1 \\
   & +1 &        &         &     &   &     &      & +1 & 693.29 & 1442.40 & 0.2 & 3 & ---  \\
26 &  0 & 376.03 & 2659.36 & 0.3 & 5 & --- &   52 &  0 & 696.83 & 1435.07 & 0.4 & 5 & ---  \\
   & -1 & 376.65 & 2654.98 & 0.7 & 2 & --- &      & -1 &        &         &     &   &      \\
   & -2 & 377.73 & 2647.29 & 0.6 & 3 & --- &      & -2 & 705.80 & 1416.83 & 0.7 & 1 & 1,+1 \\
\hline
   & +1 & 386.93 & 2584.45 & 0.4 & 1 & --- &      & +1 & 705.93 & 1416.57 & 0.7 & 1 & 1,+1 \\
27 &  0 & 387.47 & 2580.84 & 0.4 & 5 & 2,0 &   53 &  0 & 709.87 & 1408.71 & 0.2 & 5 & ---  \\
   & -1 &        &         &     &   &     &      & -1 & 713.80 & 1400.95 & 0.2 & 5 & ---  \\
   & -2 & 390.30 & 2562.13 & 1.5 & 1 & 2,-2&      & -2 &        &         &     &   &      \\
\hline
   & +2 & 397.23 & 2517.43 & 0.4 & 3 & --- &      & +2 & 713.23 & 1402.07 & 0.5 & 3 & ---  \\
   & +1 & 398.91 & 2506.83 & 0.3 & 2 & 2,? &      & +1 &        &         &     &   &      \\
28 &  0 & 400.06 & 2499.63 & 1.4 & 1 & 2,? &   54 &  0 &        &         &     &   &      \\
   & -2 & 402.36 & 2485.34 & 0.5 & 1 & --- &      & -2 &        &         &     &   &      \\
\hline
   & +2 & 410.43 & 2436.47 & 0.7 & 1 & ---  &     & +2 &        &         &     &   &      \\
   & +1 & 412.00 & 2427.18 & 0.6 & 1 & 2,+1:&     & +1 & 729.72 & 1370.39 & 0.3 & 5 & 1,0: \\
29 &  0 & 413.14 & 2420.49 & 0.1 & 5 & 2,0: &  55 &  0 &        &         &     &   &      \\
   & -1 & 414.37 & 2413.30 & 0.6 & 1 & 2,-1:&     & -1 &        &         &     &   &      \\
   & -2 & 415.59 & 2406.22 & 1.2 & 1 & 2,-2:&     & -2 & 742.95 & 1345.99 & 0.1 & 5 & ---  \\
\hline
   & +2 & 422.55 & 2366.58 & 2.0 & 1 & 2,+2 &     & +2 & 737.79 & 1355.40 & 1.0 & 1 & --- \\
   & +1 & 423.81 & 2359.55 & 0.8 & 1 & 2,+1 &     & +1 &  &  &  &  &  \\
30 &  0 & 425.04 & 2352.72 & 0.4 & 5 & 2,0  &  56 &  0 & 746.38 & 1339.80 & 0.8 & 1 & --- \\
   & -1 & 426.29 & 2345.82 & 0.9 & 1 & 2,-1 &     & -1 &  &  &  &  &  \\
   & -2 & 427.53 & 2339.02 & 1.5 & 1 & 2,-2 &     & -2 &  &  &  &  &  \\
\hline
   & +2 & 434.96 & 2299.06 & 0.1 & 5 & ---  &     & +2 &        &         &     &   &     \\
   & +1 & 436.56 & 2290.64 & 0.5 & 1 & 2,+1 &     & +1 &        &         &     &   &     \\
31 &  0 &        &         &     &   &      &  57 &  0 &        &         &     &   &     \\
   & -1 & 439.25 & 2276.61 & 0.5 & 1 & 2,-1:&     & -1 & 763.90 & 1309.07 & 0.3 & 3 & --- \\
   & -2 & 440.66 & 2269.32 & 0.1 & 4 & 2,-2:&     & -2 & 768.72 & 1300.86 & 0.3 & 3 & --- \\
\hline
   & +2 & 446.52 & 2239.54 & 0.4 & 3 & ---  &     & +2 & 762.11 & 1312.15 & 0.5 & 3 & --- \\
   & +1 & 447.89 & 2232.69 & 0.2 & 5 & ---  &     & +1 &  &  &  &  &  \\
32 &  0 & 449.43 & 2225.04 & 0.2 & 5 & ---  &  58 &  0 &        &         &     &   &      \\
   & -1 & 452.03 & 2212.24 & 2.0 & 1 & ---  &     & -1 & 776.63 & 1287.61 & 0.3 & 3 & 1,-1 \\
   & -2 & 453.26 & 2206.24 & 2.0 & 1 & (2),?&     & -2 & 780.97 & 1280.46 & 0.9 & 1 & ---  \\
\hline
   & +2 & 458.88 & 2179.22 & 0.7 & 3 & --- &     & +2 & 773.73 & 1292.44 & 0.3 & 3 & 1,0 \\
   & +1 & 460.71 & 2170.56 & 0.3 & 5 & --- &     & +1 &        &         &     &   &     \\
33 &  0 &        &         &     &   &     &  59 &  0 & 783.19 & 1276.83 & 0.3 & 3 & --- \\
\hline
35 &  0 &        &         &     &   &     &  60 &  0 &        &         &     &   &       \\
   & -1 & 488.89 & 2045.45 & 0.3 & 3 & --- &     & -1 &        &         &     &   &       \\
   & -2 &        &         &     &   &     &     & -2 & 819.95 & 1219.59 & 0.8 &   &  1,-1 \\
\hline
   & +2 & 494.85 & 2020.81 & 0.5 & 1 & --- &     & +2 &        &         &     &   &     \\
   & +1 &        &         &     &   &     &     & +1 &        &         &     &   &     \\
36 &  0 & 498.73 & 2005.09 & 0.6 & 1 & --- &  62 &  0 & 820.90 & 1218.18 & 1.8 & 1 & --- \\
   & -1 & 500.91 & 1996.37 & 0.3 & 3 & --- &     & -1 &        &         &     &   &     \\
\hline
   & +2 & 507.58 & 1970.13 & 0.3 & 3 & --- &     & +2 & 821.69 & 1217.00 & 0.6 & 3 & --- \\
   & +1 & 510.06 & 1960.55 & 0.4 & 3 & --- &     & +1 &        &         &     &   &     \\
37 &  0 & 511.98 & 1953.20 & 0.4 & 1 & --- &  63 &  0 &        &         &     &   &     \\
   & -1 & 514.06 & 1945.30 & 0.4 & 3 & --- &     & -1 & 838.65 & 1192.39 & 0.6 & 2 & --- \\
   & -2 & 516.03 & 1937.87 & 2.0 & 1 & 1,+1&     & -2 & 844.78 & 1183.74 & 0.9 & 1 & --- \\
\hline
   & +2 & 519.30 & 1925.67 & 0.7 & 1 & --- &     & +2 &        &         &     &   &     \\
   & +1 &        &         &     &   &     &     & +1 &        &         &     &   &     \\
38 &  0 & 524.03 & 1908.29 & 0.2 & 5 & --- &  64 &  0 &        &         &     &   &     \\
   & -1 & 526.43 & 1899.59 & 0.4 & 1 & --- &     & -1 &        &         &     &   &     \\
   & -2 &        &         &     &   &     &     & -2 & 857.36 & 1166.37 & 0.4 & 3 & --- \\
\hline
   & +2 & 531.83 & 1880.30 & 1.0 & 1 & --- &     & +2 &        &         &     &   &     \\
   & +1 &        &         &     &   &     &     & +1 & 852.08 & 1173.60 & 0.6 &   & --- \\
39 &  0 & 536.37 & 1864.38 & 0.2 & 5 & --- &  65 &  0 & 858.84 & 1164.36 & 0.2 & 4 & --- \\
   & -2 & 540.96 & 1848.57 & 0.4 & 1 & --- &     & -2 &        &         &     &   &     \\
\hline
   & +2 & 544.31 & 1837.19 & 0.6 & 1 & --- &     & +2 & 859.67 & 1163.24 & 0.5 & 3 & --- \\
   & +1 & 546.05 & 1831.33 & 0.9 & 1 & --- &     & +1 &        &  &  &  &  \\
40 &  0 & 547.00 & 1828.15 & 0.2 & 3 & --- &  66 &  0 &        &  &  &  &  \\
   & -1 & 550.52 & 1816.46 & 0.3 & 3 & --- &     & -1 & 877.10 & 1140.12 & 0.3 & 3 & --- \\
\hline
   & +2 & 556.64 & 1796.49 & 0.3 & 3 & --- &     & +2 &        &  &  &  &  \\
   & +1 & 558.98 & 1788.97 & 0.4 & 3 & --- &     & +1 &        &  &  &  &  \\
41 &  0 & 561.99 & 1779.39 & 0.5 & 2 & --- &  67 &  0 &        &  &  &  &  \\
   & -1 & 563.48 & 1774.68 & 0.3 & 3 & --- &     & -1 & 889.67 & 1124.01 & 0.4 & 5 & --- \\
\hline
   & +1 & 571.19 & 1750.73 & 0.5 & 3 & (2),?&     & +1& 901.07 & 1109.79 & 0.7 & 3 & --- \\
42 &  0 & 573.69 & 1743.10 & 1.0 & 1 & --- &  69  & 0 &  &  &  &  &  \\
   & -1 & 576.02 & 1736.05 & 0.2 & 5 & --- &      & -1&  &  &  &  &  \\
   & -2 & 579.11 & 1726.79 & 0.3 & 3 & 2,-1:&     & -2&  &  &  &  &  \\
\hline
   & +2 & 580.34 & 1723.13 & 0.4 & 3 & --- &      & +2 &  &  &  &  &  \\
43 &  0 & 585.26 & 1708.64 & 0.6 & 3 & --- &   70 &  0  &  &  &  &  &  \\
   & -2 &        &         &     &   &     &      & -2 & 934.05 & 1070.61 & 0.5 &     & 2,? \\
\hline
   & +1 &        &         &     &   &     &     & +1 & 924.94 & 1081.15 & 0.3 &   & --- \\
46 &  0 &        &         &     &   &     &  71 &  0 &        &         &     &   &     \\
   & -1 & 626.47 & 1596.25 & 0.3 & 5 & --- &     & -1 & 939.68 & 1064.19 & 0.3 & 5 & --- \\
   & -2 & 629.54 & 1588.46 & 0.5 & 5 & --- &     & -2 & 947.45 & 1055.46 & 0.5 & 1 & --- \\
\hline
47 &  0 &        &         &     &   &    &  72 &  0 & 945.01 & 1058.19 & 0.3 &   3 & --- \\
   & -1 &        &         &     &   &    &     & -1 &        &         &     &     &  \\
   & -2 & 641.90 & 1557.88 & 0.8 & 3 & -- &     & -2 &        &         &     &     &  \\
\hline
   & +1 & 644.04 & 1552.70 & 0.4 & 3 & --- &     & +1 & 961.09 & 1040.49 & 0.2 & 5 & --- \\
48 &  0 &        &         &     &   &     &  74 &  0 &  &  &  &  &  \\
   & -1 & 650.83 & 1536.50 & 0.5 & 3 & --- &     & -1 &  &  &  &  &  \\
\hline
  & +2           &         &     &   &     &       &        & +2 & 966.95 & 1034.18 & 0.7 &   & ---   \\
49 & 0 & 660.46  & 1514.10 & 0.4 & 5 & --- &   75 &  0 & 982.68 & 1017.63 & 0.2 & 7 & 2.?   \\
  & -1 &         &         &     &   &     &      & -1 & 988.70 & 1011.43 & 0.1 &   & 2,-1: \\
  & -2 & 666.86  & 1499.57 & 0.3 & 3 & --- &      & -2 &        &         &     &   &       \\
\hline
\end{longtable}
 

%% file: 3470tb13.tex
 
\scriptsize
 \begin{longtable} {cccc} 
 \caption{\label{tab_T1983} Detected pulsation modes in the 1983 dataset
	 ($T_{\rm max}$ computed from $T_o=244\,5346.0$ BCT).} \\
\hline \hline
Frequency  & Period    &  Amplitude &  ${\rm T}_{\rm max}$\\
 ($\mu$Hz)&  (sec)  &   (mma)  &  (s)     \\
 \hline 
\endfirsthead
\caption{continued.}\\
 \hline \hline 
Frequency  & Period    &  Amplitude &  ${\rm T}_{\rm max}$\\
 ($\mu$Hz)&  (sec)  &   (mma)  &  (s)     \\
 \hline 
\endhead
\hline
\endfoot
$  1011.9365 \pm    0.0079$ & $ 988.204333 \pm    0.007711$ & $  1.1791 \pm   0.1256$ & $  17.62 \pm   49.50$ \\
$  1112.6100 \pm    0.0111$ & $ 898.787523 \pm    0.008955$ & $  0.9190 \pm   0.1380$ & $ 638.41 \pm   63.26$ \\
$  1151.0329 \pm    0.0119$ & $ 868.784910 \pm    0.008981$ & $  0.7887 \pm   0.1259$ & $ 527.90 \pm   65.72$ \\
$  1161.1547 \pm    0.0101$ & $ 861.211712 \pm    0.007524$ & $  0.9217 \pm   0.1259$ & $ 844.50 \pm   55.52$ \\
$  1181.1207 \pm    0.0090$ & $ 846.653521 \pm    0.006440$ & $  1.0447 \pm   0.1270$ & $  32.58 \pm   48.27$ \\
$  1186.4182 \pm    0.0046$ & $ 842.873106 \pm    0.003243$ & $  2.2830 \pm   0.1418$ & $ 493.04 \pm   24.39$ \\
$  1216.9900 \pm    0.0160$ & $ 821.699418 \pm    0.010776$ & $  0.6305 \pm   0.1409$ & $ 486.90 \pm   84.62$ \\
$  1218.1499 \pm    0.0053$ & $ 820.917047 \pm    0.003587$ & $  1.9605 \pm   0.1416$ & $ 386.19 \pm   27.72$ \\
$  1219.8514 \pm    0.0073$ & $ 819.772012 \pm    0.004880$ & $  1.3999 \pm   0.1380$ & $ 766.46 \pm   37.77$ \\
$  1226.9416 \pm    0.0099$ & $ 815.034743 \pm    0.006564$ & $  1.0191 \pm   0.1410$ & $ 172.52 \pm   51.98$ \\
$  1243.3324 \pm    0.0141$ & $ 804.290153 \pm    0.009095$ & $  0.7829 \pm   0.1757$ & $ 654.30 \pm   73.35$ \\
$  1254.5377 \pm    0.0133$ & $ 797.106396 \pm    0.008461$ & $  0.8206 \pm   0.1757$ & $ 349.41 \pm   68.93$ \\
$  1260.5005 \pm    0.0119$ & $ 793.335677 \pm    0.007504$ & $  0.8284 \pm   0.1307$ & $ 338.19 \pm   60.24$ \\
$  1280.4246 \pm    0.0109$ & $ 780.990943 \pm    0.006675$ & $  0.9026 \pm   0.1304$ & $ 386.32 \pm   54.35$ \\
$  1352.4929 \pm    0.0170$ & $ 739.375432 \pm    0.009270$ & $  0.5584 \pm   0.1279$ & $ 314.78 \pm   79.60$ \\
$  1357.6442 \pm    0.0107$ & $ 736.570013 \pm    0.005828$ & $  0.9506 \pm   0.1378$ & $ 379.29 \pm   50.19$ \\
$  1391.4715 \pm    0.0134$ & $ 718.663683 \pm    0.006941$ & $  0.6877 \pm   0.1249$ & $  70.92 \pm   61.31$ \\
$  1448.5587 \pm    0.0148$ & $ 690.341362 \pm    0.007061$ & $  0.6323 \pm   0.1270$ & $ 329.19 \pm   64.92$ \\
$  1455.0232 \pm    0.0092$ & $ 687.274290 \pm    0.004341$ & $  1.1181 \pm   0.1390$ & $ 379.94 \pm   40.07$ \\
$  1529.1025 \pm    0.0108$ & $ 653.978381 \pm    0.004640$ & $  0.8726 \pm   0.1285$ & $ 292.69 \pm   45.02$ \\
$  1535.1914 \pm    0.0202$ & $ 651.384588 \pm    0.008592$ & $  0.4664 \pm   0.1278$ & $ 327.07 \pm   83.68$ \\
$  1611.2702 \pm    0.0107$ & $ 620.628356 \pm    0.004117$ & $  0.9618 \pm   0.1389$ & $ 465.32 \pm   42.10$ \\
$  1700.9494 \pm    0.0157$ & $ 587.906960 \pm    0.005411$ & $  0.6505 \pm   0.1376$ & $ 421.02 \pm   58.42$ \\
$  1755.9298 \pm    0.0198$ & $ 569.498851 \pm    0.006418$ & $  0.5359 \pm   0.1432$ & $  64.88 \pm   71.52$ \\
$  1764.1313 \pm    0.0116$ & $ 566.851230 \pm    0.003729$ & $  0.9206 \pm   0.1446$ & $ 359.75 \pm   41.72$ \\
$  1779.4498 \pm    0.0090$ & $ 561.971446 \pm    0.002833$ & $  1.2115 \pm   0.1473$ & $ 412.12 \pm   31.99$ \\
$  1809.1382 \pm    0.0178$ & $ 552.749380 \pm    0.005437$ & $  0.6271 \pm   0.1496$ & $ 526.53 \pm   62.80$ \\
$  1823.7553 \pm    0.0113$ & $ 548.319169 \pm    0.003389$ & $  0.9871 \pm   0.1492$ & $ 495.74 \pm   39.35$ \\
$  1829.6743 \pm    0.0135$ & $ 546.545366 \pm    0.004023$ & $  0.7892 \pm   0.1433$ & $ 234.94 \pm   46.77$ \\
$  1854.0516 \pm    0.0021$ & $ 539.359305 \pm    0.000601$ & $  5.9342 \pm   0.1617$ & $ 532.55 \pm    7.14$ \\
$  1858.1887 \pm    0.0027$ & $ 538.158481 \pm    0.000793$ & $  4.4800 \pm   0.1618$ & $ 140.06 \pm    9.44$ \\
$  1862.7482 \pm    0.0106$ & $ 536.841215 \pm    0.003062$ & $  1.0068 \pm   0.1447$ & $  51.93 \pm   36.17$ \\
$  1877.5834 \pm    0.0193$ & $ 532.599526 \pm    0.005468$ & $  0.5525 \pm   0.1444$ & $ 413.36 \pm   65.14$ \\
$  1901.2433 \pm    0.0099$ & $ 525.971619 \pm    0.002744$ & $  1.1041 \pm   0.1462$ & $ 482.25 \pm   33.19$ \\
$  1928.0004 \pm    0.0141$ & $ 518.672104 \pm    0.003794$ & $  0.6720 \pm   0.1285$ & $ 359.36 \pm   46.42$ \\
$  1932.6317 \pm    0.0088$ & $ 517.429174 \pm    0.002354$ & $  1.2554 \pm   0.1492$ & $ 274.07 \pm   28.90$ \\
$  1933.7778 \pm    0.0085$ & $ 517.122487 \pm    0.002267$ & $  1.3028 \pm   0.1473$ & $ 499.16 \pm   27.89$ \\
$  1937.8870 \pm    0.0013$ & $ 516.025972 \pm    0.000359$ & $  8.7643 \pm   0.1604$ & $ 223.87 \pm    4.42$ \\
$  1941.4769 \pm    0.0117$ & $ 515.071793 \pm    0.003093$ & $  0.9146 \pm   0.1442$ & $ 156.07 \pm   38.11$ \\
$  1963.7950 \pm    0.0164$ & $ 509.218114 \pm    0.004257$ & $  0.5715 \pm   0.1269$ & $ 218.49 \pm   53.05$ \\
$  2005.9249 \pm    0.0159$ & $ 498.523151 \pm    0.003947$ & $  0.6724 \pm   0.1444$ & $ 409.06 \pm   50.25$ \\
$  2006.4344 \pm    0.0145$ & $ 498.396553 \pm    0.003601$ & $  0.7311 \pm   0.1429$ & $ 255.65 \pm   45.92$ \\
$  2020.5927 \pm    0.0047$ & $ 494.904296 \pm    0.001156$ & $  2.3447 \pm   0.1491$ & $   3.71 \pm   14.83$ \\
$  2025.2590 \pm    0.0022$ & $ 493.764000 \pm    0.000527$ & $  5.4731 \pm   0.1600$ & $ 182.52 \pm    6.78$ \\
$  2026.0712 \pm    0.0095$ & $ 493.566068 \pm    0.002309$ & $  1.1630 \pm   0.1476$ & $ 313.03 \pm   29.76$ \\
$  2029.6308 \pm    0.0244$ & $ 492.700450 \pm    0.005930$ & $  0.3990 \pm   0.1330$ & $ 115.65 \pm   76.54$ \\
$  2051.9225 \pm    0.0119$ & $ 487.347832 \pm    0.002833$ & $  0.8185 \pm   0.1320$ & $ 321.17 \pm   36.91$ \\
$  2072.0988 \pm    0.0149$ & $ 482.602478 \pm    0.003471$ & $  0.6545 \pm   0.1332$ & $ 170.74 \pm   45.76$ \\
$  2087.8452 \pm    0.0110$ & $ 478.962716 \pm    0.002533$ & $  0.8871 \pm   0.1326$ & $ 309.38 \pm   33.55$ \\
$  2119.5632 \pm    0.0125$ & $ 471.795314 \pm    0.002781$ & $  0.7708 \pm   0.1301$ & $  20.26 \pm   37.44$ \\
$  2128.4479 \pm    0.0152$ & $ 469.825928 \pm    0.003348$ & $  0.6276 \pm   0.1293$ & $ 153.32 \pm   45.21$ \\
$  2133.1093 \pm    0.0109$ & $ 468.799230 \pm    0.002404$ & $  0.9720 \pm   0.1438$ & $ 118.38 \pm   32.56$ \\
$  2141.8612 \pm    0.0138$ & $ 466.883655 \pm    0.003019$ & $  0.7126 \pm   0.1336$ & $ 296.95 \pm   41.04$ \\
$  2171.5971 \pm    0.0126$ & $ 460.490575 \pm    0.002680$ & $  0.7510 \pm   0.1284$ & $ 124.18 \pm   36.93$ \\
$  2177.4422 \pm    0.0079$ & $ 459.254450 \pm    0.001671$ & $  1.3725 \pm   0.1472$ & $ 352.71 \pm   23.09$ \\
$  2178.4015 \pm    0.0110$ & $ 459.052210 \pm    0.002324$ & $  0.8789 \pm   0.1314$ & $ 424.17 \pm   32.15$ \\
$  2206.2713 \pm    0.0076$ & $ 453.253413 \pm    0.001554$ & $  1.4114 \pm   0.1444$ & $  79.20 \pm   21.76$ \\
$  2210.2990 \pm    0.0041$ & $ 452.427481 \pm    0.000838$ & $  2.7826 \pm   0.1520$ & $  75.60 \pm   11.74$ \\
$  2212.5858 \pm    0.0162$ & $ 451.959885 \pm    0.003305$ & $  0.5801 \pm   0.1270$ & $ 289.54 \pm   46.42$ \\
$  2214.3841 \pm    0.0025$ & $ 451.592837 \pm    0.000511$ & $  4.5395 \pm   0.1519$ & $ 345.00 \pm    7.16$ \\
$  2222.8546 \pm    0.0115$ & $ 449.871985 \pm    0.002336$ & $  0.9125 \pm   0.1429$ & $ 251.75 \pm   32.95$ \\
$  2249.1975 \pm    0.0077$ & $ 444.603024 \pm    0.001515$ & $  1.4051 \pm   0.1457$ & $  56.58 \pm   21.63$ \\
$  2259.0124 \pm    0.0159$ & $ 442.671322 \pm    0.003109$ & $  0.6013 \pm   0.1293$ & $ 151.59 \pm   44.58$ \\
$  2273.3189 \pm    0.0115$ & $ 439.885501 \pm    0.002218$ & $  0.9272 \pm   0.1437$ & $  37.27 \pm   32.03$ \\
$  2316.7935 \pm    0.0105$ & $ 431.631037 \pm    0.001953$ & $  0.9266 \pm   0.1316$ & $ 377.13 \pm   28.75$ \\
$  2324.3126 \pm    0.0171$ & $ 430.234724 \pm    0.003171$ & $  0.6123 \pm   0.1404$ & $ 148.06 \pm   46.68$ \\
$  2334.7293 \pm    0.0076$ & $ 428.315172 \pm    0.001394$ & $  1.2879 \pm   0.1326$ & $ 416.39 \pm   20.64$ \\
$  2336.4100 \pm    0.0115$ & $ 428.007074 \pm    0.002113$ & $  0.9080 \pm   0.1405$ & $ 333.24 \pm   31.25$ \\
$  2340.6038 \pm    0.0120$ & $ 427.240179 \pm    0.002185$ & $  0.7996 \pm   0.1297$ & $ 177.87 \pm   32.45$ \\
$  2366.6416 \pm    0.0042$ & $ 422.539691 \pm    0.000754$ & $  2.3616 \pm   0.1349$ & $  81.13 \pm   11.33$ \\
$  2369.4702 \pm    0.0084$ & $ 422.035264 \pm    0.001491$ & $  1.1572 \pm   0.1303$ & $ 327.43 \pm   22.43$ \\
$  2383.2251 \pm    0.0059$ & $ 419.599475 \pm    0.001035$ & $  1.6882 \pm   0.1337$ & $ 219.72 \pm   15.66$ \\
$  2401.8652 \pm    0.0167$ & $ 416.343105 \pm    0.002891$ & $  0.5787 \pm   0.1302$ & $ 389.12 \pm   44.08$ \\
$  2415.2562 \pm    0.0100$ & $ 414.034745 \pm    0.001713$ & $  0.9684 \pm   0.1305$ & $ 266.36 \pm   26.27$ \\
$  2436.4093 \pm    0.0134$ & $ 410.440061 \pm    0.002252$ & $  0.7345 \pm   0.1315$ & $ 354.62 \pm   34.89$ \\
$  2462.0764 \pm    0.0153$ & $ 406.161236 \pm    0.002524$ & $  0.6309 \pm   0.1300$ & $  97.49 \pm   39.44$ \\
$  2512.4785 \pm    0.0170$ & $ 398.013362 \pm    0.002691$ & $  0.5859 \pm   0.1339$ & $ 397.56 \pm   42.92$ \\
$  2522.7090 \pm    0.0167$ & $ 396.399270 \pm    0.002632$ & $  0.5855 \pm   0.1312$ & $ 364.67 \pm   42.21$ \\
$  2562.2870 \pm    0.0056$ & $ 390.276339 \pm    0.000856$ & $  1.7742 \pm   0.1349$ & $ 243.50 \pm   13.92$ \\
$  2597.5527 \pm    0.0218$ & $ 384.977754 \pm    0.003231$ & $  0.4413 \pm   0.1302$ & $ 247.21 \pm   53.27$ \\
$  2640.8545 \pm    0.0144$ & $ 378.665322 \pm    0.002058$ & $  0.6903 \pm   0.1338$ & $  30.95 \pm   34.49$ \\
$  2655.0047 \pm    0.0136$ & $ 376.647173 \pm    0.001929$ & $  0.7176 \pm   0.1319$ & $ 376.05 \pm   32.52$ \\
$  2684.9228 \pm    0.0144$ & $ 372.450192 \pm    0.001996$ & $  0.6832 \pm   0.1328$ & $ 337.22 \pm   34.03$ \\
$  2700.0979 \pm    0.0215$ & $ 370.356941 \pm    0.002946$ & $  0.4454 \pm   0.1296$ & $ 146.36 \pm   50.48$ \\
$  2760.8020 \pm    0.0175$ & $ 362.213595 \pm    0.002299$ & $  0.5611 \pm   0.1328$ & $ 201.08 \pm   40.31$ \\
$  2837.1015 \pm    0.0175$ & $ 352.472408 \pm    0.002172$ & $  0.5425 \pm   0.1285$ & $ 252.36 \pm   39.09$ \\
 \hline
 \end{longtable}

%% file: 3470tb14.tex
 
\scriptsize
 \begin{longtable} {cccc}
 \caption{\label{tab_T1985} Detected pulsation modes in the 1985
   dataset ($T_{\rm max}$ computed from $T_o=244\,6147.0$ BCT). } \\
\hline \hline
Frequency  & Period    &  Amplitude &  ${\rm T}_{\rm max}$\footnote{The times of maximum $T_{\rm max}$ 
   are computed from $T_o=244\,6147.0$ (BCT).} \\
 ($\mu$Hz)&  (sec)  &   (mma)  &  (s)     \\
 \hline 
\endfirsthead
\caption{continued.}\\
 \hline \hline 
Frequency  & Period    &  Amplitude &  ${\rm T}_{\rm max}$\footnote{The times of maximum $T_{\rm max}$ 
   are computed from $T_o=244\,6147.0$ (BCT).} \\
 ($\mu$Hz)&  (sec)  &   (mma)  &  (s)     \\
 \hline 
\endhead
\hline
\endfoot


$  1034.5428 \pm    0.0217$ & $ 966.610560 \pm    0.020314$ & $  0.6849 \pm   0.1988$ & $ 736.76 \pm   94.62$ \\
$  1078.0905 \pm    0.0201$ & $ 927.565918 \pm    0.017289$ & $  0.6848 \pm   0.1815$ & $ 516.68 \pm   84.53$ \\
$  1102.9193 \pm    0.0235$ & $ 906.684649 \pm    0.019318$ & $  0.5980 \pm   0.1865$ & $ 336.53 \pm   95.88$ \\
$  1135.4811 \pm    0.0212$ & $ 880.683988 \pm    0.016460$ & $  0.6612 \pm   0.1822$ & $ 742.20 \pm   86.97$ \\
$  1160.5270 \pm    0.0124$ & $ 861.677505 \pm    0.009235$ & $  1.0889 \pm   0.1814$ & $ 189.62 \pm   48.88$ \\
$  1160.6326 \pm    0.0089$ & $ 861.599096 \pm    0.006592$ & $  1.5447 \pm   0.1821$ & $ 226.99 \pm   34.68$ \\
$  1167.8772 \pm    0.0206$ & $ 856.254435 \pm    0.015102$ & $  0.6882 \pm   0.1837$ & $ 715.95 \pm   82.30$ \\
$  1173.5780 \pm    0.0201$ & $ 852.095023 \pm    0.014619$ & $  0.6553 \pm   0.1746$ & $ 417.38 \pm   77.38$ \\
$  1186.6776 \pm    0.0131$ & $ 842.688835 \pm    0.009292$ & $  1.0753 \pm   0.1868$ & $ 595.48 \pm   49.73$ \\
$  1230.8545 \pm    0.0187$ & $ 812.443695 \pm    0.012359$ & $  0.7349 \pm   0.1812$ & $ 196.48 \pm   69.76$ \\
$  1265.0999 \pm    0.0260$ & $ 790.451391 \pm    0.016224$ & $  0.7356 \pm   0.2509$ & $ 378.00 \pm   98.51$ \\
$  1265.3800 \pm    0.0108$ & $ 790.276444 \pm    0.006735$ & $  1.3816 \pm   0.1988$ & $ 351.40 \pm   38.37$ \\
$  1274.6813 \pm    0.0261$ & $ 784.509853 \pm    0.016070$ & $  0.7337 \pm   0.2510$ & $ 224.14 \pm   98.27$ \\
$  1339.7982 \pm    0.0155$ & $ 746.381081 \pm    0.008647$ & $  0.8750 \pm   0.1805$ & $   7.76 \pm   52.24$ \\
$  1402.0384 \pm    0.0247$ & $ 713.247214 \pm    0.012565$ & $  0.5407 \pm   0.1777$ & $ 292.46 \pm   79.40$ \\
$  1416.7367 \pm    0.0189$ & $ 705.847439 \pm    0.009436$ & $  0.7605 \pm   0.1919$ & $ 161.39 \pm   60.26$ \\
$  1446.1612 \pm    0.0158$ & $ 691.485856 \pm    0.007577$ & $  0.9403 \pm   0.1985$ & $ 635.83 \pm   49.36$ \\
$  1471.8909 \pm    0.0243$ & $ 679.398170 \pm    0.011201$ & $  0.5597 \pm   0.1807$ & $ 185.42 \pm   74.34$ \\
$  1475.5162 \pm    0.0290$ & $ 677.728912 \pm    0.013319$ & $  0.4697 \pm   0.1795$ & $ 240.86 \pm   89.38$ \\
$  1550.6392 \pm    0.0104$ & $ 644.895333 \pm    0.004318$ & $  1.7487 \pm   0.2415$ & $  86.68 \pm   30.18$ \\
$  1584.5568 \pm    0.0198$ & $ 631.091302 \pm    0.007887$ & $  0.7024 \pm   0.1772$ & $ 552.05 \pm   57.01$ \\
$  1588.4141 \pm    0.0255$ & $ 629.558741 \pm    0.010126$ & $  0.5431 \pm   0.1761$ & $ 478.52 \pm   73.22$ \\
$  1617.4207 \pm    0.0278$ & $ 618.268316 \pm    0.010612$ & $  0.4886 \pm   0.1806$ & $ 466.49 \pm   77.32$ \\
$  1619.9864 \pm    0.0178$ & $ 617.289125 \pm    0.006780$ & $  0.8086 \pm   0.1921$ & $ 206.64 \pm   49.57$ \\
$  1717.5117 \pm    0.0373$ & $ 582.237654 \pm    0.012652$ & $  0.3535 \pm   0.1772$ & $ 181.18 \pm   97.57$ \\
$  1718.2909 \pm    0.0284$ & $ 581.973646 \pm    0.009610$ & $  0.5009 \pm   0.1888$ & $  77.54 \pm   74.76$ \\
$  1743.0959 \pm    0.0135$ & $ 573.691892 \pm    0.004453$ & $  1.0035 \pm   0.1809$ & $  86.98 \pm   35.00$ \\
$  1763.4590 \pm    0.0268$ & $ 567.067343 \pm    0.008618$ & $  0.4986 \pm   0.1781$ & $ 467.81 \pm   68.63$ \\
$  1780.7922 \pm    0.0130$ & $ 561.547822 \pm    0.004095$ & $  1.0315 \pm   0.1798$ & $ 307.14 \pm   32.91$ \\
$  1790.2400 \pm    0.0252$ & $ 558.584319 \pm    0.007858$ & $  0.5249 \pm   0.1782$ & $ 131.33 \pm   63.25$ \\
$  1800.4563 \pm    0.0139$ & $ 555.414752 \pm    0.004286$ & $  1.0264 \pm   0.1885$ & $ 189.10 \pm   34.92$ \\
$  1807.1851 \pm    0.0253$ & $ 553.346743 \pm    0.007737$ & $  0.5230 \pm   0.1784$ & $ 302.81 \pm   62.89$ \\
$  1823.6754 \pm    0.0108$ & $ 548.343187 \pm    0.003262$ & $  1.3674 \pm   0.1927$ & $ 292.15 \pm   27.08$ \\
$  1826.2281 \pm    0.0179$ & $ 547.576729 \pm    0.005362$ & $  0.7463 \pm   0.1748$ & $ 128.89 \pm   44.42$ \\
$  1853.9013 \pm    0.0033$ & $ 539.403043 \pm    0.000971$ & $  4.4803 \pm   0.1985$ & $ 430.19 \pm    8.12$ \\
$  1858.2488 \pm    0.0033$ & $ 538.141067 \pm    0.000959$ & $  4.8752 \pm   0.2149$ & $ 131.71 \pm    8.03$ \\
$  1893.5625 \pm    0.0131$ & $ 528.105086 \pm    0.003655$ & $  1.1711 \pm   0.1998$ & $ 352.02 \pm   31.27$ \\
$  1911.4126 \pm    0.0160$ & $ 523.173287 \pm    0.004367$ & $  0.8856 \pm   0.1879$ & $ 380.39 \pm   37.72$ \\
$  1933.7114 \pm    0.0054$ & $ 517.140243 \pm    0.001440$ & $  3.0092 \pm   0.2152$ & $ 485.22 \pm   12.55$ \\
$  1937.8440 \pm    0.0024$ & $ 516.037412 \pm    0.000650$ & $  7.4480 \pm   0.2414$ & $  28.37 \pm    5.68$ \\
$  1941.6301 \pm    0.0123$ & $ 515.031147 \pm    0.003263$ & $  1.2133 \pm   0.1991$ & $ 387.92 \pm   28.38$ \\
$  1967.5851 \pm    0.0217$ & $ 508.237238 \pm    0.005617$ & $  0.6510 \pm   0.1824$ & $ 480.79 \pm   50.88$ \\
$  2005.0693 \pm    0.0112$ & $ 498.735878 \pm    0.002793$ & $  1.3306 \pm   0.1987$ & $  81.86 \pm   25.24$ \\
$  2020.9467 \pm    0.0121$ & $ 494.817607 \pm    0.002974$ & $  1.1327 \pm   0.1808$ & $  93.61 \pm   27.12$ \\
$  2025.3204 \pm    0.0054$ & $ 493.749041 \pm    0.001320$ & $  2.9996 \pm   0.2154$ & $ 290.75 \pm   12.07$ \\
$  2064.3576 \pm    0.0122$ & $ 484.412195 \pm    0.002853$ & $  1.1559 \pm   0.1871$ & $ 147.21 \pm   26.56$ \\
$  2103.2286 \pm    0.0176$ & $ 475.459501 \pm    0.003969$ & $  0.7866 \pm   0.1817$ & $ 216.98 \pm   37.66$ \\
$  2176.4901 \pm    0.0128$ & $ 459.455332 \pm    0.002696$ & $  1.1370 \pm   0.1942$ & $ 431.99 \pm   26.61$ \\
$  2179.1563 \pm    0.0196$ & $ 458.893196 \pm    0.004136$ & $  0.7141 \pm   0.1872$ & $ 346.20 \pm   40.61$ \\
$  2195.4017 \pm    0.0292$ & $ 455.497513 \pm    0.006050$ & $  0.4674 \pm   0.1814$ & $ 312.52 \pm   59.94$ \\
$  2210.5733 \pm    0.0109$ & $ 452.371342 \pm    0.002224$ & $  1.3724 \pm   0.1986$ & $ 143.63 \pm   22.16$ \\
$  2214.4590 \pm    0.0060$ & $ 451.577571 \pm    0.001213$ & $  2.7234 \pm   0.2149$ & $ 308.42 \pm   12.13$ \\
$  2219.4165 \pm    0.0123$ & $ 450.568874 \pm    0.002497$ & $  1.1831 \pm   0.1943$ & $ 323.79 \pm   25.16$ \\
$  2219.5751 \pm    0.0058$ & $ 450.536692 \pm    0.001180$ & $  3.1264 \pm   0.2419$ & $  44.14 \pm   11.81$ \\
$  2269.4384 \pm    0.0189$ & $ 440.637642 \pm    0.003674$ & $  0.6945 \pm   0.1754$ & $ 338.86 \pm   37.59$ \\
$  2302.3939 \pm    0.0160$ & $ 434.330549 \pm    0.003024$ & $  0.8768 \pm   0.1869$ & $ 306.06 \pm   31.42$ \\
$  2331.3456 \pm    0.0354$ & $ 428.936832 \pm    0.006513$ & $  0.4683 \pm   0.2034$ & $ 409.67 \pm   68.48$ \\
$  2334.0609 \pm    0.0108$ & $ 428.437841 \pm    0.001982$ & $  1.2824 \pm   0.1813$ & $  65.21 \pm   20.93$ \\
$  2354.4629 \pm    0.0461$ & $ 424.725320 \pm    0.008317$ & $  0.3600 \pm   0.2032$ & $  46.07 \pm   88.33$ \\
$  2364.9175 \pm    0.0198$ & $ 422.847741 \pm    0.003540$ & $  0.6717 \pm   0.1771$ & $  49.45 \pm   37.74$ \\
$  2371.2417 \pm    0.0103$ & $ 421.719989 \pm    0.001832$ & $  1.4474 \pm   0.1986$ & $ 260.20 \pm   19.58$ \\
$  2389.6684 \pm    0.0109$ & $ 418.468103 \pm    0.001911$ & $  1.4818 \pm   0.2150$ & $ 218.43 \pm   20.59$ \\
$  2407.1592 \pm    0.0171$ & $ 415.427442 \pm    0.002959$ & $  0.8055 \pm   0.1809$ & $  30.17 \pm   32.22$ \\
$  2433.7686 \pm    0.0149$ & $ 410.885402 \pm    0.002517$ & $  0.9678 \pm   0.1914$ & $ 195.63 \pm   27.65$ \\
$  2439.2207 \pm    0.0131$ & $ 409.967002 \pm    0.002198$ & $  1.0734 \pm   0.1868$ & $  73.11 \pm   24.18$ \\
$  2511.8868 \pm    0.0261$ & $ 398.107114 \pm    0.004131$ & $  0.5253 \pm   0.1749$ & $ 231.33 \pm   47.30$ \\
$  2517.4640 \pm    0.0267$ & $ 397.225150 \pm    0.004212$ & $  0.4874 \pm   0.1734$ & $ 374.09 \pm   47.73$ \\
$  2562.3537 \pm    0.0135$ & $ 390.266188 \pm    0.002054$ & $  1.1989 \pm   0.2149$ & $ 175.43 \pm   23.71$ \\
$  2584.3824 \pm    0.0304$ & $ 386.939638 \pm    0.004556$ & $  0.4486 \pm   0.1752$ & $ 380.99 \pm   53.65$ \\
$  2605.7493 \pm    0.0241$ & $ 383.766768 \pm    0.003543$ & $  0.6205 \pm   0.1995$ & $ 144.87 \pm   41.66$ \\
$  2654.4085 \pm    0.0235$ & $ 376.731762 \pm    0.003330$ & $  0.6139 \pm   0.1913$ & $ 150.64 \pm   39.87$ \\
$  2659.2838 \pm    0.0402$ & $ 376.041099 \pm    0.005684$ & $  0.3491 \pm   0.1871$ & $ 269.52 \pm   68.11$ \\
$  2717.2780 \pm    0.0238$ & $ 368.015347 \pm    0.003220$ & $  0.6278 \pm   0.1993$ & $  90.92 \pm   39.46$ \\
$  2773.3897 \pm    0.0482$ & $ 360.569597 \pm    0.006269$ & $  0.3378 \pm   0.1764$ & $ 227.32 \pm   82.35$ \\
$  2775.9713 \pm    0.0459$ & $ 360.234271 \pm    0.005954$ & $  0.3546 \pm   0.1762$ & $ 147.01 \pm   78.27$ \\
$  2786.2589 \pm    0.0308$ & $ 358.904199 \pm    0.003964$ & $  0.4555 \pm   0.1867$ & $ 277.29 \pm   49.77$ \\
$  2851.0184 \pm    0.0275$ & $ 350.751857 \pm    0.003387$ & $  0.5890 \pm   0.2152$ & $ 199.07 \pm   43.54$ \\
$  2903.0088 \pm    0.0634$ & $ 344.470198 \pm    0.007524$ & $  0.2267 \pm   0.1912$ & $ 181.87 \pm   98.45$ \\
$  2972.3323 \pm    0.0278$ & $ 336.436139 \pm    0.003150$ & $  0.5829 \pm   0.2150$ & $ 191.96 \pm   42.22$ \\
 \hline
 \end{longtable}

%% file: 3470tb15.tex
 

\scriptsize
 \begin{longtable} {cccc}
 \caption{\label{tab_T1989} Detected pulsation modes in the 1989 dataset
   ($T_{\rm max}$ computed from $T_o=244\,7593.0$ BCT). . } \\
\hline \hline 
Frequency  & Period    &  Amplitude &  ${\rm T}_{\rm max}$ \\
 ($\mu$Hz)&  (sec)  &   (mma)  &  (s)     \\
 \hline 
\endfirsthead
\caption{continued.}\\
 \hline \hline 
Frequency  & Period    &  Amplitude &  ${\rm T}_{\rm max}$ \\
 ($\mu$Hz)&  (sec)  &   (mma)  &  (s)     \\
 \hline 
\endhead
\hline
\endfoot

$  1017.6244 \pm    0.1505$ & $ 982.680884 \pm    0.145369$ & $  0.2355 \pm   0.0527$ & $ 835.02 \pm   83.61$ \\
$  1081.1478 \pm    0.1171$ & $ 924.942960 \pm    0.100212$ & $  0.2865 \pm   0.0518$ & $ 685.45 \pm   61.49$ \\
$  1083.0795 \pm    0.0576$ & $ 923.293233 \pm    0.049073$ & $  0.5750 \pm   0.0519$ & $ 578.15 \pm   30.48$ \\
$  1155.9663 \pm    0.1244$ & $ 865.077117 \pm    0.093109$ & $  0.2678 \pm   0.0522$ & $  49.06 \pm   62.12$ \\
$  1164.3414 \pm    0.1462$ & $ 858.854597 \pm    0.107846$ & $  0.2277 \pm   0.0523$ & $ 183.33 \pm   72.72$ \\
$  1186.3982 \pm    0.0364$ & $ 842.887344 \pm    0.025827$ & $  0.9129 \pm   0.0520$ & $ 621.86 \pm   17.59$ \\
$  1194.5258 \pm    0.0694$ & $ 837.152258 \pm    0.048667$ & $  0.4734 \pm   0.0514$ & $ 413.69 \pm   33.26$ \\
$  1219.5889 \pm    0.0412$ & $ 819.948424 \pm    0.027698$ & $  0.8064 \pm   0.0526$ & $ 736.66 \pm   19.47$ \\
$  1223.4084 \pm    0.1365$ & $ 817.388508 \pm    0.091173$ & $  0.2412 \pm   0.0514$ & $ 121.26 \pm   63.84$ \\
$  1227.6156 \pm    0.0757$ & $ 814.587257 \pm    0.050229$ & $  0.4336 \pm   0.0514$ & $ 645.39 \pm   35.33$ \\
$  1230.6595 \pm    0.0814$ & $ 812.572474 \pm    0.053740$ & $  0.4082 \pm   0.0526$ & $ 308.64 \pm   38.12$ \\
$  1276.8218 \pm    0.1238$ & $ 783.194636 \pm    0.075956$ & $  0.2729 \pm   0.0522$ & $ 447.87 \pm   55.98$ \\
$  1287.5897 \pm    0.0941$ & $ 776.644911 \pm    0.056747$ & $  0.3601 \pm   0.0521$ & $ 216.57 \pm   42.35$ \\
$  1292.4168 \pm    0.1075$ & $ 773.744170 \pm    0.064379$ & $  0.3091 \pm   0.0515$ & $ 256.44 \pm   47.87$ \\
$  1300.8603 \pm    0.1254$ & $ 768.722034 \pm    0.074088$ & $  0.2620 \pm   0.0514$ & $ 109.60 \pm   55.25$ \\
$  1323.8489 \pm    0.1114$ & $ 755.373222 \pm    0.063550$ & $  0.3060 \pm   0.0523$ & $ 522.84 \pm   47.98$ \\
$  1345.9740 \pm    0.2271$ & $ 742.956428 \pm    0.125357$ & $  0.1515 \pm   0.0529$ & $ 498.10 \pm   95.94$ \\
$  1359.3972 \pm    0.2012$ & $ 735.620174 \pm    0.108857$ & $  0.1695 \pm   0.0540$ & $ 610.86 \pm   83.76$ \\
$  1365.8060 \pm    0.1254$ & $ 732.168389 \pm    0.067237$ & $  0.2618 \pm   0.0514$ & $  72.66 \pm   52.63$ \\
$  1367.1423 \pm    0.0350$ & $ 731.452732 \pm    0.018711$ & $  0.9771 \pm   0.0524$ & $ 711.78 \pm   14.87$ \\
$  1370.8007 \pm    0.1154$ & $ 729.500659 \pm    0.061392$ & $  0.2903 \pm   0.0526$ & $ 237.90 \pm   48.05$ \\
$  1373.6916 \pm    0.1595$ & $ 727.965433 \pm    0.084543$ & $  0.2100 \pm   0.0523$ & $ 251.61 \pm   66.06$ \\
$  1375.3251 \pm    0.0473$ & $ 727.100800 \pm    0.025032$ & $  0.7227 \pm   0.0524$ & $   5.58 \pm   19.98$ \\
$  1416.5690 \pm    0.0441$ & $ 705.931037 \pm    0.021974$ & $  0.7513 \pm   0.0519$ & $ 314.76 \pm   17.84$ \\
$  1442.3849 \pm    0.1604$ & $ 693.296238 \pm    0.077092$ & $  0.2145 \pm   0.0523$ & $ 614.41 \pm   64.41$ \\
$  1449.7783 \pm    0.0676$ & $ 689.760631 \pm    0.032140$ & $  0.5101 \pm   0.0521$ & $ 157.26 \pm   26.92$ \\
$  1454.0459 \pm    0.0900$ & $ 687.736194 \pm    0.042572$ & $  0.3689 \pm   0.0524$ & $ 587.77 \pm   35.55$ \\
$  1458.1900 \pm    0.1912$ & $ 685.781680 \pm    0.089917$ & $  0.1725 \pm   0.0514$ & $ 340.92 \pm   75.00$ \\
$  1487.6353 \pm    0.2375$ & $ 672.207780 \pm    0.107307$ & $  0.1438 \pm   0.0536$ & $  21.18 \pm   91.14$ \\
$  1495.7875 \pm    0.0999$ & $ 668.544150 \pm    0.044649$ & $  0.3496 \pm   0.0547$ & $ 535.93 \pm   38.06$ \\
$  1496.8471 \pm    0.0556$ & $ 668.070922 \pm    0.024808$ & $  0.5948 \pm   0.0515$ & $ 417.71 \pm   21.26$ \\
$  1499.5242 \pm    0.1066$ & $ 666.878186 \pm    0.047388$ & $  0.3122 \pm   0.0522$ & $ 524.33 \pm   40.83$ \\
$  1505.0529 \pm    0.1092$ & $ 664.428487 \pm    0.048187$ & $  0.3120 \pm   0.0529$ & $ 387.82 \pm   41.10$ \\
$  1550.4193 \pm    0.0396$ & $ 644.986828 \pm    0.016481$ & $  0.8661 \pm   0.0532$ & $  47.76 \pm   14.54$ \\
$  1554.4871 \pm    0.0614$ & $ 643.298997 \pm    0.025423$ & $  0.5349 \pm   0.0515$ & $ 509.52 \pm   22.66$ \\
$  1558.7631 \pm    0.0334$ & $ 641.534310 \pm    0.013752$ & $  1.0265 \pm   0.0533$ & $ 553.31 \pm   12.20$ \\
$  1719.1686 \pm    0.0339$ & $ 581.676519 \pm    0.011483$ & $  0.9750 \pm   0.0519$ & $   0.77 \pm   11.31$ \\
$  1726.7450 \pm    0.1159$ & $ 579.124301 \pm    0.038885$ & $  0.2812 \pm   0.0511$ & $  26.09 \pm   38.52$ \\
$  1786.6284 \pm    0.0396$ & $ 559.713468 \pm    0.012405$ & $  0.8354 \pm   0.0519$ & $ 197.14 \pm   12.71$ \\
$  1790.6783 \pm    0.0139$ & $ 558.447616 \pm    0.004335$ & $  2.4897 \pm   0.0545$ & $ 275.26 \pm    4.46$ \\
$  1794.8854 \pm    0.0111$ & $ 557.138625 \pm    0.003434$ & $  3.1440 \pm   0.0546$ & $ 135.76 \pm    3.54$ \\
$  1804.0014 \pm    0.1315$ & $ 554.323292 \pm    0.040413$ & $  0.2488 \pm   0.0512$ & $  16.04 \pm   41.71$ \\
$  1824.8078 \pm    0.2048$ & $ 548.002915 \pm    0.061511$ & $  0.1593 \pm   0.0513$ & $ 293.96 \pm   64.32$ \\
$  1854.0658 \pm    0.0068$ & $ 539.355194 \pm    0.001982$ & $  5.3727 \pm   0.0564$ & $ 392.63 \pm    2.10$ \\
$  1858.2018 \pm    0.0091$ & $ 538.154698 \pm    0.002640$ & $  3.8176 \pm   0.0548$ & $ 173.63 \pm    2.83$ \\
$  1862.3605 \pm    0.0260$ & $ 536.952969 \pm    0.007486$ & $  1.4123 \pm   0.0563$ & $ 322.28 \pm    7.97$ \\
$  1899.8434 \pm    0.2042$ & $ 526.359180 \pm    0.056574$ & $  0.1600 \pm   0.0512$ & $ 353.43 \pm   61.65$ \\
$  1929.4276 \pm    0.0103$ & $ 518.288425 \pm    0.002764$ & $  3.5757 \pm   0.0562$ & $  60.53 \pm    3.05$ \\
$  1933.6139 \pm    0.0089$ & $ 517.166333 \pm    0.002383$ & $  3.8907 \pm   0.0548$ & $ 414.10 \pm    2.65$ \\
$  1937.7787 \pm    0.0060$ & $ 516.054807 \pm    0.001586$ & $  6.1725 \pm   0.0562$ & $  15.74 \pm    1.76$ \\
$  1953.1183 \pm    0.1119$ & $ 512.001750 \pm    0.029321$ & $  0.2939 \pm   0.0512$ & $ 192.75 \pm   32.84$ \\
$  1957.9498 \pm    0.1110$ & $ 510.738317 \pm    0.028949$ & $  0.2960 \pm   0.0513$ & $ 167.15 \pm   32.55$ \\
$  1994.3999 \pm    0.1183$ & $ 501.403964 \pm    0.029750$ & $  0.2910 \pm   0.0537$ & $ 499.42 \pm   33.86$ \\
$  1996.3986 \pm    0.1183$ & $ 500.901973 \pm    0.029689$ & $  0.2761 \pm   0.0511$ & $ 343.39 \pm   34.00$ \\
$  2005.0984 \pm    0.0683$ & $ 498.728633 \pm    0.016994$ & $  0.5148 \pm   0.0538$ & $ 253.86 \pm   19.45$ \\
$  2016.1480 \pm    0.1777$ & $ 495.995325 \pm    0.043722$ & $  0.1959 \pm   0.0539$ & $ 391.98 \pm   50.50$ \\
$  2020.7602 \pm    0.0728$ & $ 494.863279 \pm    0.017839$ & $  0.4635 \pm   0.0525$ & $ 276.16 \pm   20.54$ \\
$  2025.1181 \pm    0.0099$ & $ 493.798373 \pm    0.002425$ & $  3.4706 \pm   0.0541$ & $ 451.15 \pm    2.82$ \\
$  2033.1382 \pm    0.1143$ & $ 491.850489 \pm    0.027658$ & $  0.2931 \pm   0.0521$ & $ 452.71 \pm   32.04$ \\
$  2206.3680 \pm    0.0492$ & $ 453.233543 \pm    0.010101$ & $  0.6726 \pm   0.0519$ & $ 306.31 \pm   12.79$ \\
$  2210.2818 \pm    0.0103$ & $ 452.431007 \pm    0.002103$ & $  3.3597 \pm   0.0544$ & $ 288.90 \pm    2.68$ \\
$  2212.1173 \pm    0.0916$ & $ 452.055590 \pm    0.018725$ & $  0.3552 \pm   0.0510$ & $ 145.16 \pm   23.77$ \\
$  2214.3950 \pm    0.0093$ & $ 451.590617 \pm    0.001906$ & $  3.6952 \pm   0.0544$ & $ 112.64 \pm    2.43$ \\
$  2269.1586 \pm    0.0316$ & $ 440.691991 \pm    0.006138$ & $  1.0518 \pm   0.0521$ & $  54.68 \pm    8.00$ \\
$  2276.4452 \pm    0.0695$ & $ 439.281377 \pm    0.013402$ & $  0.4861 \pm   0.0528$ & $ 387.34 \pm   17.33$ \\
$  2290.8154 \pm    0.0842$ & $ 436.525798 \pm    0.016050$ & $  0.4088 \pm   0.0523$ & $ 181.89 \pm   20.83$ \\
$  2328.2752 \pm    0.0938$ & $ 429.502487 \pm    0.017313$ & $  0.3710 \pm   0.0539$ & $ 237.35 \pm   23.03$ \\
$  2338.8610 \pm    0.0181$ & $ 427.558544 \pm    0.003300$ & $  1.9225 \pm   0.0539$ & $ 202.62 \pm    4.42$ \\
$  2345.4819 \pm    0.1009$ & $ 426.351623 \pm    0.018341$ & $  0.3333 \pm   0.0519$ & $ 415.13 \pm   24.80$ \\
$  2352.7107 \pm    0.0825$ & $ 425.041635 \pm    0.014898$ & $  0.4074 \pm   0.0519$ & $ 409.23 \pm   20.20$ \\
$  2366.7507 \pm    0.0436$ & $ 422.520217 \pm    0.007785$ & $  0.7702 \pm   0.0525$ & $  14.13 \pm   10.51$ \\
$  2404.1168 \pm    0.0894$ & $ 415.953171 \pm    0.015460$ & $  0.3869 \pm   0.0536$ & $ 409.35 \pm   21.31$ \\
$  2406.1507 \pm    0.0287$ & $ 415.601567 \pm    0.004960$ & $  1.1933 \pm   0.0527$ & $  82.93 \pm    6.87$ \\
$  2408.5641 \pm    0.2037$ & $ 415.185136 \pm    0.035118$ & $  0.1609 \pm   0.0513$ & $ 151.70 \pm   48.51$ \\
$  2413.1890 \pm    0.0542$ & $ 414.389421 \pm    0.009312$ & $  0.6461 \pm   0.0539$ & $  62.29 \pm   12.94$ \\
$  2420.4775 \pm    0.1766$ & $ 413.141629 \pm    0.030138$ & $  0.1854 \pm   0.0513$ & $ 137.51 \pm   41.84$ \\
$  2427.1272 \pm    0.0547$ & $ 412.009726 \pm    0.009277$ & $  0.6025 \pm   0.0515$ & $ 123.19 \pm   12.90$ \\
$  2485.0991 \pm    0.1152$ & $ 402.398437 \pm    0.018661$ & $  0.2825 \pm   0.0510$ & $  91.61 \pm   26.60$ \\
$  2499.6344 \pm    0.0241$ & $ 400.058512 \pm    0.003859$ & $  1.4193 \pm   0.0532$ & $  69.66 \pm    5.53$ \\
$  2506.8053 \pm    0.0880$ & $ 398.914115 \pm    0.013999$ & $  0.3728 \pm   0.0514$ & $ 272.35 \pm   20.13$ \\
$  2562.1587 \pm    0.0329$ & $ 390.295893 \pm    0.005015$ & $  1.0058 \pm   0.0519$ & $ 269.36 \pm    7.37$ \\
$  2580.6640 \pm    0.1095$ & $ 387.497178 \pm    0.016437$ & $  0.2971 \pm   0.0510$ & $ 156.59 \pm   24.35$ \\
 \hline
 \end{longtable}

%% file: 3470tb16.tex
 

\scriptsize
 \begin{longtable} {cccc}
 \caption{\label{tab_T1993} Detected pulsation modes in the 1993 dataset
   	($T_{\rm max}$ computed from  $T_o=244\,6147.0$ BCT). } \\
\hline \hline
Frequency  & Period    &  Amplitude &  ${\rm T}_{\rm max}$ \\
 ($\mu$Hz)&  (sec)  &   (mma)  &  (s)     \\
 \hline 
\endfirsthead
\caption{continued.}\\
 \hline \hline 
Frequency  & Period    &  Amplitude &  ${\rm T}_{\rm max}$ \\
 ($\mu$Hz)&  (sec)  &   (mma)  &  (s)     \\
 \hline 
\endhead
\hline
\endfoot

$  1033.2862 \pm    0.0900$ & $ 967.786090 \pm    0.084282$ & $  0.3132 \pm   0.0709$ & $ 941.51 \pm   64.90$ \\
$  1034.1774 \pm    0.0382$ & $ 966.952057 \pm    0.035763$ & $  0.7283 \pm   0.0700$ & $ 156.13 \pm   27.45$ \\
$  1039.2650 \pm    0.0916$ & $ 962.218468 \pm    0.084791$ & $  0.3065 \pm   0.0709$ & $ 729.29 \pm   65.82$ \\
$  1058.2071 \pm    0.0793$ & $ 944.994582 \pm    0.070789$ & $  0.3558 \pm   0.0701$ & $ 272.82 \pm   55.76$ \\
$  1080.2204 \pm    0.0870$ & $ 925.737038 \pm    0.074575$ & $  0.3189 \pm   0.0697$ & $ 897.77 \pm   59.84$ \\
$  1139.4529 \pm    0.0912$ & $ 877.614173 \pm    0.070280$ & $  0.3044 \pm   0.0697$ & $ 254.65 \pm   59.55$ \\
$  1140.1251 \pm    0.0981$ & $ 877.096766 \pm    0.075488$ & $  0.2824 \pm   0.0695$ & $  37.96 \pm   63.94$ \\
$  1160.2915 \pm    0.0395$ & $ 861.852395 \pm    0.029347$ & $  0.7204 \pm   0.0709$ & $ 686.17 \pm   25.33$ \\
$  1163.2534 \pm    0.0501$ & $ 859.657957 \pm    0.037026$ & $  0.5567 \pm   0.0701$ & $ 808.54 \pm   31.95$ \\
$  1166.3591 \pm    0.0735$ & $ 857.368871 \pm    0.054024$ & $  0.4417 \pm   0.0723$ & $ 197.67 \pm   47.40$ \\
$  1167.3995 \pm    0.0511$ & $ 856.604768 \pm    0.037496$ & $  0.6338 \pm   0.0714$ & $ 438.91 \pm   32.65$ \\
$  1171.6597 \pm    0.0896$ & $ 853.490112 \pm    0.065293$ & $  0.3129 \pm   0.0702$ & $ 586.99 \pm   56.99$ \\
$  1192.4088 \pm    0.0450$ & $ 838.638565 \pm    0.031625$ & $  0.6224 \pm   0.0702$ & $ 220.03 \pm   28.11$ \\
$  1197.1341 \pm    0.0983$ & $ 835.328307 \pm    0.068583$ & $  0.2826 \pm   0.0697$ & $ 144.95 \pm   60.98$ \\
$  1220.1689 \pm    0.0430$ & $ 819.558646 \pm    0.028858$ & $  0.6508 \pm   0.0701$ & $ 552.22 \pm   26.15$ \\
$  1309.0639 \pm    0.0807$ & $ 763.904662 \pm    0.047121$ & $  0.3677 \pm   0.0726$ & $ 217.47 \pm   45.58$ \\
$  1314.8173 \pm    0.0994$ & $ 760.561941 \pm    0.057486$ & $  0.2815 \pm   0.0699$ & $ 186.80 \pm   56.28$ \\
$  1320.4297 \pm    0.1316$ & $ 757.329209 \pm    0.075477$ & $  0.2261 \pm   0.0771$ & $ 711.09 \pm   74.84$ \\
$  1324.0973 \pm    0.0857$ & $ 755.231490 \pm    0.048879$ & $  0.3268 \pm   0.0700$ & $ 333.45 \pm   48.18$ \\
$  1332.3277 \pm    0.0178$ & $ 750.566122 \pm    0.010052$ & $  1.6315 \pm   0.0780$ & $ 501.89 \pm   10.08$ \\
$  1344.2132 \pm    0.1346$ & $ 743.929621 \pm    0.074506$ & $  0.2142 \pm   0.0745$ & $ 184.53 \pm   75.63$ \\
$  1366.8532 \pm    0.0523$ & $ 731.607465 \pm    0.028017$ & $  0.5452 \pm   0.0704$ & $ 176.88 \pm   28.58$ \\
$  1370.3877 \pm    0.0768$ & $ 729.720507 \pm    0.040906$ & $  0.3614 \pm   0.0698$ & $ 138.18 \pm   41.62$ \\
$  1389.6408 \pm    0.0719$ & $ 719.610424 \pm    0.037228$ & $  0.3991 \pm   0.0707$ & $ 333.50 \pm   38.71$ \\
$  1400.9626 \pm    0.1245$ & $ 713.794935 \pm    0.063452$ & $  0.2251 \pm   0.0699$ & $ 229.35 \pm   65.67$ \\
$  1408.7119 \pm    0.1060$ & $ 709.868381 \pm    0.053430$ & $  0.2643 \pm   0.0698$ & $ 269.36 \pm   55.65$ \\
$  1410.3156 \pm    0.0759$ & $ 709.061159 \pm    0.038169$ & $  0.3700 \pm   0.0703$ & $ 468.66 \pm   40.06$ \\
$  1437.1959 \pm    0.1182$ & $ 695.799355 \pm    0.057216$ & $  0.2440 \pm   0.0713$ & $  45.98 \pm   61.72$ \\
$  1449.6613 \pm    0.0492$ & $ 689.816324 \pm    0.023429$ & $  0.6045 \pm   0.0708$ & $  88.77 \pm   24.74$ \\
$  1458.0250 \pm    0.0717$ & $ 685.859316 \pm    0.033749$ & $  0.3936 \pm   0.0709$ & $ 256.25 \pm   36.73$ \\
$  1460.8938 \pm    0.1669$ & $ 684.512460 \pm    0.078199$ & $  0.1779 \pm   0.0716$ & $ 490.11 \pm   84.06$ \\
$  1469.9053 \pm    0.0962$ & $ 680.315961 \pm    0.044504$ & $  0.2948 \pm   0.0711$ & $ 446.71 \pm   48.71$ \\
$  1474.0148 \pm    0.1176$ & $ 678.419251 \pm    0.054138$ & $  0.2410 \pm   0.0711$ & $  85.46 \pm   59.29$ \\
$  1537.7154 \pm    0.0856$ & $ 650.315397 \pm    0.036210$ & $  0.3291 \pm   0.0697$ & $ 455.33 \pm   41.56$ \\
$  1552.6555 \pm    0.0727$ & $ 644.057891 \pm    0.030169$ & $  0.3895 \pm   0.0700$ & $ 436.86 \pm   34.65$ \\
$  1559.0218 \pm    0.0801$ & $ 641.427846 \pm    0.032940$ & $  0.3544 \pm   0.0700$ & $ 225.65 \pm   38.03$ \\
$  1607.6603 \pm    0.0827$ & $ 622.021937 \pm    0.031990$ & $  0.3365 \pm   0.0698$ & $ 338.49 \pm   38.21$ \\
$  1723.1638 \pm    0.0566$ & $ 580.327893 \pm    0.019058$ & $  0.4983 \pm   0.0716$ & $ 387.76 \pm   24.44$ \\
$  1736.0197 \pm    0.1180$ & $ 576.030323 \pm    0.039138$ & $  0.2388 \pm   0.0709$ & $ 321.35 \pm   51.08$ \\
$  1774.7121 \pm    0.0795$ & $ 563.471691 \pm    0.025253$ & $  0.3791 \pm   0.0768$ & $ 509.13 \pm   33.46$ \\
$  1779.3249 \pm    0.0717$ & $ 562.010920 \pm    0.022644$ & $  0.4311 \pm   0.0775$ & $ 343.25 \pm   29.57$ \\
$  1786.4773 \pm    0.0250$ & $ 559.760815 \pm    0.007836$ & $  1.3773 \pm   0.0781$ & $ 283.92 \pm   10.60$ \\
$  1787.6437 \pm    0.0881$ & $ 559.395583 \pm    0.027580$ & $  0.3756 \pm   0.0763$ & $ 414.67 \pm   37.91$ \\
$  1788.9755 \pm    0.0786$ & $ 558.979164 \pm    0.024561$ & $  0.3969 \pm   0.0765$ & $ 374.54 \pm   33.25$ \\
$  1790.6871 \pm    0.0233$ & $ 558.444856 \pm    0.007271$ & $  1.5092 \pm   0.0840$ & $  52.03 \pm    9.39$ \\
$  1794.9656 \pm    0.0443$ & $ 557.113746 \pm    0.013750$ & $  0.6912 \pm   0.0755$ & $ 313.15 \pm   18.52$ \\
$  1796.4485 \pm    0.0803$ & $ 556.653867 \pm    0.024878$ & $  0.3801 \pm   0.0746$ & $ 271.90 \pm   33.10$ \\
$  1802.5758 \pm    0.2227$ & $ 554.761699 \pm    0.068546$ & $  0.1424 \pm   0.0772$ & $   9.31 \pm   88.33$ \\
$  1816.4675 \pm    0.0899$ & $ 550.519073 \pm    0.027252$ & $  0.3188 \pm   0.0711$ & $ 380.79 \pm   36.83$ \\
$  1837.1351 \pm    0.0689$ & $ 544.325793 \pm    0.020423$ & $  0.4111 \pm   0.0699$ & $ 360.34 \pm   27.97$ \\
$  1842.7054 \pm    0.1386$ & $ 542.680350 \pm    0.040818$ & $  0.2191 \pm   0.0741$ & $  68.66 \pm   55.07$ \\
$  1848.5585 \pm    0.0511$ & $ 540.962043 \pm    0.014962$ & $  0.5506 \pm   0.0705$ & $ 385.61 \pm   20.58$ \\
$  1854.0679 \pm    0.0041$ & $ 539.354562 \pm    0.001199$ & $  7.4667 \pm   0.0780$ & $ 272.24 \pm    1.65$ \\
$  1858.1632 \pm    0.0104$ & $ 538.165870 \pm    0.003017$ & $  2.8177 \pm   0.0732$ & $  93.17 \pm    4.21$ \\
$  1859.8665 \pm    0.0614$ & $ 537.672989 \pm    0.017757$ & $  0.4629 \pm   0.0701$ & $ 453.52 \pm   24.57$ \\
$  1862.2440 \pm    0.0238$ & $ 536.986550 \pm    0.006872$ & $  1.1779 \pm   0.0703$ & $ 130.11 \pm    9.50$ \\
$  1862.8139 \pm    0.0539$ & $ 536.822283 \pm    0.015526$ & $  0.5164 \pm   0.0698$ & $ 427.05 \pm   21.49$ \\
$  1864.3898 \pm    0.1090$ & $ 536.368511 \pm    0.031348$ & $  0.2625 \pm   0.0707$ & $ 416.62 \pm   43.72$ \\
$  1866.0015 \pm    0.2279$ & $ 535.905253 \pm    0.065441$ & $  0.1287 \pm   0.0754$ & $ 403.22 \pm   92.13$ \\
$  1866.8473 \pm    0.1130$ & $ 535.662450 \pm    0.032423$ & $  0.2520 \pm   0.0705$ & $ 492.49 \pm   45.25$ \\
$  1889.2619 \pm    0.0808$ & $ 529.307254 \pm    0.022637$ & $  0.3493 \pm   0.0703$ & $ 526.40 \pm   31.80$ \\
$  1908.3014 \pm    0.1193$ & $ 524.026237 \pm    0.032768$ & $  0.2359 \pm   0.0702$ & $ 496.60 \pm   46.44$ \\
$  1919.3508 \pm    0.0539$ & $ 521.009508 \pm    0.014619$ & $  0.5170 \pm   0.0701$ & $ 455.07 \pm   20.88$ \\
$  1921.9779 \pm    0.0676$ & $ 520.297335 \pm    0.018295$ & $  0.4526 \pm   0.0755$ & $ 468.85 \pm   25.53$ \\
$  1929.3992 \pm    0.0100$ & $ 518.296046 \pm    0.002682$ & $  3.1239 \pm   0.0755$ & $ 402.29 \pm    3.89$ \\
$  1930.8374 \pm    0.0694$ & $ 517.910003 \pm    0.018616$ & $  0.4047 \pm   0.0705$ & $ 513.52 \pm   26.76$ \\
$  1933.5527 \pm    0.0076$ & $ 517.182704 \pm    0.002026$ & $  4.2646 \pm   0.0778$ & $ 259.92 \pm    2.90$ \\
$  1934.2010 \pm    0.0439$ & $ 517.009347 \pm    0.011739$ & $  0.6335 \pm   0.0699$ & $  44.63 \pm   16.89$ \\
$  1937.7357 \pm    0.0054$ & $ 516.066256 \pm    0.001429$ & $  5.9070 \pm   0.0751$ & $ 430.56 \pm    2.07$ \\
$  1940.6333 \pm    0.0879$ & $ 515.295703 \pm    0.023332$ & $  0.3166 \pm   0.0701$ & $ 501.79 \pm   33.64$ \\
$  1941.6345 \pm    0.0526$ & $ 515.029999 \pm    0.013965$ & $  0.5783 \pm   0.0742$ & $ 512.29 \pm   20.39$ \\
$  1945.3247 \pm    0.0623$ & $ 514.053010 \pm    0.016466$ & $  0.4518 \pm   0.0709$ & $  84.25 \pm   23.85$ \\
$  1948.7083 \pm    0.0780$ & $ 513.160450 \pm    0.020534$ & $  0.3801 \pm   0.0720$ & $ 117.80 \pm   29.89$ \\
$  1953.2238 \pm    0.0469$ & $ 511.974107 \pm    0.012304$ & $  0.5927 \pm   0.0699$ & $ 505.16 \pm   17.87$ \\
$  1958.2977 \pm    0.0757$ & $ 510.647580 \pm    0.019732$ & $  0.3779 \pm   0.0707$ & $ 106.62 \pm   28.83$ \\
$  1960.5431 \pm    0.0623$ & $ 510.062738 \pm    0.016211$ & $  0.4656 \pm   0.0711$ & $ 238.26 \pm   23.85$ \\
$  1965.0174 \pm    0.0716$ & $ 508.901354 \pm    0.018543$ & $  0.3990 \pm   0.0712$ & $  91.53 \pm   27.24$ \\
$  1970.0741 \pm    0.0761$ & $ 507.595116 \pm    0.019606$ & $  0.3647 \pm   0.0696$ & $ 195.12 \pm   28.70$ \\
$  2020.9997 \pm    0.0385$ & $ 494.804625 \pm    0.009422$ & $  0.7243 \pm   0.0700$ & $ 431.68 \pm   14.15$ \\
$  2025.1355 \pm    0.0103$ & $ 493.794131 \pm    0.002508$ & $  2.7496 \pm   0.0709$ & $  83.23 \pm    3.77$ \\
$  2045.3755 \pm    0.1085$ & $ 488.907782 \pm    0.025935$ & $  0.2561 \pm   0.0697$ & $ 288.79 \pm   39.40$ \\
$  2206.1419 \pm    0.0188$ & $ 453.280005 \pm    0.003857$ & $  1.5642 \pm   0.0728$ & $ 307.96 \pm    6.29$ \\
$  2210.2752 \pm    0.0107$ & $ 452.432358 \pm    0.002184$ & $  2.7965 \pm   0.0740$ & $ 339.80 \pm    3.64$ \\
$  2211.7581 \pm    0.0757$ & $ 452.129006 \pm    0.015466$ & $  0.3696 \pm   0.0701$ & $ 156.57 \pm   25.47$ \\
$  2214.3454 \pm    0.0074$ & $ 451.600738 \pm    0.001513$ & $  3.9293 \pm   0.0723$ & $ 197.64 \pm    2.51$ \\
$  2216.2287 \pm    0.0570$ & $ 451.216963 \pm    0.011612$ & $  0.4867 \pm   0.0698$ & $ 134.57 \pm   19.13$ \\
$  2226.2960 \pm    0.0781$ & $ 449.176562 \pm    0.015747$ & $  0.3706 \pm   0.0708$ & $ 155.61 \pm   26.10$ \\
$  2227.9061 \pm    0.0578$ & $ 448.851945 \pm    0.011636$ & $  0.4996 \pm   0.0709$ & $ 178.28 \pm   19.33$ \\
$  2239.4576 \pm    0.0694$ & $ 446.536693 \pm    0.013838$ & $  0.3991 \pm   0.0696$ & $ 218.07 \pm   23.03$ \\
$  2269.4107 \pm    0.0224$ & $ 440.643032 \pm    0.004344$ & $  1.2648 \pm   0.0703$ & $ 115.71 \pm    7.30$ \\
$  2276.7283 \pm    0.0460$ & $ 439.226755 \pm    0.008873$ & $  0.6139 \pm   0.0701$ & $ 136.89 \pm   14.92$ \\
$  2290.5488 \pm    0.0389$ & $ 436.576592 \pm    0.007423$ & $  0.7237 \pm   0.0704$ & $ 247.51 \pm   12.62$ \\
$  2339.1130 \pm    0.0330$ & $ 427.512477 \pm    0.006032$ & $  0.8525 \pm   0.0705$ & $ 424.71 \pm   10.48$ \\
$  2346.0971 \pm    0.0353$ & $ 426.239820 \pm    0.006406$ & $  0.8032 \pm   0.0706$ & $ 195.92 \pm   11.12$ \\
$  2359.4700 \pm    0.0328$ & $ 423.824003 \pm    0.005897$ & $  0.8627 \pm   0.0706$ & $ 339.12 \pm   10.29$ \\
$  2366.5109 \pm    0.0154$ & $ 422.563024 \pm    0.002750$ & $  1.8229 \pm   0.0704$ & $ 117.34 \pm    4.84$ \\
$  2380.0856 \pm    0.0824$ & $ 420.152952 \pm    0.014551$ & $  0.3363 \pm   0.0696$ & $ 138.69 \pm   25.73$ \\
$  2406.3307 \pm    0.0235$ & $ 415.570477 \pm    0.004063$ & $  1.1894 \pm   0.0701$ & $ 114.34 \pm    7.26$ \\
$  2413.0358 \pm    0.0499$ & $ 414.415741 \pm    0.008570$ & $  0.5608 \pm   0.0701$ & $  70.22 \pm   15.35$ \\
$  2478.6019 \pm    0.0322$ & $ 403.453250 \pm    0.005237$ & $  0.8725 \pm   0.0703$ & $ 401.63 \pm    9.62$ \\
$  2485.4890 \pm    0.0320$ & $ 402.335323 \pm    0.005185$ & $  0.8750 \pm   0.0703$ & $ 317.17 \pm    9.56$ \\
$  2499.7528 \pm    0.0355$ & $ 400.039555 \pm    0.005687$ & $  0.7880 \pm   0.0705$ & $ 151.33 \pm   10.58$ \\
$  2559.0240 \pm    0.0998$ & $ 390.773989 \pm    0.015244$ & $  0.2774 \pm   0.0695$ & $ 203.20 \pm   28.99$ \\
 \hline
 \end{longtable}

%% file: 3470tb17.tex
 

\scriptsize
 \begin{longtable} {cccc}
 \caption{\label{tab_T2002} Detected pulsation modes in the 2002 dataset
   	($T_{\rm max}$ computed from $T_o=245\,2410.0$ BCT). } \\
\hline \hline 
Frequency  & Period    &  Amplitude &  ${\rm T}_{\rm max}$ \\
 ($\mu$Hz)&  (sec)  &   (mma)  &  (s)     \\
 \hline 
\endfirsthead
\caption{continued.}\\
 \hline \hline 
Frequency  & Period    &  Amplitude &  ${\rm T}_{\rm max}$ \\
 ($\mu$Hz)&  (sec)  &   (mma)  &  (s)     \\
 \hline 
\endhead
\hline
\endfoot

$  1070.6102 \pm    0.0861$ & $ 934.046767 \pm    0.075104$ & $  0.5357 \pm   0.0871$ & $ 107.39 \pm   50.77$ \\
$  1124.0253 \pm    0.1287$ & $ 889.659728 \pm    0.101866$ & $  0.3526 \pm   0.0859$ & $ 688.36 \pm   72.72$ \\
$  1139.3026 \pm    0.0857$ & $ 877.729971 \pm    0.066030$ & $  0.5280 \pm   0.0860$ & $ 747.58 \pm   47.88$ \\
$  1179.7990 \pm    0.0944$ & $ 847.601985 \pm    0.067817$ & $  0.4948 \pm   0.0887$ & $ 765.09 \pm   50.63$ \\
$  1224.8006 \pm    0.1001$ & $ 816.459446 \pm    0.066728$ & $  0.4508 \pm   0.0853$ & $ 599.17 \pm   51.69$ \\
$  1366.5982 \pm    0.1319$ & $ 731.743955 \pm    0.070626$ & $  0.3443 \pm   0.0860$ & $  74.59 \pm   61.10$ \\
$  1371.3576 \pm    0.1032$ & $ 729.204412 \pm    0.054864$ & $  0.5059 \pm   0.0935$ & $ 101.42 \pm   44.89$ \\
$  1374.7202 \pm    0.1422$ & $ 727.420768 \pm    0.075249$ & $  0.3224 \pm   0.0867$ & $ 487.34 \pm   65.33$ \\
$  1375.5050 \pm    0.0805$ & $ 727.005714 \pm    0.042522$ & $  0.5990 \pm   0.0909$ & $ 654.32 \pm   36.96$ \\
$  1405.2834 \pm    0.1118$ & $ 711.600227 \pm    0.056631$ & $  0.5233 \pm   0.0949$ & $ 417.45 \pm   46.17$ \\
$  1412.9375 \pm    0.1605$ & $ 707.745398 \pm    0.080389$ & $  0.2813 \pm   0.0852$ & $  25.32 \pm   71.73$ \\
$  1415.7909 \pm    0.1342$ & $ 706.318985 \pm    0.066956$ & $  0.4019 \pm   0.0911$ & $ 501.06 \pm   57.43$ \\
$  1435.0568 \pm    0.1252$ & $ 696.836548 \pm    0.060794$ & $  0.3665 \pm   0.0867$ & $ 100.40 \pm   55.05$ \\
$  1495.9336 \pm    0.1321$ & $ 668.478884 \pm    0.059015$ & $  0.3456 \pm   0.0853$ & $  55.45 \pm   55.49$ \\
$  1514.0709 \pm    0.1250$ & $ 660.471055 \pm    0.054510$ & $  0.3845 \pm   0.0893$ & $ 499.27 \pm   50.25$ \\
$  1536.4736 \pm    0.0983$ & $ 650.841008 \pm    0.041623$ & $  0.4947 \pm   0.0891$ & $ 540.22 \pm   38.64$ \\
$  1596.2317 \pm    0.1287$ & $ 626.475476 \pm    0.050510$ & $  0.3591 \pm   0.0876$ & $  53.34 \pm   50.99$ \\
$  1728.2481 \pm    0.1071$ & $ 578.620634 \pm    0.035841$ & $  0.4318 \pm   0.0868$ & $ 375.10 \pm   39.29$ \\
$  1780.0778 \pm    0.0828$ & $ 561.773192 \pm    0.026133$ & $  0.5547 \pm   0.0886$ & $  18.20 \pm   29.57$ \\
$  1786.5244 \pm    0.0824$ & $ 559.746053 \pm    0.025824$ & $  0.5639 \pm   0.0871$ & $ 434.82 \pm   29.19$ \\
$  1790.7357 \pm    0.0211$ & $ 558.429697 \pm    0.006571$ & $  2.2240 \pm   0.0886$ & $ 366.77 \pm    7.44$ \\
$  1790.9382 \pm    0.1458$ & $ 558.366561 \pm    0.045466$ & $  0.3097 \pm   0.0851$ & $ 366.77 \pm   51.36$ \\
$  1794.0625 \pm    0.0981$ & $ 557.394184 \pm    0.030472$ & $  0.4683 \pm   0.0865$ & $ 509.78 \pm   34.56$ \\
$  1794.8864 \pm    0.0447$ & $ 557.138323 \pm    0.013887$ & $  1.0415 \pm   0.0878$ & $ 315.93 \pm   15.75$ \\
$  1809.4518 \pm    0.0892$ & $ 552.653564 \pm    0.027232$ & $  0.5176 \pm   0.0875$ & $  94.29 \pm   31.06$ \\
$  1831.3343 \pm    0.0487$ & $ 546.049955 \pm    0.014514$ & $  0.9362 \pm   0.0861$ & $ 536.28 \pm   16.78$ \\
$  1836.2799 \pm    0.0951$ & $ 544.579286 \pm    0.028195$ & $  0.4844 \pm   0.0864$ & $  66.81 \pm   32.62$ \\
$  1837.1193 \pm    0.0794$ & $ 544.330474 \pm    0.023538$ & $  0.6341 \pm   0.0963$ & $ 125.39 \pm   27.38$ \\
$  1846.8182 \pm    0.0742$ & $ 541.471815 \pm    0.021759$ & $  0.6357 \pm   0.0904$ & $  88.89 \pm   24.95$ \\
$  1848.8445 \pm    0.2050$ & $ 540.878378 \pm    0.059977$ & $  0.2298 \pm   0.0869$ & $ 283.10 \pm   69.56$ \\
$  1853.6162 \pm    0.0910$ & $ 539.486009 \pm    0.026483$ & $  0.5071 \pm   0.0872$ & $ 154.66 \pm   31.02$ \\
$  1854.0242 \pm    0.0060$ & $ 539.367285 \pm    0.001760$ & $  8.3568 \pm   0.0943$ & $ 430.10 \pm    2.05$ \\
$  1858.1503 \pm    0.0161$ & $ 538.169592 \pm    0.004669$ & $  2.9824 \pm   0.0910$ & $ 199.88 \pm    5.48$ \\
$  1861.8588 \pm    0.0845$ & $ 537.097655 \pm    0.024364$ & $  0.5706 \pm   0.0946$ & $ 142.98 \pm   28.37$ \\
$  1862.1695 \pm    0.0440$ & $ 537.008057 \pm    0.012676$ & $  1.1613 \pm   0.0958$ & $  78.57 \pm   14.79$ \\
$  1880.2493 \pm    0.0451$ & $ 531.844376 \pm    0.012771$ & $  1.0293 \pm   0.0867$ & $ 405.53 \pm   15.23$ \\
$  1881.1465 \pm    0.0896$ & $ 531.590703 \pm    0.025327$ & $  0.5358 \pm   0.0916$ & $ 171.05 \pm   29.55$ \\
$  1896.0151 \pm    0.0791$ & $ 527.421973 \pm    0.022008$ & $  0.6424 \pm   0.0973$ & $ 117.94 \pm   25.63$ \\
$  1899.5102 \pm    0.0596$ & $ 526.451492 \pm    0.016518$ & $  0.7695 \pm   0.0866$ & $ 445.05 \pm   19.82$ \\
$  1924.3662 \pm    0.0694$ & $ 519.651606 \pm    0.018740$ & $  0.7102 \pm   0.0903$ & $ 197.67 \pm   23.19$ \\
$  1925.7273 \pm    0.1793$ & $ 519.284327 \pm    0.048361$ & $  0.2580 \pm   0.0864$ & $ 186.83 \pm   58.67$ \\
$  1925.7573 \pm    0.0850$ & $ 519.276242 \pm    0.022910$ & $  0.5794 \pm   0.0905$ & $ 174.21 \pm   28.38$ \\
$  1926.9414 \pm    0.1290$ & $ 518.957139 \pm    0.034731$ & $  0.3572 \pm   0.0863$ & $  49.20 \pm   42.41$ \\
$  1928.1807 \pm    0.1029$ & $ 518.623600 \pm    0.027670$ & $  0.4430 \pm   0.0862$ & $ 328.53 \pm   33.71$ \\
$  1929.3921 \pm    0.0287$ & $ 518.297980 \pm    0.007708$ & $  1.7091 \pm   0.0921$ & $ 100.33 \pm    9.27$ \\
$  1933.6410 \pm    0.0191$ & $ 517.159088 \pm    0.005107$ & $  2.4516 \pm   0.0887$ & $ 198.49 \pm    6.24$ \\
$  1937.1805 \pm    0.0951$ & $ 516.214164 \pm    0.025330$ & $  0.4936 \pm   0.0879$ & $  27.52 \pm   30.81$ \\
$  1937.5984 \pm    0.0070$ & $ 516.102831 \pm    0.001862$ & $  7.2537 \pm   0.0946$ & $ 434.69 \pm    2.28$ \\
$  1941.9776 \pm    0.0730$ & $ 514.938987 \pm    0.019356$ & $  0.6935 \pm   0.0946$ & $ 137.30 \pm   23.83$ \\
$  1942.7075 \pm    0.0851$ & $ 514.745523 \pm    0.022550$ & $  0.5492 \pm   0.0881$ & $ 139.88 \pm   27.53$ \\
$  2008.9093 \pm    0.0996$ & $ 497.782543 \pm    0.024687$ & $  0.4577 \pm   0.0863$ & $  32.45 \pm   31.29$ \\
$  2136.8990 \pm    0.1556$ & $ 467.967828 \pm    0.034076$ & $  0.2920 \pm   0.0860$ & $ 270.93 \pm   46.03$ \\
$  2170.5511 \pm    0.1300$ & $ 460.712495 \pm    0.027588$ & $  0.3551 \pm   0.0890$ & $  83.61 \pm   38.07$ \\
$  2172.6801 \pm    0.2086$ & $ 460.261035 \pm    0.044183$ & $  0.2266 \pm   0.0867$ & $ 188.87 \pm   60.42$ \\
$  2195.1020 \pm    0.1247$ & $ 455.559702 \pm    0.025882$ & $  0.3727 \pm   0.0895$ & $ 230.03 \pm   36.05$ \\
$  2198.5986 \pm    0.1632$ & $ 454.835188 \pm    0.033755$ & $  0.3698 \pm   0.1013$ & $ 219.14 \pm   41.48$ \\
$  2199.7496 \pm    0.1770$ & $ 454.597206 \pm    0.036576$ & $  0.3172 \pm   0.0900$ & $ 151.45 \pm   50.00$ \\
$  2203.3154 \pm    0.0909$ & $ 453.861480 \pm    0.018723$ & $  0.6159 \pm   0.0987$ & $ 211.71 \pm   25.04$ \\
$  2206.1699 \pm    0.0319$ & $ 453.274258 \pm    0.006552$ & $  1.4671 \pm   0.0886$ & $ 243.59 \pm    9.15$ \\
$  2208.6948 \pm    0.1069$ & $ 452.756075 \pm    0.021920$ & $  0.4326 \pm   0.0876$ & $ 280.92 \pm   30.53$ \\
$  2209.8242 \pm    0.1304$ & $ 452.524684 \pm    0.026693$ & $  0.4560 \pm   0.1021$ & $ 204.12 \pm   33.23$ \\
$  2210.2000 \pm    0.0173$ & $ 452.447752 \pm    0.003544$ & $  2.7767 \pm   0.0911$ & $  75.01 \pm    4.95$ \\
$  2212.4681 \pm    0.1416$ & $ 451.983909 \pm    0.028921$ & $  0.3813 \pm   0.0901$ & $ 422.65 \pm   40.78$ \\
$  2214.4441 \pm    0.0146$ & $ 451.580605 \pm    0.002985$ & $  3.3943 \pm   0.0941$ & $ 171.35 \pm    4.18$ \\
$  2215.3478 \pm    0.1491$ & $ 451.396388 \pm    0.030380$ & $  0.3748 \pm   0.0987$ & $ 328.11 \pm   40.86$ \\
$  2224.9708 \pm    0.1838$ & $ 449.444093 \pm    0.037129$ & $  0.2551 \pm   0.0880$ & $ 227.54 \pm   51.97$ \\
$  2232.6552 \pm    0.1923$ & $ 447.897205 \pm    0.038583$ & $  0.2390 \pm   0.0869$ & $ 317.15 \pm   54.29$ \\
$  2270.2367 \pm    0.0637$ & $ 440.482714 \pm    0.012354$ & $  0.7243 \pm   0.0872$ & $  50.70 \pm   17.74$ \\
$  2276.6719 \pm    0.0774$ & $ 439.237640 \pm    0.014935$ & $  0.5979 \pm   0.0876$ & $ 254.15 \pm   21.51$ \\
$  2290.7585 \pm    0.1186$ & $ 436.536633 \pm    0.022598$ & $  0.3902 \pm   0.0876$ & $  42.62 \pm   32.67$ \\
$  2303.3819 \pm    0.0951$ & $ 434.144256 \pm    0.017933$ & $  0.4816 \pm   0.0866$ & $ 142.89 \pm   26.22$ \\
$  2312.8515 \pm    0.1140$ & $ 432.366722 \pm    0.021303$ & $  0.4549 \pm   0.0872$ & $  87.10 \pm   29.16$ \\
$  2323.5281 \pm    0.2197$ & $ 430.379995 \pm    0.040693$ & $  0.2365 \pm   0.0863$ & $ 238.33 \pm   56.16$ \\
$  2338.6337 \pm    0.0376$ & $ 427.600100 \pm    0.006869$ & $  1.3643 \pm   0.0940$ & $ 150.88 \pm   10.07$ \\
$  2339.5492 \pm    0.2815$ & $ 427.432772 \pm    0.051422$ & $  0.1616 \pm   0.0861$ & $ 260.01 \pm   76.26$ \\
$  2344.2511 \pm    0.1114$ & $ 426.575472 \pm    0.020270$ & $  0.4168 \pm   0.0878$ & $ 163.77 \pm   30.01$ \\
$  2345.7248 \pm    0.0492$ & $ 426.307468 \pm    0.008938$ & $  0.9784 \pm   0.0909$ & $ 327.08 \pm   13.27$ \\
$  2358.0435 \pm    0.0825$ & $ 424.080388 \pm    0.014837$ & $  0.5594 \pm   0.0872$ & $ 316.62 \pm   22.14$ \\
$  2366.4338 \pm    0.0313$ & $ 422.576783 \pm    0.005580$ & $  1.6362 \pm   0.0941$ & $ 105.47 \pm    8.29$ \\
$  2376.0473 \pm    0.1284$ & $ 420.867035 \pm    0.022749$ & $  0.3506 \pm   0.0855$ & $ 245.65 \pm   34.21$ \\
$  2405.2427 \pm    0.0894$ & $ 415.758461 \pm    0.015461$ & $  0.5093 \pm   0.0860$ & $ 105.21 \pm   23.51$ \\
$  2413.5025 \pm    0.0662$ & $ 414.335592 \pm    0.011373$ & $  0.6977 \pm   0.0871$ & $ 130.27 \pm   17.33$ \\
$  2418.2414 \pm    0.0802$ & $ 413.523652 \pm    0.013719$ & $  0.5829 \pm   0.0887$ & $ 228.47 \pm   20.99$ \\
$  2426.9190 \pm    0.0718$ & $ 412.045069 \pm    0.012195$ & $  0.6427 \pm   0.0876$ & $  61.36 \pm   18.76$ \\
$  2454.1297 \pm    0.1375$ & $ 407.476435 \pm    0.022832$ & $  0.3532 \pm   0.0881$ & $  52.49 \pm   36.61$ \\
$  2466.5843 \pm    0.1430$ & $ 405.418952 \pm    0.023503$ & $  0.3630 \pm   0.0930$ & $ 308.39 \pm   36.59$ \\
$  2477.0475 \pm    0.0917$ & $ 403.706433 \pm    0.014943$ & $  0.5014 \pm   0.0871$ & $ 395.12 \pm   23.41$ \\
$  2493.0801 \pm    0.1250$ & $ 401.110256 \pm    0.020111$ & $  0.3771 \pm   0.0896$ & $ 275.03 \pm   31.96$ \\
$  2494.2523 \pm    0.1651$ & $ 400.921744 \pm    0.026537$ & $  0.2889 \pm   0.0877$ & $ 253.69 \pm   41.98$ \\
$  2499.7609 \pm    0.0753$ & $ 400.038258 \pm    0.012052$ & $  0.6148 \pm   0.0875$ & $ 213.83 \pm   19.05$ \\
$  2505.3706 \pm    0.0700$ & $ 399.142542 \pm    0.011157$ & $  0.6596 \pm   0.0870$ & $  67.48 \pm   17.59$ \\
$  2517.2488 \pm    0.1433$ & $ 397.259101 \pm    0.022617$ & $  0.3349 \pm   0.0902$ & $ 226.23 \pm   36.19$ \\
$  2518.7564 \pm    0.1643$ & $ 397.021324 \pm    0.025902$ & $  0.2769 \pm   0.0860$ & $ 294.01 \pm   41.27$ \\
$  2527.2090 \pm    0.1941$ & $ 395.693427 \pm    0.030384$ & $  0.2460 \pm   0.0872$ & $ 329.32 \pm   47.98$ \\
$  2547.8652 \pm    0.1658$ & $ 392.485442 \pm    0.025543$ & $  0.2790 \pm   0.0877$ & $  79.07 \pm   41.68$ \\
$  2557.9851 \pm    0.3118$ & $ 390.932688 \pm    0.047647$ & $  0.1483 \pm   0.0877$ & $ 232.22 \pm   77.72$ \\
$  2561.8125 \pm    0.2072$ & $ 390.348636 \pm    0.031567$ & $  0.2224 \pm   0.0872$ & $  49.50 \pm   51.04$ \\
$  2580.7262 \pm    0.1065$ & $ 387.487827 \pm    0.015996$ & $  0.4336 \pm   0.0871$ & $  61.83 \pm   26.00$ \\
$  2642.6779 \pm    0.1131$ & $ 378.404050 \pm    0.016191$ & $  0.4069 \pm   0.0871$ & $ 212.25 \pm   27.04$ \\
$  2647.3574 \pm    0.0767$ & $ 377.735169 \pm    0.010950$ & $  0.6011 \pm   0.0871$ & $ 225.95 \pm   18.32$ \\
$  2675.6539 \pm    0.0842$ & $ 373.740420 \pm    0.011765$ & $  0.5884 \pm   0.0939$ & $ 161.95 \pm   19.92$ \\
$  2799.7842 \pm    0.1911$ & $ 357.170391 \pm    0.024379$ & $  0.2453 \pm   0.0885$ & $  87.83 \pm   43.21$ \\
$  2829.5995 \pm    0.1772$ & $ 353.406907 \pm    0.022129$ & $  0.2588 \pm   0.0868$ & $  74.56 \pm   39.59$ \\
$  2881.6572 \pm    0.1588$ & $ 347.022543 \pm    0.019122$ & $  0.2872 \pm   0.0864$ & $ 244.89 \pm   34.85$ \\
$  3560.1456 \pm    0.1034$ & $ 280.887389 \pm    0.008155$ & $  0.4369 \pm   0.0857$ & $  96.01 \pm   18.38$ \\
 \hline
 \end{longtable}

%% file: 3470.bbl
\begin{thebibliography}{}

  \bibitem[1986]{barstow86} Barstow, M. A., Holberg, J. B., Grauer, A. D. and Winget, D. E. 1986,
      \apj, 306, L25

  \bibitem[1991]{bradley91} Bradley, P. A. and Winget, D.~E. 1991, \apjs, 75, 463

  \bibitem[1992]{brassard92} Brassard, P., Fountaine, G., Wesemael, F. and Tassoul, M. 1999, \apjs, 81, 747

  \bibitem[1993]{breger93} Breger, M., Handler, G. 1993, Balt. Ast., 2, 468

  \bibitem[1990]{brickhill90} Brickhill, A. J. 1990, \mnras, 246, 510

  \bibitem[1992]{brickhill92} Brickhill, A. J. 1992, \mnras, 259, 519

  \bibitem[1993]{bruvold93} Bruvold, A., 1993, Balt. Ast., 2, 530

  \bibitem[2006]{corsico2006} C\'orsico, A. H., Althaus, L. G. and Miller Bertolami, M. M. 2006, \aap, 458, 259

  \bibitem[2007]{corsico2007} C\'orsico, A. H. et al.  in preparation.

  \bibitem[1995]{costa95} Costa, J. E. S. and Kepler, S. O., 1995, Balt. Ast., 4, 334

  \bibitem[1999]{costa99} Costa, J. E. S. and Kepler, S. 0., 1999, Balt. Ast., 9, 451

  \bibitem[1999]{costa99b} Costa, J. E. S., Kepler, S. O. and Winget, D. E. 1999, \apj, 522, 973

  \bibitem[2003]{costa2003} Costa, J. E. S., Kepler, S. O., Winget, D. E., O'Brien, M. S., Bond, H. E.,
       Kawaler, S. D. and Dreizler, S. 2003, Balt. Ast. , 12, 23

  \bibitem[2007]{costa2007} Costa, J. E. S. and Kepler, S. O. 2007, in preparation.

  \bibitem[1998]{dreizler98} Dreizler, S. and Heber, U. 1998, \aap, 334, 618


  \bibitem[1999]{goldreich99a} Goldreich, P. and Wu, Y. 1999a, AJ, 511, 904

  \bibitem[1999]{goldreich99B} Goldreich, P. and Wu, Y. 1999B, AJ, 523, 805

  \bibitem[1986]{green} Green, R. F., Schmidt, M. and Liebert, J. 
      1986, \apj, 61, 305

  \bibitem[2007]{jahn2007} Jahn, D., Rauch, T., Reiff, E., Werner, K., Kruk, J. W. and Herwig, F.
      2007, \aap, 462, 281 

  \bibitem[1989]{jones1989} Jones, P.W., Pesnell, W.D., Hansen, C.J. and Kawaler, S.D.
      1989, \apj, 336, 403

 \bibitem[1986]{kawaler86} Kawaler, S. D. 1986, PhD. Thesis. Univ. Texas at Austin

  \bibitem[1988]{kawaler88} Kawaler, S. D. 1988, \apj, 334, 220

  \bibitem[1990]{kawaler90} Kawaler, S. D. and Weiss, P. 1990, in {\it Lecture Notes in Physics},
      367, 431

  \bibitem[1994]{kawaler94} Kawaler, S. D. and Bradley, P. A. 1994, \apj, 427, 415 (KB94)

  \bibitem[1999]{kawaler99} Kawaler, S. D., Sekii, T., and Gough, D. 1999, \apj, 516, 349

  \bibitem[1993]{kepler93} Kepler, S. 0. 1993, Balt. Ast., 2, 515

  \bibitem[1995]{kepler95} Kepler, S. O.,  Giovannini, O., Wood, M. A., Nather, R. E., 
      Winget, D. E., Kanaan, A., Kleinman, S. J., Bradley, P. A., Provencal, J. L., 
      Clemens, J. C., Claver, C. F., Watson, T. K., Yanagida, K., Krisciunas, K., 
      Marar, T. M. K., Seetha, S., Ashoka, B. N., Leibowitz, E., Mendelson, H., 
      Mazeh, T., Moskalik, P., Krzesinski, J., Pajdosz, G., Zola, S., Solheim, J.-E., 
      Emanuelsen, P.-I., Dolez, N., Vauclair, G., Chevreton, M., Fremy, J.-R., 
      Barstow, M. A., Sansom, A. E., Tweedy, R. W., Wickramasinghe, D. T., 
      Ferrario, L., Sullivan, D. J., van der Peet, A. J., Buckley, D. A. H. and Chen, A.-L.
      1995, \apj, 447, 874

  \bibitem[2003]{kepler2003} Kepler, S. O., Nather, R. E., Winget, D. E., Nitta, A., Kleinman, S. J., 
      Metcalfe, T., Sekiguchi, K., Xiaojun, Jiang, Sullivan, D., Sullivan, T., Janulis, R., Meistas, E., 
      Kalytis, R., Krzesinski, J., Ogoza, W., Zola, S., O'Donoghue, D., Romero-Colmenero, E., Martinez, P., 
      Dreizler, S., Deetjen, J., Nagel, T., Schuh, S. L., Vauclair, G., Ning, Fu Jian, Chevreton, M., 
      Solheim, J.-E., Gonzalez Perez, J. M., Johannessen, F., Kanaan, A., Costa, J. E. S., Murillo Costa, A. F., 
      Wood, M. A., Silvestri, N., Ahrens, T. J., Jones, A. K., Collins, A. E., Boyer, M., Shaw, J. S., 
      Mukadam, A., Klumpe, E. W., Larrison, J., Kawaler, S., Riddle, R., Ulla, A. and Bradley, P.
      2003, \aap, 401, 639 

  \bibitem[1997]{Kuschnig1997} Kuschnig, R., Weiss, W. W., Gruber, R., Bely, P. Y. and Jenkner, H.
      1997, \aap, 328, 544

  \bibitem[1979]{macgraw} McGraw, J. T., Starrfield, S. G., Liebert, J. and Green, R. F.
      1979, in: White dwarfs and variable degenerate stars.
      (Rochester, NY) 377

  \bibitem[2006]{miller2006} Miller Bertolami, M. M. and Althaus, L. G. 2006, \aap, 454, 845


  \bibitem[2005]{montgomery2005} Montgomery, M. H. 2005, \apj, 633, 1142

  \bibitem[1990]{nather90} Nather, R. E., Winget, D. E., Clemens, J. C., Hansen, C. J. and Hine, B. P.
      1990, \apj, 361, 309

  \bibitem[1985]{pesnell85} Pesnell, W. D. 1985, \apj, 292, 238

  \bibitem[1996]{pfeiffer96} Pfeiffer, B., Vauclair, G., Dolez, N. et al. 1996, \aap, 314, 182

  \bibitem[1986]{press86} Press, W. H., Teukolsky, S. A., Vetterling, W. T. and Flannery, B. P.
      1996, in {\it Numerical recipes in FORTRAN: the art of scientific computing}, 2.ed.

  \bibitem[1982]{robinson82} Robinson, E. L., Kepler, S. O. and Nather, R. E.
      1982, \apj, 259, 219

  \bibitem[1994]{sakurai94} Sakurai, J. J. 1994, Modern  Quantum Mechanics. Addison-Wesley.
      ISBN 0-201-53929-2. pp. 104-109.

  \bibitem[1982]{scargle82} Scargle, J. D. 1982, \apj, 263, 835

  \bibitem[1991]{schwarzenberg91} Schwarzenberg-Czerny, A. 1991, \mnras, 253, 198

  \bibitem[1999]{schwarzenberg99} Schwarzenberg-Czerny, A. 1999, \mnras, 516, 315

  \bibitem[1988]{shibahashi88} Shibahashi, H. 1988, in Advances in Helio and Asteroseismology.
    IAU Symp. 123, p.133

  \bibitem[1985]{sion85} Sion, E. M., Liebert, J. and Starrfield, S. G. 1985, \apj, 292, 471

  \bibitem[1989]{unno89} Unno, W., Osaki, Y., Ando, H., Saio, H., Shibahashi, H. 1989 
   Nonradial Oscillations of Stars, 2nd ed., Univ. Tokyo, Tokyo, p. 378

  \bibitem[2002]{vauclair02} Vauclair, G., Moskalik, P., Pfeiffer, 
    B., Chevreton, M., Dolez, N., Serre, B., Kleinman, S. J., Barstow, M., Sansom, A. E., 
    Solheim, J.-E., Belmonte, J. A., Kawaler, S. D., Kepler, S. O., Kanaan, A., Giovannini, O., 
    Winget, D. E., Watson, T. K., Nather, R. E., Clemens, J. C., Provencal, J., Dixson, J. S., 
    Yanagida, K., Nitta Kleinman, A., Montgomery, M., Klumpe, E. W., Bruvold, A., O'Brien, M. S., 
    Hansen, C. J., Grauer, A. D., Bradley, P. A., Wood, M. A., Achilleos, N., Jiang, S. Y., 
    Fu, J. N., Marar, T. M. K., Ashoka, B. N., Meistas, E. G., Chernyshev, A. V., Mazeh, T., 
    Leibowitz, E., Hemar, S., Krzezski, J., Pajdosz, G. and Zola, S.  
    2002, \aap, 381, 122

  \bibitem[1999]{vuille00} Vuille, F. 2000, Balt. Ast., 9, 33

  \bibitem[1982]{wegner82} Wegner, G., Barry, D. C., Holberg, J. B., Forrester, W. T. and McGraw, J. T. 
      1982, Bull. \aaps, 14, 914

  \bibitem[2003]{weidner03} Weidner, C. and Koester, D. 2003, \aap, 405, 657

  \bibitem[1995]{werner95} Werner, K. 1995, Balt. Ast., 4, 340 


  \bibitem[2006]{werner06} Werner, K. and Herwig, F. 2006, \pasp, 118, 183

  \bibitem[1985]{winget85} Winget, D.E., Kepler, S.O., Robinson, E.L. and Nather, R.E. 1985, \apj, 292, 606

  \bibitem[1991]{winget91} Winget, D.E., Nather, R.E., Clemens, J.C., Provencal, J.L., Kleinman, S.J., 
      Bradley, P.A., Wood, M.A., Claver, C.F., Frueh, M.L., Grauer, A.D., Hine, B.P., Hansen, C.J., 
      Fontaine, G., Achilleos, N., Wickramasinghe, D.T., Marar, T.M.K., Seetha, S., Ashoka, B.N., 
      O'Donoghue, D., Warner, B., Kurtz, D.W., Buckley, D.A., Brickhill, J., Vauclair, G., Dolez, N., 
      Chevreton,M., Barstow,M.A., Solheim, J.E., Kanaan, A., 
      Kepler, S.O., Henry, G.W. and Kawaler, S.D.
      1991, \apj, 378, 326 (W91) 


\end{thebibliography}
